\documentclass[a4paper,fleqn,usenatbib,useAMS,times]{mnras}
\usepackage{graphicx}	
\usepackage{amsmath}	
\usepackage{amssymb}	
\usepackage{multicol}   
\usepackage{bm}		
\usepackage{pdflscape}	
\usepackage[T1]{fontenc}
\usepackage{ae,aecompl}
\usepackage{newtxtext,newtxmath} 
\usepackage{fancyhdr}  
\usepackage{supertabular}
\usepackage{longtable}
\usepackage{url}
\usepackage{dcolumn}
\usepackage{morefloats}	 
\usepackage{rotating,caption}  	 
\usepackage{tikz}    	 
\usepackage{adjustbox}
\usepackage{blindtext} 
\usepackage{natbib} 
\usepackage{threeparttable} 
\usepackage{pdflscape} 
\usepackage{tabularx}
\usepackage{cleveref}    
\newcommand{\sqcm}{cm$^{-2}$}  
\newcommand{\lya}{Ly$\alpha$}
\newcommand{\lyb}{Ly$\beta$}

\newcommand{\HI}{\mbox{H\,{\sc i}}}

\newcommand{\OII}{\mbox{O\,{\sc ii}}}
\newcommand{\OIII}{\mbox{O\,{\sc iii}}}

\newcommand{\OVI}{\mbox{O\,{\sc vi}}}

\newcommand{\CIV}{\mbox{C\,{\sc iv}}}

\newcommand{\CIII}{\mbox{C\,{\sc iii}}}

\newcommand{\SiII}{\mbox{Si\,{\sc ii}}}

\newcommand{\SiIV}{\mbox{Si\,{\sc iv}}}

\newcommand{\MgII}{\mbox{Mg\,{\sc ii}}}

\newcommand{\zlae}{$z_{\rm peak}$} 
\newcommand{\zcorr}{$z_{\rm corr}$} 
\newcommand{\voff}{$V_{\rm offset}$}

\newcommand{\degree}{\ensuremath{^\circ}}
\newcommand{\angstrom}{\text{\normalfont\AA}} 

\newcommand{\be}{\begin{equation}}
\newcommand{\en}{\end{equation}}

\def\kms{km~s$^{-1}$}

\def\Msun{$\rm M_{\odot}$} 
  
\title[MUSEQuBES: Characterizing the CGM of $z\approx3.3$ Ly$\alpha$ Emitters]{     
\vskip-0.5cm 
{\LARGE MUSEQuBES: Characterizing the circumgalactic medium of redshift $\approx3.3$ Ly$\alpha$ emitters}}           
\author[S. Muzahid et al.]
{
\parbox{\textwidth}{\vskip-0.5cm  
Sowgat Muzahid,$^{1,2}$\thanks{E-mail: sowgat@iucaa.in}    
Joop Schaye,$^{3}$
Sebastiano Cantalupo,$^{4,5}$ 
Raffaella Anna Marino,$^{5}$ 
Nicolas F. Bouch{\'e},$^{6}$   
Sean Johnson,$^{7}$  
Michael Maseda,$^{3}$ 
Martin Wendt,$^{8, 2}$  
Lutz Wisotzki,$^{2}$ 
and  
Johannes Zabl$^{6,9}$         
} 
\vspace*{8pt}\\   
$^{1}$ IUCAA, Post Bag 04, Ganeshkhind, Pune-411007, India \\  
$^{2}$ Leibniz-Institute for Astrophysics Potsdam (AIP), An der Sternwarte 16, D-14482 Potsdam, Germany \\ 
$^{3}$ Leiden Observatory, Leiden University, PO Box 9513, NL-2300 RA Leiden, the Netherlands  \\  
$^{4}$ Department of Physics, University of Milan Bicocca, Piazza della Scienza 3, I-20126 Milano, Italy \\ 
$^{5}$ Department of Physics, ETH Z$\ddot{u}$rich, Wolfgang-Pauli-Strasse 27, 8093 Z$\ddot{u}$rich, Switzerland \\
$^{6}$ Univ Lyon, Univ Lyon1, Ens de Lyon, CNRS, Centre de Recherche Astrophysique de Lyon UMR5574, F-69230, Saint-Genis-Laval, France \\ 
$^{7}$ Department of Astronomy, University of Michigan, 1085 S. University Ave, Ann Arbor, MI 48109, USA \\ 
$^{8}$ Institut f$\ddot{u}$r Physik und Astronomie, Universit$\ddot{a}$t Potsdam, Karl-Liebknecht-Str 24/25, D-14476 Golm, Germany \\   
$^{9}$ Department of Astronomy \& Physics, Saint Mary's University, 923 Robie Street, Halifax, Nova Scotia, B3H 3C3, Canada  \\
} 
\date{Accepted. Received; in original form} 
\pubyear{2021}
\begin{document}
\label{firstpage}
\pagerange{\pageref{firstpage}--\pageref{lastpage}}   
\maketitle   
\begin {abstract} \\ 
\noindent
We present the first characterization of the circumgalactic medium of \lya\ emitters (LAEs), using a sample of 96 $z\approx3.3$ LAEs detected with the VLT/MUSE in fields centered on 8 bright background quasars. The LAEs have low Ly$\alpha$ luminosities ($\sim 10^{42}\,\text{erg}\,\text{s}^{-1}$) and star formation rates (SFRs) $\sim 1~\text{M}_\odot\,\text{yr}^{-1}$, which for main sequence galaxies corresponds to stellar masses of only $\sim 10^{8.6}\,\text{M}_\odot$. The median transverse distance between the LAEs and the quasar sightlines is 165 proper kpc (pkpc). We stacked the high-resolution quasar spectra and measured significant excess \HI\ and \CIV\ absorption near the LAEs out to 500~$\text{km}\,\text{s}^{-1}$ and at least $\approx 250$ pkpc (corresponding to $\approx 7$ virial radii). At $\lesssim 30~\text{km}\,\text{s}^{-1}$ from the galaxies the median \HI\ and \CIV\ optical depths are enhanced by an order of magnitude. The absorption is significantly stronger around the $\approx 1/3$ of our LAEs that are part of `groups', which we attribute to the large-scale structures in which they are embedded. We do not detect any strong dependence of either the \HI\ or \CIV\ absorption on transverse distance (over the range $\approx 50-250$~pkpc), redshift, or the properties of the \lya\ emission line (luminosity, full width at half maximum, or equivalent width). However, for \HI, but not \CIV, the absorption at $\lesssim 100\,\text{km}\,\text{s}^{-1}$ from the LAE does increase with the SFR. This suggests that LAEs surrounded by more \HI\ tend to have higher SFRs.

\end {abstract}

\section{Introduction} 
\label{sec:intro}

The circumgalactic medium (CGM) is the reservoir of gas and metals in the immediate vicinity of galaxies. The physical and chemical properties of the CGM retain imprints of gas infall and outflow processes that regulate galaxy evolution. Owing to its extremely low density, direct detection of the bulk of the CGM in emission is not yet possible. However, it has recently become feasible to study the densest part of the CGM of high-redshift galaxies. For example, extended \lya\ emission is detected over a few tens of kpc around normal $z>3$ \lya\ emitting galaxies \citep[e.g.,][]{Wisotzki16,Leclercq17} and over a few hundreds of kpc around quasar host-galaxies \citep[e.g.,][]{Cantalupo14,Borisova16,Cai19}, a protocluster \citep{Umehata19} and groups of \lya\ emitters \citep{Bacon21}. 

Absorption line spectroscopy of bright background sources (usually quasars) is a proven technique to probe the diffuse circumgalactic gas \citep[see][for a review]{Tumlinson17}. Thanks to the Cosmic Origins Spectrograph (COS) on board Hubble Space Telescope, it is now well-established that the metal-enriched, ionized, multiphase CGM harbors gas and metal masses that are at least comparable to those in galaxies themselves and can potentially account for the `missing baryons' and `missing metals' in low $z$ galaxies \citep[e.g.,][]{Tumlinson11,Werk14,Peeples14,Keeney17,Johnson17}. 

Despite the availability of a large repository of high-resolution, optical spectra of high-$z$ quasars \citep[e.g.,][]{OMeara15,Murphy19}, CGM studies at high $z$ have been more limited. This is primarily because detecting typical high-$z$ galaxies is challenging owing to their relative faintness.  

On large scales, cross-correlation analyses between \HI, traced by the \lya\ forest absorption, and high-$z$ galaxies have been carried out \citep[e.g.,][]{Adelberger05,Crighton11,Bielby17,Mukae17}. \citet{Crighton11} reported an excess \HI\ absorption within $\approx5h^{-1}$ comoving Mpc (cMpc\footnote{For physical (comoving) distances we use the prefix `p' (`c') before the unit.}) of $z\approx 3$ Lyman Break Galaxies (LBGs). The spatial correlation between photo-$z$ galaxies and \lya\ forest absorption lines have been studied at $z\approx2-3$ by \citet{Mukae17}. They reported a positive correlation between galaxy overdensities and \lya\ forest absorption lines in space, which indicates that high-$z$ galaxies are surrounded by an excess of \HI\ gas. \citet{Momose21b} performed a cross-correlation analysis between 570 galaxies with spectroscopic redshifts and the \lya\ forest transmission at $z\approx2.3$ and found a correlation out to 50~$h^{-1}$ cMpc. In addition, they showed that the signal varies with the galaxy population, depending on scale. For scales $r> 5h^{-1}$~cMpc, the strongest signals are exhibited by active galactic nuclei (AGNs) and submillimeter galaxies (SMGs). \lya\ emitters (LAEs), on the contrary, are more strongly correlated with the absorption on small scales ($r<5h^{-1}$~cMpc). \cite{Bielby20} detected LAEs within 250~pkpc and 1000~\kms\ of 6 out of 9 strong absorption systems at redshift 4--5, which they showed implies the absorbers are associated with galaxy overdensities. 

The bulk of the information on the CGM at $z\approx 2-3$ comes from the Keck Baryonic Structure Survey (KBSS) of UV-color-selected LBGs \citep[e.g.,][]{Adelberger05,Steidel10,Rudie12a,Rudie19,Rakic12,Turner14,YChen20}. In a pioneering work, \citet{Steidel10} demonstrated that the composites of background galaxy spectra can be used to characterize the CGM. They presented the dependence of \lya\ and other metal line (e.g., \SiII, \SiIV, and \CIV) absorption on the galactocentric radius (impact parameter, $\rho$) out to 125~pkpc using 512 close foreground-background angular pairs constructed of 89 LBGs from the KBSS survey. 

Using 679 KBSS galaxies probed by 15 background quasars, \citet{Rakic12} produced the first 2D maps of \lya\ absorption as a function of impact parameter and velocity difference around high-$z$ star forming galaxies by stacking the quasars' spectra. They found that the median \lya\ optical depth around LBGs is enhanced compared to randomly chosen regions out to hundreds of kpc. From a comparison to hydrodynamical simulations, \citet{Rakic13} showed that hydrodynamical simulations can reproduce the data if the LBGs reside in haloes of mass $M_{h}\sim 10^{12}$~\Msun, a mass consistent with estimated based on galaxy clustering \citep{Trainor12}. 

\citet{Rudie12a} performed Voigt profile decomposition of the full \lya\ forest region of the same 15 quasars, and studied the absorber-galaxy connection. They found that nearly half of the strong \HI\ absorbers with column density $N(\HI)>10^{15.5}$~\sqcm, arise from the CGM, which they defined as the region with $\rho<300$~pkpc and within $\pm300$~\kms\ of galaxy's systemic redshift. The median Doppler parameters for the CGM absorbers were found to be larger than for randomly chosen absorbers, which they attributed to accretion shocks and galactic winds around the LBGs. 

Using a sample of 854 KBSS galaxies (35 with $\rho<$~250~pkpc) as well as a subsample of 340 objects with accurate redshifts based on nebular emission lines, \citet{Turner14}  confirmed the results of \citet{Rakic12} for \HI\ and demonstrated that the detected redshift space distortions are robust to redshift errors. They also produced the first 2D maps of the distribution of metal ions, finding a strong enhancement of absorption relative to random regions for \CIII, \CIV, \SiIV, and \OVI, which extends out to at least 180~pkpc in the transverse direction and $\pm240$~\kms\ along the line of sight (LOS). The absorption signals for \HI\ and \CIV\ extend to 2~pMpc in the transverse direction, corresponding to the maximum impact parameter in their sample. \cite{Turner17} compared the results for \HI, \CIV\ and \SiIV\ to the EAGLE simulation \citep{Schaye15}, finding excellent agreement, and showed that in EAGLE the redshift-space distortions detected for impact parameters similar to the virial radii are due to infall rather than outflows. \cite{Turner15} presented a statistical detection of \OVI\ near the KBSS galaxies that is not associated with significant \HI\ and extends out to velocities exceeding the virial velocity, suggesting the presence of fast, hot and metal-rich outflows. 

Recently, \citet{YChen20} have extended the sample of \citet{Steidel10} to $\sim$200,000 pairs, and presented a 2D apparent \lya\ optical depth map as a function of LOS velocity ($\rm v_{LOS}$) and impact parameter, finding results consistent with \cite{Turner14}. They reported a significant asymmetry for $\rm 50 \lesssim \rho (pkpc) \lesssim 200$ and $\rm 200 \lesssim |v_{LOS}| (km~s^{-1}) \lesssim 500$, and argued that contamination of the \lya\ absorption by extended, diffuse \lya\ emission from the foreground galaxy is the most likely reason for the asymmetry. Studies with much brighter background sources, such as quasars, will however not be affected by such phenomena. A simple, two component analytic model with purely radial inflow and outflow can reproduce the overall features of the observed map. The model suggests that outflows dominate in the inner parts ($\rho\lesssim50$~pkpc), whereas inflows dominates the outer regions ($\rho \gtrsim100$~pkpc) of the CGM. This leads to a sudden dip in the \lya\ rest-frame equivalent width ($W_r$) profile, which otherwise can be  approximated as a power law with a slope of $-0.4$.   

None of these studies have explored whether the diffuse circumgalactic gas shows any trend with the properties of the host galaxies such as the star formation rate (SFR). Moreover, the KBSS is limited to relatively massive ($M_h\sim10^{12}$~\Msun) galaxies at $z\approx 2-3$ with SFR $\gtrsim 10$~\Msun~$\rm yr^{-1}$. Such objects represent the tip of the iceberg of the typical high-$z$ galaxy population, which is difficult to probe with broad/narrow-band color selection techniques such as used by the KBSS. 

\begin{figure}
\includegraphics[trim=0.6cm 6.2cm 0.0cm 8.0cm,width=0.5\textwidth,angle=00]{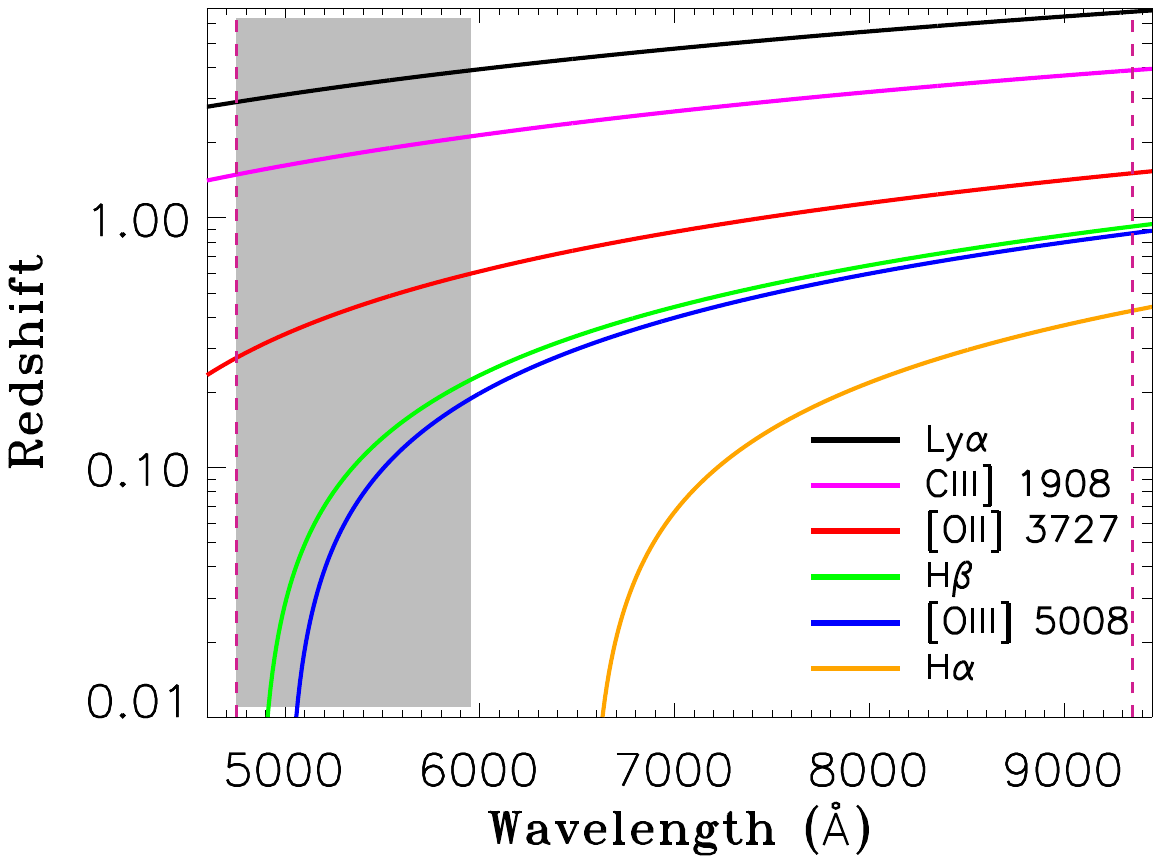} 
\caption{Redshift ranges covered by MUSE for different lines. The vertical dashed lines indicate the MUSE wavelength coverage in the nominal mode. We searched for LAEs in the redshift range 2.9--3.9 (below the redshifts of the background quasars). The corresponding wavelength range is indicated by the grey shaded region. The prominent emission lines from low-$z$ objects that can contaminate our LAE sample are shown. The redshifts at which a given curve enters and leaves the shaded region define the range for which the corresponding line can contaminate the LAE sample. For example, $[\OII]$ emitters between $z=0.27-0.52$ are potential contaminants for our sample. Note that low-$z$ $\rm H\alpha$ emitters cannot contaminate our sample.       
} 
\label{fig:muse-cov}      
\end{figure}

With the advent of powerful IFUs such as the Multi Unit Spectroscopic Explorer \citep[MUSE;][]{Bacon10} on the Very Large Telescope (VLT), it is now possible to achieve high spectroscopic completeness of faint galaxies. The high throughput (35\% at around 7000~\AA), large $1'\times1'$ field of view (FoV), and $0.2''$/pixel spatial sampling of MUSE are ideally suited to identify foreground galaxies around bright background quasars. MUSE allows one to find galaxies and obtain their redshifts simultaneously. Most importantly, galaxies that are too faint to be detected in the continuum emission can still be picked up through their line emission. For example, the continuous spectral coverage of 4750--9350~\AA\ in the nominal mode of MUSE allows one to detect \lya\ emission from galaxies in the redshift range $\approx$~2.9--6.6 (see Fig.~\ref{fig:muse-cov}). Therefore, MUSE fields around bright high-$z$ quasars with high-quality optical spectra provide a new opportunity to probe the connection between low-mass galaxies and their gaseous environments at $z>3$. Using $\approx50$~h of MUSE guaranteed time observations, we present here the first statistical study of the CGM of 96 $z\approx3.3$, low-mass ($M_{h}\sim10^{11}$~\Msun), star forming ($\lesssim1$~\Msun~$\rm yr^{-1}$) \lya\ emitting galaxies. We note here that a number of studies in the literature performed the reverse experiment, i.e., searched for LAEs around different types of absorbers such as \CIV\ absorbers \citep[]{Diaz15,Zahedy19b,Diaz21}, Lyman limit systems \citep[]{Fumagalli16,Lofthouse20,Nielsen20}, and damped \lya\ absorbers \citep[]{Mackenzie19} using narrowband/long-slit/IFU spectroscopy, but only for a handful of systems.         

The LAEs whose CGM is probed here are detected in 8 MUSE fields, centered on 8 high-$z$ quasars, as part of the MUSE  Quasar-field  Blind  Emitters  Survey \citep[MUSEQuBES;][]{Muzahid20}. Since MUSE allows us to detect LAEs only at $z>2.9$, the redshift of the background quasars must be $>2.9$ in order to probe the CGM of the foreground LAEs. For redshift $z>4$ the \lya-forest absorption in quasar spectra become too severe for accurate measurements of \lya\ equivalent widths and/or column densities. When constructing our sample, we thus restrict ourselves to UV-bright quasars ($V<19$) in the redshift range $3.5-4.0$ that are accessible with the VLT. We finally chose 8 UV-bright quasars in the redshift range 3.6--3.9 for which significant amounts of data were already available in the public archive, which we however augmented with new VLT/UVES observations, yielding $\approx 250$~h of UVES data in total. The LAEs in our sample thus have redshifts in the range 2.9--3.9. For this redshift range night-sky emission lines have only a minimal impact in the corresponding \lya\ observed wavelength range (4750--5960~\AA). Minimizing the effects of sky-line contamination is crucial for correctly identifying galaxies that are detected via standalone emission line such as the \lya\ line for $z>3$ LAEs.  

Using the LAE sample from the MUSEQuBES survey, we showed in \cite{Muzahid20} that the stacked CGM absorption can be used to calibrate the \lya\ redshifts, at least statistically, as first done for KBSS galaxies by \cite{Rakic11}. This is important, because \lya\ lines are known to be affected by resonant scattering. Here we used the empirical calibration from \citet[]{Muzahid20} to correct the \lya\ redshifts, and subsequently study the trends between the LAE properties and the stacked CGM absorption. This paper is organized as follows: In Section~\ref{sec:observations} we describe the MUSE and UVES/HIRES observation details and data reduction procedures. In Section~\ref{sec:analysis} we describe the MUSE data analysis and the LAE sample, followed by the pixel optical depth recovery from the quasar spectra. The main results of this study are presented in Section~\ref{sec:results}. These findings are discussed in the context of existing results from the literature in Section~\ref{sec:discussion}. Finally, a summary of the paper is presented in Section~\ref{sec:summary}. Throughout this study, we adopt a flat $\Lambda$CDM cosmology with $H_0=70$~\kms~Mpc$^{-1}$, $\Omega_{\rm M}=0.3$ and $\Omega_{\Lambda}=0.7$.

\section{Observations}       
\label{sec:observations}  

\subsection{MUSE observations and data reduction}  
\label{subsec:museobs} 

The high redshift MUSEQuBES survey exploits $\approx50$ h of MUSE guaranteed time observations (GTO) centered on 8 bright, $3.6 < z < 3.9$ quasars. The MUSE data for the 8 quasar fields were obtained in dark time with seeing $<0.9''$ between period P94 and P97 over the periods of 2 years. All fields were observed for at least 2 h and 4 fields have been observed for 10 h each. Each observation block of 1 h was divided into $4\times900$ s exposures. We conducted the observations in the WFM-NOAO-N mode which offers excellent spatial sampling of $0.2''$ per (square) pixel throughout the large $1'\times1'$ FoV and spectral resolution ($R$) ranging from $\approx1800$ (FWHM~$\approx$168~\kms) to $\approx3600$ (FWHM~$\approx86$~\kms) in the optical (4750--9350~\AA) with a spectral sampling of 1.25~\AA\ per pixel. Further details of the MUSE observations are given in Table~\ref{tab:observations}.

\begin{table*}  
\begin{threeparttable}[b] 
\caption{Summary of UVES and MUSE observations}    
\begin{tabular}{lcccrlrlrc}   
\hline  
Quasar Field   &  $\rm RA_{QSO}$  &  $\rm Dec_{QSO}$   &  $z_{\rm QSO}$   & $V$  & UVES observations &  $t_{\rm UVES}$  &  MUSE observations  &  $t_{\rm MUSE}$ & Seeing   \\   
               &  (hh:mm:ss)      &  (dd:mm:ss)        &                  &      &                   &     (h)          &                     &  (h)     & ($''$)   \\ \hline 
(1)            &  (2)             &  (3)               &  (4)             &  (5) &     (6)           &  (7)             &  (8)                &  (9)     & (10)      \\  
\hline  
Q1422$+$23$^{\alpha}$    &  14:24:38.09 & $+$22:56:00.59   &      3.631 &   15.84  &   091.A-0833(A)$^a$  &   5.4  &   60.A-9100(B)$^{b}$   &    1.0  & 0.77 \\  
	                 &              &                  &            &          &   093.A-0575(A)$^a$  &   7.1  &   095.A-0200(A)$^a$    &    3.0  &      \\  
	                 &              &                  &            &          &                      &        &   099.A-0159(A)$^a$    &    1.0  &      \\  \\                                                                                                                                      
Q0055$-$269$^{\alpha}$   &  00:57:58.02 & $-$26:43:14.75   &      3.662 &   17.47  &  65.O-0296(A)$^{c}$  &  26.1  &   094.A-0131(B)$^a$    &    3.0  & 0.72 \\ 
	                 &              &                  &            &          &  092.A-0011(A)$^a$   &  21.0  &   096.A-0222(A)$^a$    &    7.2  &      \\  \\                                                                                                                                        
Q1317$-$0507             &  13:20:29.97 & $-$05:23:35.37   &	  3.700 &   16.54  &  075.A-0464(A)$^d$   &  13.4  &   095.A-0200(A)$^a$    &    4.0  & 0.62 \\ 
	                 &              &                  &            &          &  093.A-0575(A)$^a$   &  21.7  &   096.A-0222(A)$^a$    &    4.0  &      \\    
	                 &              &                  &            &          &                      &        &   097.A-0089(A)$^a$    &    2.0  &      \\  \\                                                                                                                                       
Q1621$-$0042             &  16:21:16.92 & $-$00:42:50.91   &	  3.709 &   17.97  &  075.A-0464(A)$^d$   &  15.6  &   095.A-0200(A)$^a$    &    4.8  & 0.64 \\ 
                         &              &                  &            &          &  091.A-0833(A)$^a$   &  27.5  &   097.A-0089(A)$^a$    &    5.0  &      \\               
                         &              &                  &            &          &  093.A-0575(A)$^a$   &   5.0  &                        &         &      \\  \\                                                                                                                                        
QB2000$-$330$^{\alpha}$  &  20:03:24.12 & $-$32:51:45.14   &	  3.783 &   18.40   &  65.O-0299(A)$^{e}$  &   1.0  &   094.A-0131(B)$^a$    &   10.0  & 0.73 \\       
	                 &              &                  &            &          &  166.A-0106(A)$^{f}$ &  40.0  &                        &         &      \\  \\                                                                                                                                        
PKS1937$-$101$^{\alpha}$ &  19:39:57.26 & $-$10:02:41.52   &	  3.787 &   19.00   &  077.A-0166(A)$^{g}$ &  15.0  &   094.A-0131(B)$^a$    &    3.0  & 0.81 \\   
                         &              &                  &            &          &                      &        &   197.A-0384(C)$^{h}$  &    1.9  &      \\  \\      	                                                                                                                             
J0124$+$0044  		 &  01:24:03.78 & $+$00:44:32.74   &	  3.840 &   18.71  &  69.A-0613(A)$^{i}$  &  1.0   &   096.A-0222(A)$^a$    &    2.0  & 0.83 \\ 
	 		 &              &                  &            &          &  71.A-0114(A)$^{i}$  &  1.0   &   197.A-0384(D)$^{h}$  &    1.6  &      \\ 
             		 &              &                  &            &          &  073.A-0653(C)$^{j}$ &  3.3   &                        &         &      \\  
              		 &              &                  &            &          &  092.A-0011(A)$^a$   & 35.6   &                        &         &      \\  \\                                                                                                                                       
BRI1108$-$07$^{\alpha}$  &  11:11:13.60 & $-$08:04:02.00   &	  3.922 &   18.10   &  67.A-0022(A)$^{c}$  &  1.3   &   095.A-0200(A)$^a$    &    2.0  & 0.72 \\       
                         &              &                  &            &          &  68.B-0115(A)$^{k}$  &  7.7   &   197.A-0384(C)$^{h}$  &    2.4  &      \\  
                         &              &                  &            &          &  68.A-0492(A)$^{c}$  &  5.6   &                        &         &      \\  
\hline 
\end{tabular} 
\label{tab:observations}    
Notes-- (1) Name of the quasar field (2) Quasar right ascension (J2000) (3) Quasar declination (J2000) (4) Quasar redshift (5) Quasar $V$-band magnitude (6) PID of UVES observations (7) UVES exposure time (8) PID of MUSE observations (9) MUSE exposure time of the field. (10) Seeing (Moffat FWHM) of the final coadded data cube at around 7000~\AA. Columns 2--5 are obtained from SIMBAD (https://cds.u-strasbg.fr/) database. $^{a}$PI: Schaye. $^{b}$MUSE Commissioning data-- not used in this study. $^{c}$PI: D'Odorico. $^{d}$PI: Kim. $^{e}$PI: D'Odorico, V. $^{f}$PI: Bergeron. $^g$PI: Carswell. $^{h}$PI: Fumagalli; data not used in this study. $^{i}$PI: P\'eroux. $^{j}$PI: Bouch{\'e}. $^{k}$PI: Molaro. $^\alpha$Keck/HIRES data available.        
\end{threeparttable}  
\end{table*} 


The standard ESO MUSE Data Reduction Software \citep[DRS;][]{Weilbacher20} was used to reduce the individual raw science frames using default (recommended) sets of parameters. The four 10~h fields Q0055--269, Q1317--0507, Q1621--0042, and Q2000--330 were reduced using pipeline version {\tt v1.6}. The remaining four fields were reduced using version {\tt v2.4}. First, we prepared the master calibration files (bias, dark, flat, wavelength calibration, and twilight) for all 24 IFUs and removed the instrumental signatures from the science exposures and from the standard star exposures using recipes available in the DRS. The {\tt muse\_scibasic} recipe was then used for the basic science reduction using an illumination flat-field typically observed within $\pm1$ hour around the science exposures and using a proper geometry table. This produces the pre-reduced pixel tables for each IFUs for each science frame. Finally, the {\tt muse\_scipost} recipe was used to merge all the pixel tables from all IFUs using the proper astrometry table. Flux calibration was done during this step using the response curve and telluric absorption correction from a spectrophotometric standard star observed during the same night.

In order to minimize the impact of bad pixels and cosmic rays, two subsequent science exposures were taken with small offsets ($<1''$) and 90\degree\ rotation. Note that a small frame-to-frame spatial offset can also occur owing to a `derotator wobble' \citep[]{Bacon15}. Different frames thus do not cover the exact same sky location. The {\tt muse\_exp\_align} recipe was used to automatically compute the offsets among the exposures, and then to apply the correction to the reduced pixel tables. The reduced offset-corrected pixel tables were then used to obtain the reduced, flux-calibrated data drizzled onto a 3D ($\rm ra, dec, \lambda$) grid.

We did not use the default auto/self-calibration and sky-subtraction recipes offered by the DRS. Instead, we used the {\tt CubeFix} and {\tt CubeSharp} routines within {\tt CubEx} (v1.6) software developed by \citet{Cantalupo19} for self-calibration and sky-subtraction since they provide a better final data product in terms of both illumination and flux homogeneity compared to the standard pipeline \citep[see e.g., Fig.~1 of][]{Lofthouse20}. We refer the reader to Section~2.3 of \citet{Borisova16} for details about the self-calibration and sky-subtraction using {\tt CubEx}. The illumination corrected and sky-subtracted cubes are then combined using an average $3\sigma$-clipping algorithm using the {\tt CubeCombine} routine available within {\tt CubEx}. In order to further improve the illumination correction, we iterated the above steps after masking the bright continuum objects detected in the white-light image created from the initial combined cube. As noted by \citet{Borisova16}, typically one iteration is sufficient for adequate improvement before the individual cubes can be combined to obtain the final, science-ready data product.

The effective seeing, obtained from fitting a Moffat profile to a point source at $\lambda \approx 7000$~\AA, of the final datacubes varies between $0.62''$ and $0.83''$. The wavelengths in the MUSE data cubes are given in air. However, all redshifts given in this work correspond to vacuum redshifts, as we apply the appropriate air-to-vacuum corrections while building the LAE catalog.

\begin{figure*}
\centerline{\hbox{ 
\centerline{\vbox{
\includegraphics[trim=0.0cm 0.0cm 0.0cm 0.0cm,width=0.80\textwidth,angle=00]{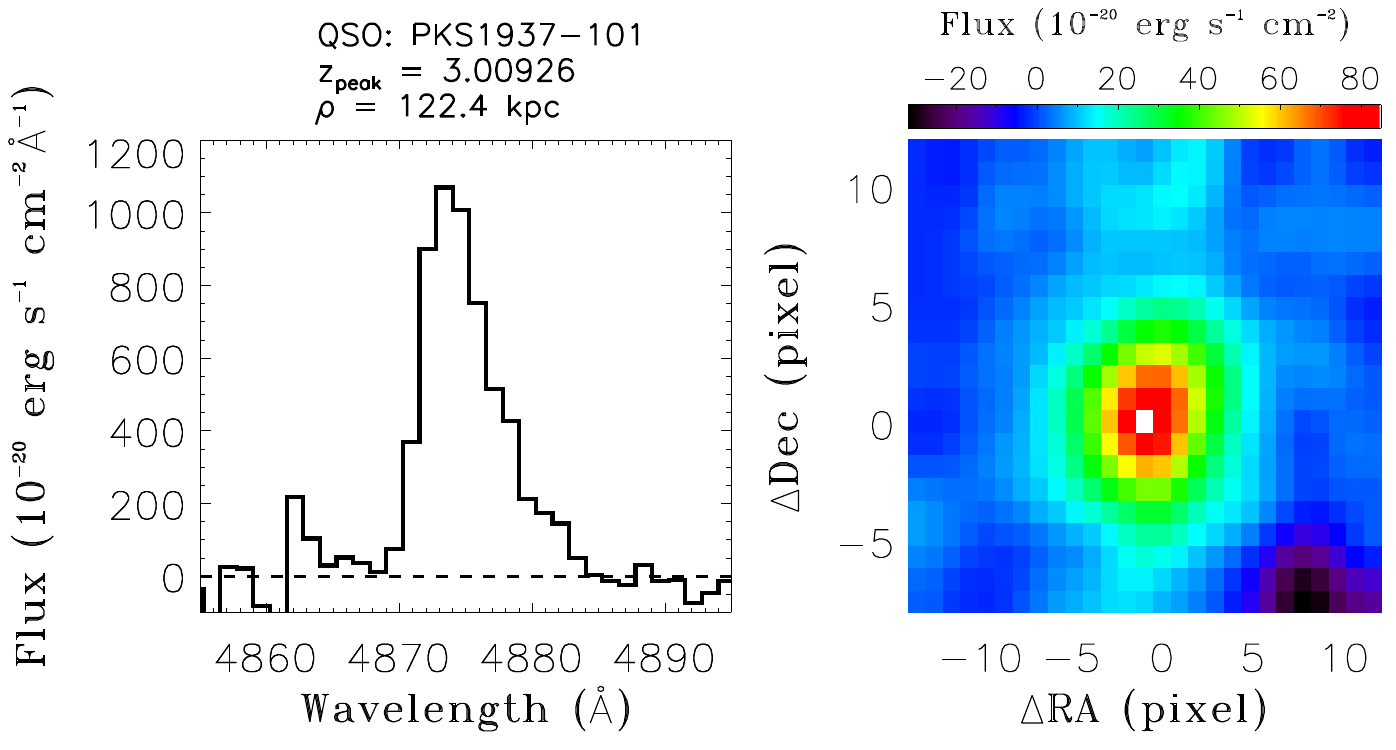} 
\includegraphics[trim=0.0cm 0.0cm 0.0cm 0.0cm,width=0.80\textwidth,angle=00]{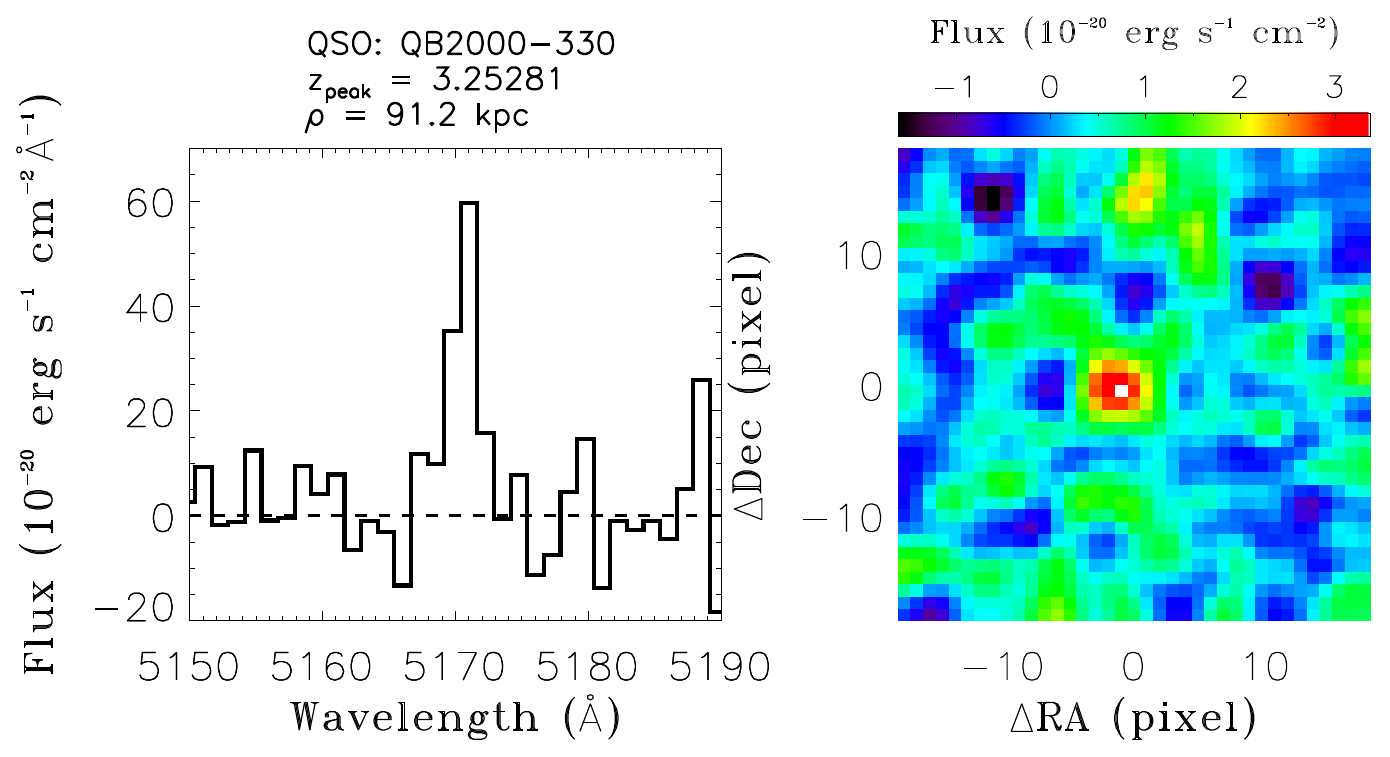} 
}}
}}   
	\caption{{\tt Top Left:} \lya\ emission from an object detected at \zlae~$=3.00926$ at an impact parameter of 122.4~pkpc towards quasar PKS~1937--101. The asymmetric profile with a prominent red wing as seen here is an unambiguous feature of \lya\ emitters. {\tt Top Right:} The pseudo-NB image, smoothed by 3 pixels, of the LAE in the left. {\tt Bottom:} The bottom panels show another example of a LAE with weak \lya\ emission. In such cases, we adopted a `method of elimination' as our LAE classification strategy (see text).}          
\label{fig:lae_example}     
\end{figure*} 

\subsection{UVES and HIRES observations and data reduction}  
\label{subsec:uvesobs} 

As mentioned in \S~\ref{sec:intro}, the 8 quasars chosen in this survey already had exquisite quality optical spectra with spectral resolution, $R\approx45,000$ obtained with the VLT/UVES (and Keck/HIRES) that were publicly available. However, sometimes the existing spectra did not cover the full wavelength range required for this study. For 5 out of 8 quasars, we augmented the existing data with new observations in order to fill in spectral gaps and/or to improve the signal-to-noise ratio ($S/N$). The spectral coverage and $S/N$ of the remaining three quasar spectra were sufficient for this study except for PKS1937--101, for which we have spectral coverage up to 6813~\AA, whereas the \CIV\ region extends out to 7493~\AA. Column~6 of Table~\ref{tab:observations} lists the proposal identifiers (PIDs) under which the data were taken. We checked the title of the each program and found that only two quasars (J~0124+0044 and BRI1108--07) were observed because of the presence of damped \lya\ absorbers (DLAs) in their spectra.

The ancillary VLT/UVES data for the 5 quasars were obtained in ESO periods P91 -- P93 in grey/dark time ($\approx123$~h in total) with a seeing $\approx1''$ (PI: Schaye). A slit width of $1''$ was used for these observations yielding a spectral resolution $R\sim$~40,000. Except for the quasar Q1422+23, the final coadded and continuum-normalized spectra were downloaded from the SQUAD database \citep[]{Murphy19}.\footnote{The spectrum of the quasar Q1422+23 was not available in the SQUAD repository.} The raw frames of the quasar Q1422+23 were reduced using the Common Pipeline Language ({\tt CPL v6.3}) of the UVES pipeline. After the initial reduction, we used the {\tt UVES Popler} \footnote{https://doi.org/10.5281/zenodo.44765} software to combine the extracted echelle orders into a single 1D spectrum. Note that the spectra in the SQUAD database are also generated using {\tt UVES Popler}. Continuum normalization of the final coadded spectrum was done by fitting smooth low-order spline polynomials in line-free regions determined by iterative sigma-clipping.

Finally, Keck/HIRES data are available for 5/8 quasars (see Table~\ref{tab:observations}). Since the $S/N$ of the UVES spectra (Table~\ref{tab:uvescov}) are more than sufficient for the stacking exercise presented here, we generally did not use the HIRES spectra except for two quasars. We used the HIRES spectra of PKS1937--101 (PIs: Cowie, Crighton, Tytler) and Q1422+23 (PIs: Cowie, Sargent) from the KODIAQ data release \citep[]{OMeara15} to fill in the gaps in the UVES spectra. The continuum-normalized UVES and HIRES spectra are combined using inverse-variance weighting.

\section{Data Analysis} 
\label{sec:analysis}

\subsection{MUSE data analysis and the LAE sample}

Here we describe the procedures for extraction and classification of objects from the final, combined MUSE data cubes obtained in Section~\ref{subsec:museobs}. Different routines within the {\tt CubEx} package are used for automatic extraction of emission line sources, for extracting 1D spectra, and for measuring \lya\ line fluxes using the curve-of-growth method.

Before extracting emission line sources from the data cubes, we first removed the quasars' point spread functions (PSFs). Galaxies close to the quasar lines of sight (small impact parameters) are more likely to produce strong absorption in the quasar spectrum. Finding such small-impact-parameter galaxies is thus important. The essential step to finding such small-impact-parameter galaxies is to get rid of the contamination due to the quasar's PSF. We used the {\tt CubePSFSub} routine which determines and subtracts the PSF empirically \citep[see Section 3.1 of][for details]{Borisova16}. Briefly, a pseudo-narrow band (NB) image is created for each wavelength layer at the position of the quasar with a bandwidth of 150 layers ($\approx187$~\AA). These images are treated as the empirical PSFs of the quasar. The adopted bandwidth is a trade off between maximizing the $S/N$ of the pseudo-NB images and minimizing the wavelength variations of the PSF. Each PSF image is then scaled to match the flux in the central $1''\times1''$ area centered on the peak of the quasar emission at that layer. This scaled empirical PSFs are subsequently subtracted from the corresponding layers up to $5''$ radius. As noted by \citet{Borisova16}, this method of PSF subtraction cannot provide any meaningful information on the central $1''\times1''$ region used for PSF rescaling which sets the lower limit on the impact parameter of $\approx8$~pkpc at $z\approx3$.

In order to remove the remaining continuum emitting sources, we next used the {\tt CubeBKGSub} routine within the {\tt CubEx} package \citep[see Section~3.2 of][for details]{Borisova16}. In brief, the spectrum of each spaxel of the PSF-subtracted cube is first binned using a bin size of 40 layers ($\approx50$~\AA). This resampled spectrum is then smoothed with a median filter with a size of three bins (radius). The smoothed cube is subsequently subtracted from the original cube.

Finally, we run {\tt CubEx} (v1.6) on the continuum-subtracted cube for automatic extraction of emission line sources. First the datacube and its associated variance are filtered (smoothed) by a Gaussian kernel with a size of 2 pixels (radius) at each wavelength layer. The contiguous voxels with $S/N>4$ are then grouped together. Following \citet{Marino18}, a group of $>40$ connected voxels is regarded a detection. In addition, we require a spectral $S/N>4.5$ measured on the 1D \lya\ emission line spectrum for an unambiguous detection. The voxels that are part of a group satisfying these conditions are used to create a 3D segmentation map (object mask). This segmentation map is subsequently used for extracting 1D spectra and for creating pseudo-NB images around the emission features. We typically masked the noisy pixels (up to 20 pixels) at the edge of the field. We also masked the first 8 wavelength layers of each cube and about 10 wavelength layers near the two strong skylines at $\approx5578$~\AA\ and $\approx5895$~\AA\ to suppress the number of spurious detections. Note that we only search for emitters with redshifts lower than the redshift of the background quasar, by masking the wavelengths redwards of the \lya\ emission.

Proper characterization of the noise is crucial for any source extraction procedure based on $S/N$ threshold. The detector noise propagated through different reduction steps cannot fully capture the correlated noise introduced by resampling and interpolation. The propagated variance produced by the MUSE pipeline and subsequently post-processed by {\tt CubFix} and {\tt CubeSharp} thus underestimates the `true' noise. In order to account for this, {\tt CubEx} rescales the propagated variance by a constant factor, determined from the spatial average of the propagated variance, for each wavelength layer. This rescaling factor typically varies between $\approx1.5$ and $2.0$.

The 1D spectra and the pseudo-NB images of all the extracted objects were inspected visually. These objects were then classified by two members of the team (SM and RAM) independently. Note that building a LAE catalog from MUSE data alone is not straightforward, since no other prominent emission lines are covered in the MUSE passband. Thus, identification of a LAE is, almost always, solely based on the \lya\ emission. This is particularly problematic for weak lines. The strong lines usually show the characteristic asymmetric profile with an occasional blue bump, and are therefore easy to classify. An example of such a strong \lya\ emitter is shown in the top panel of Fig.~\ref{fig:lae_example}. However, as shown in the bottom panel, it is not always straightforward, particularly for the weaker lines. We followed a `method of elimination' as our strategy to classify \lya\ emitters in such cases. We consider all the different possible redshift solutions for a given emitter and eliminate them systematically until we find the best/most likely solution. As demonstrated in Fig.~\ref{fig:muse-cov}, there are several prominent emission lines (e.g., $\rm \CIII]~\lambda\lambda1907,1909$, $\rm [\OII]~\lambda\lambda3727,3729$, $\rm H\beta~\lambda4862$, and $\rm [\OIII]~\lambda5008$) that can contaminate our LAE sample. It is known that the $\rm [\OII]~\lambda\lambda3727,3729$ doublet from low-$z$ objects is the major contaminant for high-$z$ LAE samples. The [\OII] emitters in the redshift range 0.27--0.52 are potential contaminants for our sample (see Fig.~\ref{fig:muse-cov}). Nevertheless, the doublet nature of the $[\OII]$ line provides important clues. In addition, other strong emission lines such as [\OIII], $\rm H\beta$, and/or $\rm H\alpha$ are always covered by MUSE for an [\OII] emitter in the redshift range 0.27--0.52. The presence or absence of additional lines helps tremendously to distinguish an $[\OII]$ emitter from a LAE \citep[e.g.,][]{Inami17}.    

In total we classified 96 objects as \lya\ emitters. For each LAE we measured the peak redshift (\zlae), impact parameter ($\rho$), line luminosity ($L$(\lya)), full width at half maximum (FWHM), UV luminosity ($L_{\rm UV}$), star formation rate (SFR), and rest-frame equivalent with ($EW_0$) as discussed in detail in Section~3 of \citet{Muzahid20}. Briefly, the \zlae\ and FWHM are measured directly from the 1D spectrum. The red peak is used in case of a double-peaked \lya\ profile. We did not apply any modeling to the emission line profile, since \lya\ is not well behaved owing to its resonant nature. The FWHM values are not corrected for instrumental broadening. The $L$(\lya) is calculated from the Galactic extinction corrected line flux, $f$(\lya), determined from the pseudo-NB images using the curve-of-growth method following \citet{Marino18}. Briefly, the total \lya\ flux was determined from the direct sum of $0.2''$ annular fluxes out to the radius where the annular flux is equal to or less than zero.

The $L_{\rm UV}$ is calculated from the Galactic extinction corrected UV continuum flux, determined by direct integration of the 1D spectrum between 1410--1640~\AA\ (FWHM of the $GALEX$ far-UV transmission curve) in the rest-frame. The dust-uncorrected SFRs are calculated from the $L_{\rm UV}$ using the local calibration relation of \citet{Kennicutt98} corrected to the \citet{Chabrier03} stellar initial mass function. Lastly, $EW_0$ are calculated by dividing $f$(\lya) by the continuum flux density, estimated by extrapolating the rest-frame 1500~\AA\ continuum assuming a spectral index of $-2.0$, and correcting for the 1+\zlae\ factor. Note that the UV continuum is detected for only 39 LAEs with $>5\sigma$ significance. For 42 LAEs we could only place conservative ($5\sigma$) upper limits. The remaining 15 objects are blended with low-$z$ continuum sources. 

\begin{table}
\begin{center}  
\begin{threeparttable}[b] 
\caption{Spectral coverage and $S/N$ of the quasar spectra}      
\begin{tabular}{lrrrr}    
\hline 
Quasar  &  $\lambda_{\rm min}^{a}$  & $\lambda_{\rm max}^{b}$  & $S/N_{5230~\angstrom}^{c}$   & $S/N_{6660~\angstrom}^{d}$ \\ 
\hline 
       Q1422+23 &   3400 &   7290 &    174  &   194 \\ 
      Q0055-269 &   3140 &   9767 &     41  &    80 \\ 
     Q1317-0507 &   3238 &   9467 &     36  &    96 \\ 
     Q1621-0042 &   3735 &   9467 &     50  &   103 \\ 
     QB2000-330 &   4152 &  10432 &     72  &    68 \\ 
    PKS1937-101 &   4000 &   6813 &     77  &    86 \\ 
     J0124+0044 &   3700 &   9467 &     28  &    71 \\ 
     BRI1108-07 &   4394 &  10429 &     14  &    24 \\  
\hline 
\end{tabular}
\label{tab:uvescov}  
Notes-- $^{a}$The minimum wavelength (in \AA) covered by the spectrum. $^{b}$The maximum wavelength (in \AA) covered by the spectrum. $^{c}$ The median $S/N$ per pixel calculated at $5230\pm50$~\AA. $^{d}$The median $S/N$ per pixel calculated at $6660\pm50$~\AA. The $S/N$ are calculated at the expected wavelengths of \lya\ and \CIV, respectively, for the median redshift of $3.3$ of the survey.           
\end{threeparttable}  
\end{center}  
\end{table} 

\begin{figure}
\includegraphics[trim=1.0cm 0.8cm 0.5cm 1.5cm,width=0.45\textwidth,angle=00]{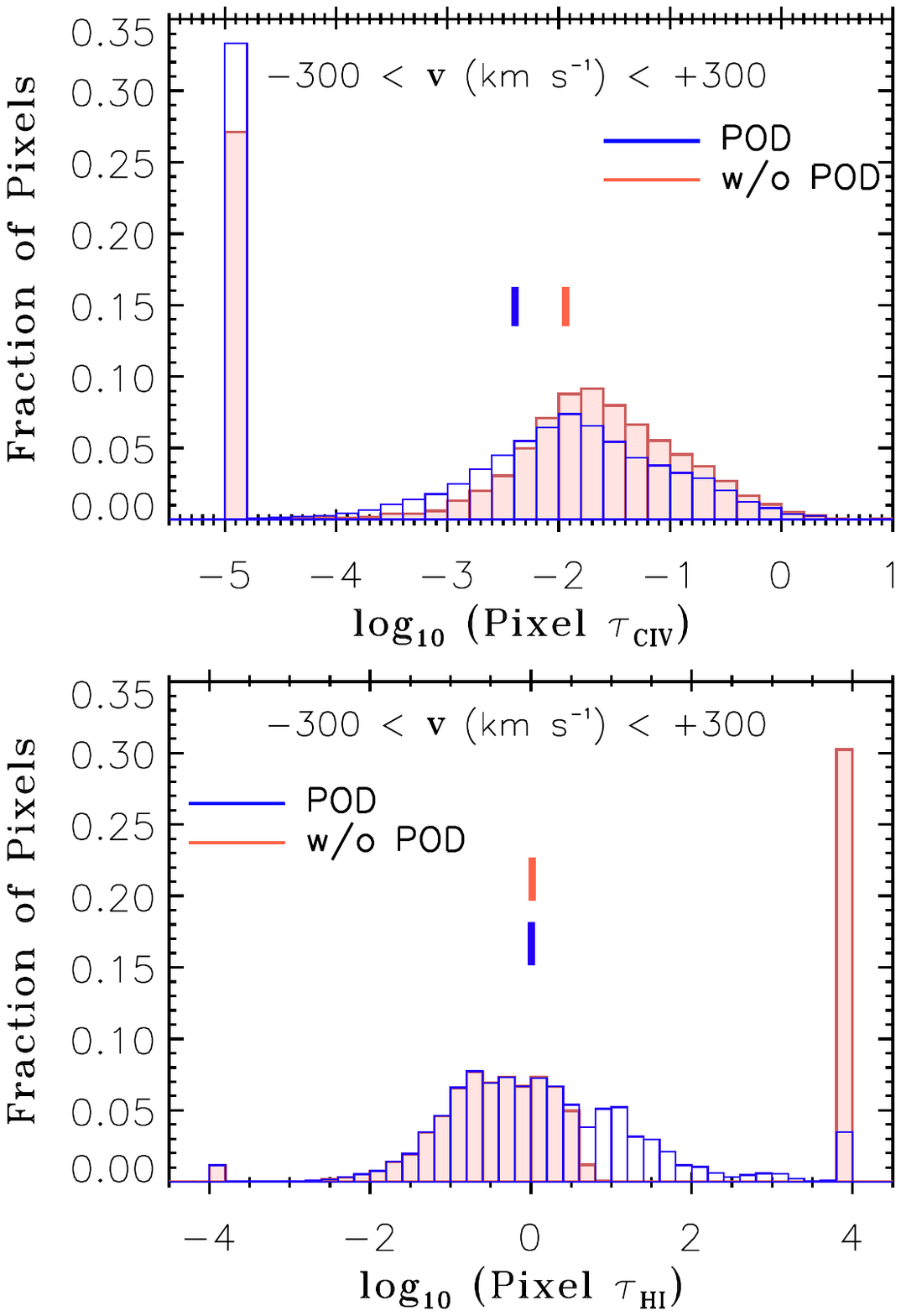}
\caption{Histograms of pixel optical depths with and without POD recovery using {\tt PODPy}. All pixels in the quasar spectra within $\pm300$~\kms\ of $z_{\rm peak}$ are used in these plots. There are 77 (92) LAEs contributing to the \HI\ (\CIV) pixel optical depth distribution in the bottom (top) panel. For \HI, the pixels with very high and low optical depths ($>10^{4}$  and $<10^{-4}$) are flagged with flag values of $10^{+4}$ and $10^{-4}$, respectively. For \CIV, the pixels with $\tau >10^{1}$ and $\tau <10^{-5}$ are flagged with flag values of $10^{+1}$ and $10^{-5}$, respectively. The median values of the distributions are indicated by the vertical ticks with corresponding colors. The effect of the correction for saturation using higher order Lyman series lines by {\tt PODPy} is evident from the bottom panel. The effects of the contamination correction and of rescaling the continua are apparent from the top panel.        
}               
\label{fig:PODvsNOPOD}       
\end{figure} 

\begin{figure*}
\centerline{\vbox{
\centerline{\hbox{ 
\includegraphics[trim=1.0cm 4.0cm 1.5cm 7cm,width=0.90\textwidth,angle=00]{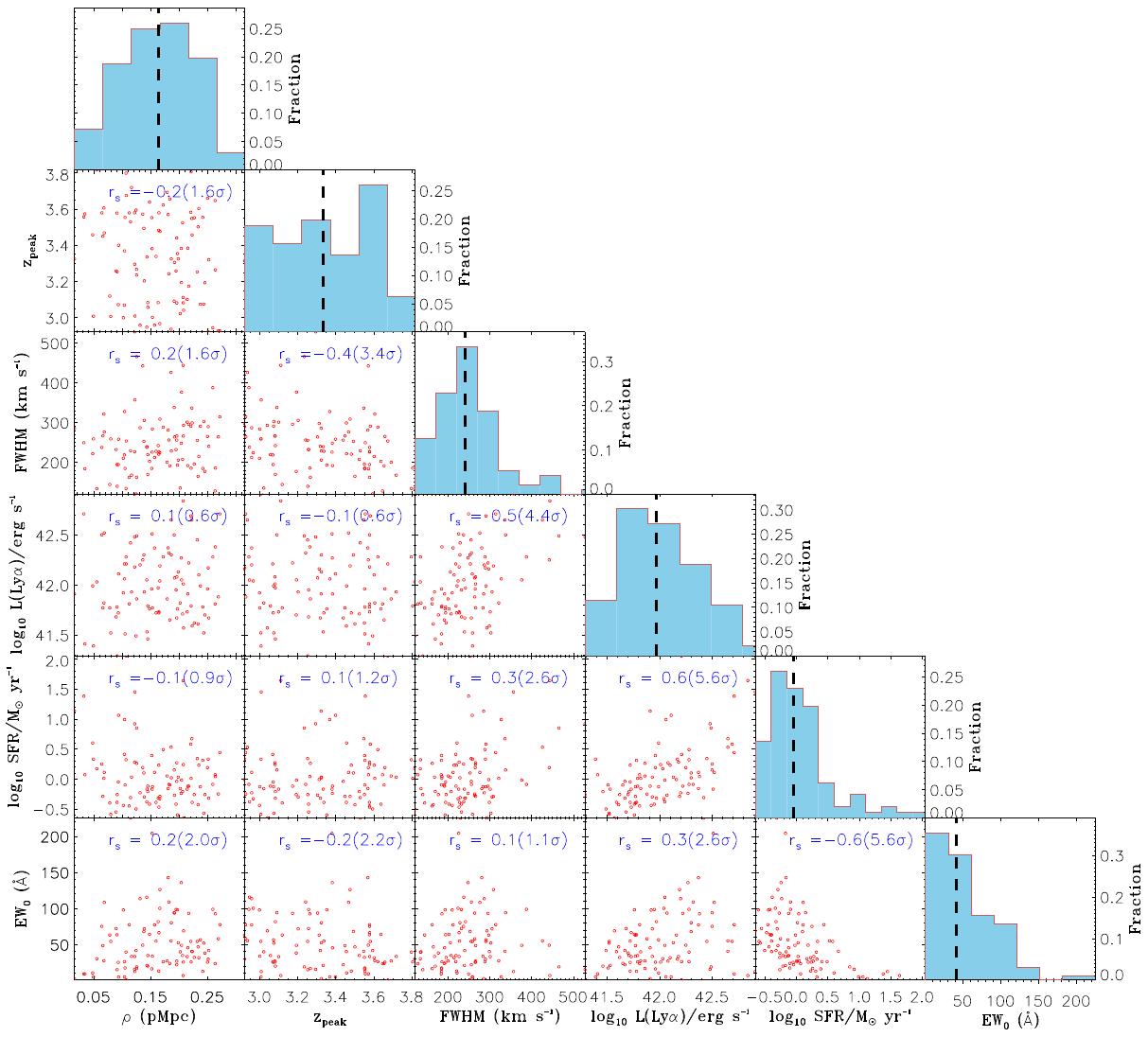} 
}}
}}   
\caption{Corner plot showing possible relations between different parameters obtained for the 96 LAEs in our sample. The Spearman rank correlation coefficients ($r_s$) and the corresponding significances, treating limits as detections, are indicated in each panel. The parameters showing a significant correlation with each other are discussed in the text. The vertical lines in the panels showing the relative frequency distributions indicate the corresponding median values.}       
\label{fig:cornerplot}     
\end{figure*} 

\subsection{Pixel optical depth recovery from the quasar spectra}    

The spectral coverage and $S/N$ of the optical spectra (FWHM $\approx$~6.6~\kms) of the quasars are given in Table~\ref{tab:uvescov}. The $S/N$ per pixels ranges from 14--174 (24--194) in the blue (red) part, indicating that these are some of best quality optical spectra of quasars ever-observed. We searched for CGM absorption in the composite spectra constructed from these high-quality quasar spectra.

Before we stack the quasar spectra, we run {\tt PODPy}\footnote{The code is available at: http://github.com/turnerm/podpy}, a python module developed by \citet{Turner14} to recover (from saturation and contamination) the correct pixel optical depths (PODs) for a given transition from each spectrum \citep[e.g.,][]{Cowie98,Schaye03,Aguirre08}. The pixel optical depth is a measure of absorption strength, and is defined as: $\tau_{\lambda} = -\ln(F_{\lambda})$, where $F_{\lambda}$ is the continuum normalized flux of a given pixel. We used the default set of parameters used in \citet{Turner14}, which is optimized for high-$z$ quasar spectra. We refer the reader to Appendix~A of \citet{Turner14} for a detailed discussion. Briefly, {\tt PODPy} iteratively examines whether the optical depth of a given pixel in a spectrum is consistent with being the transition of interest. First, we set $\tau_{\lambda} = 10^{-6}$ for pixels with $F_{\lambda}>1$. The saturated pixels with $F_{\lambda} \le 3\sigma_{\lambda}$, where $\sigma_{\lambda}$ is the error in the normalized flux, on the contrary, are set to a very high optical depth of $10^{4}$. Note that this flag affects neither the mean nor the median flux. It also does not affect the median optical depth provided it is smaller than $10^{4}$.     

For \lya, the main source of error comes from the line saturation. {\tt PODPy} takes advantage of all the available Lyman series lines (up to Ly--$\lambda914.9$) to correct the pixel optical depths of all \lya\ pixels previously deemed saturated. If a correction is not possible, the \lya\ optical depth is set to $10^{4}$. The effect of the saturation correction is evident from the bottom panel of Fig.~\ref{fig:PODvsNOPOD}. There is a sharp cutoff at $\tau \approx 0.5$ for the uncorrected optical depth distribution (red histogram) owing to the saturation of the \lya\ line. The {\tt PODpy} recovered optical depth distribution does not show such a feature, with a significantly lower number of pixels flagged as saturated ($\log_{10} \tau=4.0$).

Note that we do not use the pixels in the \lyb-forest. This sets the lower limit on redshift for the \lya\ stack as: 
\begin{equation} 
z_{\rm min}^{\rm Ly\alpha} = ((1+z_{\rm qso})\times \lambda_{\rm Ly\beta})/\lambda_{\rm Ly\alpha}-1,               
\end{equation} 
where, $\lambda_{\rm Ly\alpha}$ and $\lambda_{\rm Ly\beta}$ are the rest-frame wavelengths of \lya\ and \lyb, respectively. Moreover, we do not use the proximate regions (3000~\kms\ blueward of $z_{\rm qso}$) of the background quasar for any of the lines considered here. This sets the maximum redshift for the all transitions as:  
\begin{equation} 
z_{\rm max} = z_{\rm qso} - (1+z_{\rm qso})\times\frac{3000}{c}~,                    
\end{equation} 
where $c$ is the speed of light in vacuum in \kms. 83/96 LAEs in our sample satisfy these redshift bounds for the \lya\ stack. We further masked the \lya\ regions of 6 LAEs that are showing strong \HI\ absorption with damping wings as recommended by \citet{Turner14}. This is to avoid features dominated by a few strong absorbers whose velocity spread far exceeds that resulting from the peculiar and thermal velocities of the absorbing gas. 

For optical depth recovery of metal line doublets such as \CIV~$\lambda\lambda$1548,1550, we first refit the continuum redward of the quasar's \lya\ emission, using the {\tt fit\_continuum} routine available within {\tt PODPy}. This allows us to homogenize continuum fitting errors between different quasars. Next the algorithm checks for contamination by unrelated absorption (such as \MgII\ from low-$z$) at each pixel by testing whether the optical depth in a given pixel is higher than what is expected from the weaker member of the doublet (e.g., the $\lambda$1550 transition for the \CIV\ doublet). Next, an iterative doublet subtraction algorithm is used to remove self-contamination (i.e., eliminating any contribution from the $\lambda$1550 transition). As demonstrated in the top panel of Fig.~\ref{fig:PODvsNOPOD}, the recovery of the \CIV~$\lambda$1548 pixel optical depth is dominated by the contamination correction. While the median optical depth for the {\tt PODPy} recovered distribution is somewhat lower compared to the uncorrected distribution, it reduces the absorption unrelated to the galaxies leading to a significant suppression of noise. 

For \CIV\ stacks, we do not use pixels in the \lya\ forest region, this sets the lower limit in redshift: $z_{\rm min}^{\CIV} = ((1+z_{\rm qso})\times \lambda_{\rm Ly\alpha})/\lambda_{\CIV}-1$. All 96 LAEs in our sample satisfy the redshift bounds ([$z_{\rm min}^{\CIV}$, $z_{\rm max}$]). However, the \CIV\ regions are affected by skylines for three LAEs ($z = 3.455$ towards Q1317--0507, $z = 3.453$ towards Q1621--0042, and  $z = 3.446$ towards QB2000--330). There is no spectral coverage for the $z = 3.553$ LAE in the PKS1937--101 field. These LAEs are removed from the sample for the \CIV\ stacks.


\section{Results}     
\label{sec:results}

Scatter plots of the different parameters (i.e., $\rho$, $z_{\rm peak}$, FWHM, $L$(\lya), SFR, and $EW_0$) measured for the LAEs are shown in Fig.~\ref{fig:cornerplot}. The details on the measurements of these parameters are described in Section~3 of \citet{Muzahid20}. There is no significant correlation between the impact parameter (ranging from 16 -- 315~pkpc, median $\rho = $~165~pkpc) and any of the other parameters. This is expected, since the impact parameter, the projected separation between the LAE and the background quasar, is not governed by any physical process in the LAE. Using Spearman rank correlation tests, no significant correlation is seen between \zlae\ (ranging from 2.9 -- 3.8, median \zlae~$ = 3.3$) and other parameters except for the FWHM (ranging from 120 -- 528~\kms, median FWHM = 240~\kms). As noted in \citet{Muzahid20}, the $3.4\sigma$ anti-correlation between \zlae\ and FWHM is likely due the fact that the MUSE resolution improves with wavelength (redshift) and our FWHM values are not corrected for the MUSE resolution. The FWHM also shows a significant trend with $L$(\lya) (ranging from $10^{41.3}-10^{42.9}$~erg~s$^{-1}$, median $L$(\lya)~$=10^{42.0}$~erg~s$^{-1}$) which is expected since both parameters are related to the area under the emission line, provided the line is spectrally resolved.    

\begin{table*}  
\begin{threeparttable}[b] 
\caption{List of sub/samples generated for stacking}     
\begin{tabular}{lclcl}   
\hline  
Sample	       & N$_{\rm LAE}$~(\HI)          & Median property value	& N$_{\rm LAE}$~(\CIV)	& Median property value        \\ 		 
(1)            &  (2)		              &  (3)            &  (4)            	&   	(5)        \\  
\hline  
all				& 77	     &   ---			                   	& 92	& ---		\\	\\ 
all\_environment\_iso 		& 54	     &   ---                                       	& 62	& --- 		\\	
all\_environment\_group 	& 23	     &   ---                                       	& 30	& --- 		\\	\\ 
all\_$\rho$\_low 		& 26	     &   $\rho=$~90.8 (49.5, 107.1)~pkpc           	& 31	& $\rho=$~103.8 ( 64.2, 124.4)~pkpc  \\	
all\_$\rho$\_mid 		& 25	     &   $\rho=$~152.6 (136.1, 180.7)~pkpc   	     	& 30	& $\rho=$~166.1 (145.0, 192.0)~pkpc  \\
all\_$\rho$\_high  		& 26	     &   $\rho=$~222.4 (197.1, 249.9)~pkpc          	& 31	& $\rho=$~231.9 (206.1, 263.3)~pkpc  \\	\\ 
iso\_$\rho$\_low 		& 18	     &   $\rho=$~78.2 (48.1, 104.1)~pkpc	        & 21	& $\rho=$~93.7 (64.2, 116.5)~pkpc     \\	 
iso\_$\rho$\_mid 		& 18	     &   $\rho=$~143.5 (122.0, 172.1)~pkpc     	   	& 20	& $\rho=$~159.5 (138.9, 180.7)~pkpc   \\	
iso\_$\rho$\_high 		& 18	     &   $\rho=$~216.9 (192.0, 263.3)~pkpc          	& 21	& $\rho=$~215.4 (202.0, 259.3)~pkpc   \\	 \\ 
iso\_$z_{\rm peak}$\_low 	& 27	     &   $z_{\rm peak} =$~3.254 (3.008, 3.338)	   	& 31    & $z_{\rm peak} =$~3.091 (3.000, 3.295)  \\	
iso\_$z_{\rm peak}$\_high   	& 27	     &   $z_{\rm peak} =$~3.553 (3.456, 3.647)	   	& 31    & $z_{\rm peak} =$~3.529 (3.381, 3.647)  \\  \\ 
iso\_$L ({\rm Ly\alpha})$\_low	& 27	     &   $\log_{\rm 10} L\rm (Ly\alpha)=$~41.76 (42.52, 41.91)$^{a}$  	& 31    & $\log_{\rm 10} L\rm (Ly\alpha)=$~41.76 (41.52, 41.91)$^{a}$	\\	
iso\_$L ({\rm Ly\alpha})$\_high & 27	     &   $\log_{\rm 10} L\rm (Ly\alpha)=$~42.28 (42.06, 42.50)$^{a}$  	& 31 	& $\log_{\rm 10} L\rm (Ly\alpha)=$~42.28 (42.06, 42.52)$^{a}$	\\	\\ 
iso\_FWHM\_low			& 27	     &   $\rm FWHM=$~199.7 (143.6, 228.6)~\kms     	& 31	& $\rm FWHM=$~199.7 (139.1, 228.6)~\kms		\\	
iso\_FWHM\_high     		& 27	     &   $\rm FWHM=$~285.1 (255.5, 327.2)~\kms     	& 31	& $\rm FWHM=$~292.7 (255.7, 387.6)~\kms		\\	\\ 
all\_SFR\_non-det		& 36	     &   $\rm \log_{\rm 10} SFR=$~--0.30 (--0.50, --0.09)$^{b,c}$       & 41	& $\rm \log_{\rm 10} SFR=$~--0.26 (--0.45, --0.09)$^{b,c}$  	\\	
all\_SFR\_low      		& 15	     &   $\rm \log_{\rm 10} SFR=$~--0.12 (--0.32, +0.07)$^{b}$         	& 20	& $\rm \log_{\rm 10} SFR=$~--0.04 (--0.26, +0.07)$^{b}$ 	        \\	
all\_SFR\_high       		& 15	     &   $\rm \log_{\rm 10} SFR=$~+0.28 (+0.19, +0.58)$^{b}$          	& 18	& $\rm \log_{\rm 10} SFR=$~+0.33 (+0.19, +0.58)$^{b}$ 		\\	\\ 
all\_$EW_0$\_low        	& 15	     &   $EW_0=$~25.3 (16.6, 33.2)~\AA	   	& 19 	& $EW_0=$~26.1 (18.8, 38.0)~\AA		\\	
all\_$EW_0$\_high        	& 15	     &   $EW_0=$~63.0 (48.9, 93.0)~\AA         	& 19    & $EW_0=$~66.3 (54.1, 93.0)~\AA		\\	\\ 
\hline  
\end{tabular} 
\label{tab:sample_summary}     
	Notes-- (1) List of different sub/samples generated for stacking. Except for the full sample (`all'), the name of the subsamples have three parts. The first part indicates the parent sample (`all' or `iso') from which the subsample is created. The second part indicates the property based on which the subsample is generated (i.e., impact parameter ($\rho$), \lya\ redshift ($z_{\rm peak}$), \lya\ line luminosity ($L ({\rm Ly\alpha})$), FWHM, SFR, and \lya\ emission rest equivalent width ($EW_0$)). The third part (low/high etc.) provides the unique identifier. The `all\_environment\_iso' is the `isolated only' subsample. The `all\_SFR\_non-det' is the subsample of all the UV--continuum undetected objects. (2) Number of LAEs contributing to the \HI\ stack. (3) The median value of the property for the sub--sample. The values in parentheses provide the corresponding 68 percentile ranges. (4) Number of LAEs contributing to the \CIV\ stack. (5) Same as (3) but for \CIV\ stacks. \\ 
$^{a}L\rm (Ly\alpha)$ in $\rm erg~s^{-1}$ \\ 
$^{b}\rm SFR$ in $\rm M_\odot ~\rm yr^{-1}$ \\ 
$^{c}$The median and 68 percentile range are calculated treating limits as detections              
\end{threeparttable}  
\end{table*} 	 
 

The anti-correlation seen between SFR and $EW_0$ follows from the fact that the SFR is directly proportional to the continuum luminosity whereas the $EW_0$ is inversely proportional to the continuum flux density. Finally, we notice a significant ($5.6\sigma$) and strong ($r_s=0.6$) correlation between SFR and $L$(\lya). This is interesting since the SFRs are calculated from the UV continuum emission, whereas $L$(\lya) are measured from the line emission. Owing to the resonant scattering and susceptibility to dust, the \lya\ line is not considered to be a good SFR indicator. The correlation between SFR and $L$(\lya), nonetheless, suggests that the \lya\ line emission is dominated by the recombination radiation from atoms that were ionized by the UV photons generated via star formation activity, and that the escape fraction of \lya\ photons does not depend strongly on $L$(\lya).  

The dust-uncorrected median SFR ($\sim1$~$\rm M_\odot$~yr$^{-1}$) of our sample corresponds to $M_*\sim10^{8.6}~\rm M_{\odot}$ for main sequence galaxies \citep[]{Behroozi19}. This is similar to the dynamical masses of the continuum-faint $z\approx2.5$ LAEs studied by \citet{Trainor15}, estimated assuming a spherical geometry and a radius of 0.6~pkpc. On such scales, the dynamical mass is likely dominated by stars. Using the abundance matching relation of \citet{Moster13} we obtain the corresponding halo mass of $\sim10^{11}~\rm M_{\odot}$, and a virial radius, $R_{200}\approx35$~pkpc, where $R_{200}$ is the radius at which the mean enclosed density is 200 times the critical density of the universe.   

Next, we investigate possible trends between the stacked \HI\ and \CIV\ CGM absorption and the different properties (e.g., $z$, $\rho$, SFR) of the LAEs discussed in the previous section. The details of the subsamples based on these properties are presented in Table~\ref{tab:sample_summary}. Note that \lya\ and \CIV\ are the only two transitions for which the stacked absorption is detected at $>3\sigma$ significance. We do not consider the higher order Lyman series lines (\lyb\ and above) in our analysis owing to the high degree of blending in the \lya\ forest blueward of \lyb\ emission. However, the higher order Lyman series lines are used by {\sc PODPy}, whenever possible, to correct the \lya\ optical depths for the saturated pixels.

Since the \lya\ redshifts (\zlae) do not represent the systemic redshifts, we obtained the corrected redshifts, \zcorr, by blue-shifting each \zlae\ value by the amount \voff\ determined from the empirical relation (i.e., \voff~$\rm =0.89\times FWHM - 85$~\kms) obtained in \citet{Muzahid20} for our sample of LAEs. For the entire analysis presented here, we used {\tt PODpy} recovered spectra to improve the continuum $S/N$ of the stacked profiles. However, the results remain nearly unchanged if we use the original spectra instead.

\begin{figure}
\includegraphics[trim=0.5cm 3.0cm 0.0cm 4.0cm,width=0.45\textwidth,angle=00]{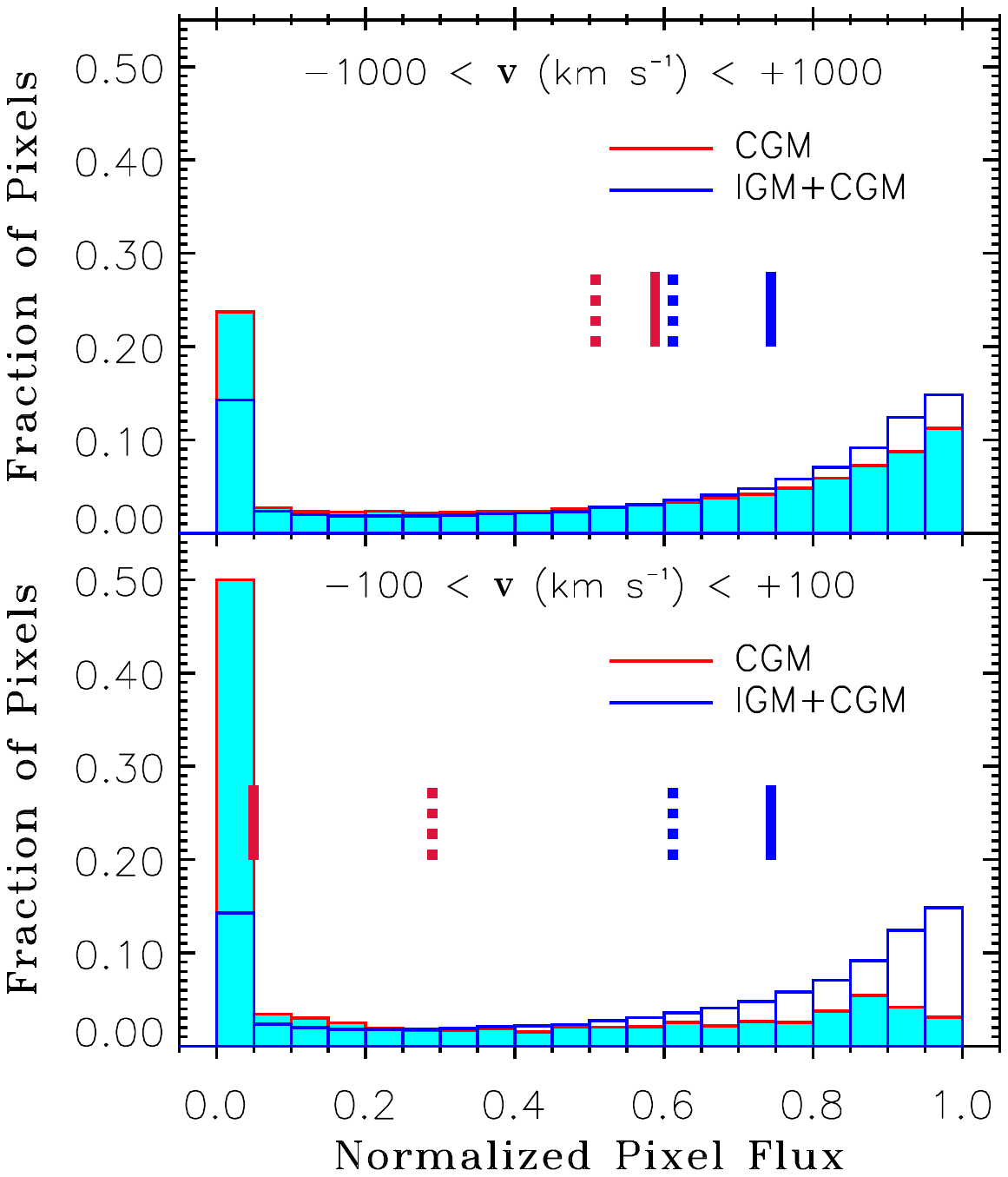} 
\vskip0.2cm 
\caption{Histograms of \lya\ pixel flux, labelled as CGM, within $\pm100$~\kms\ (bottom) or $\pm1000$~\kms\ (top) of the LAE redshift \zcorr\ are compared with the histogram of pixel flux in the \lya--forest regions of all the 8 quasars (labelled as IGM+CGM). The median and mean values are indicated, respectively, by the solid and dashed vertical lines with corresponding colors. The distributions for the CGM are right-skewed showing a sharp peak in the lowest flux bin. The \lya--forest flux distribution (IGM+CGM) shows the exact opposite trend. A two-sided KS--test suggest that the pixel flux distribution for the CGM is significantly different compared to that of the \lya--forest. As expected, the difference is somewhat reduced in the top panel in which a $10$ times larger velocity window around each \zcorr\ is used.}              
\label{fig:CGMvsIGM}      
\end{figure} 

\begin{figure*}
\centerline{\vbox{
\centerline{\hbox{ 
\includegraphics[trim=0cm 2cm 0cm 2.5cm,width=0.35\textwidth,angle=00]{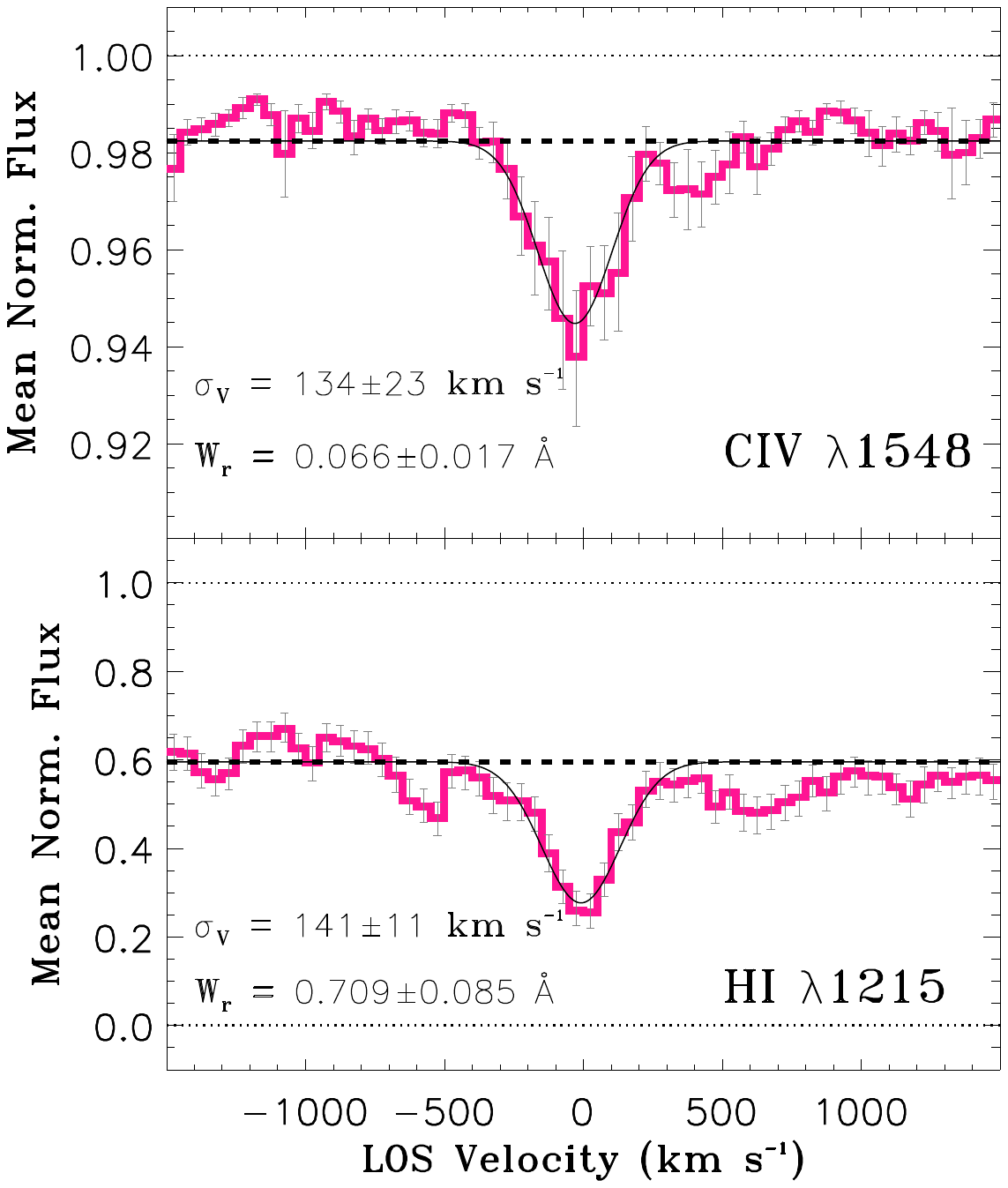} 
\includegraphics[trim=0cm 2cm 0cm 2.5cm,width=0.35\textwidth,angle=00]{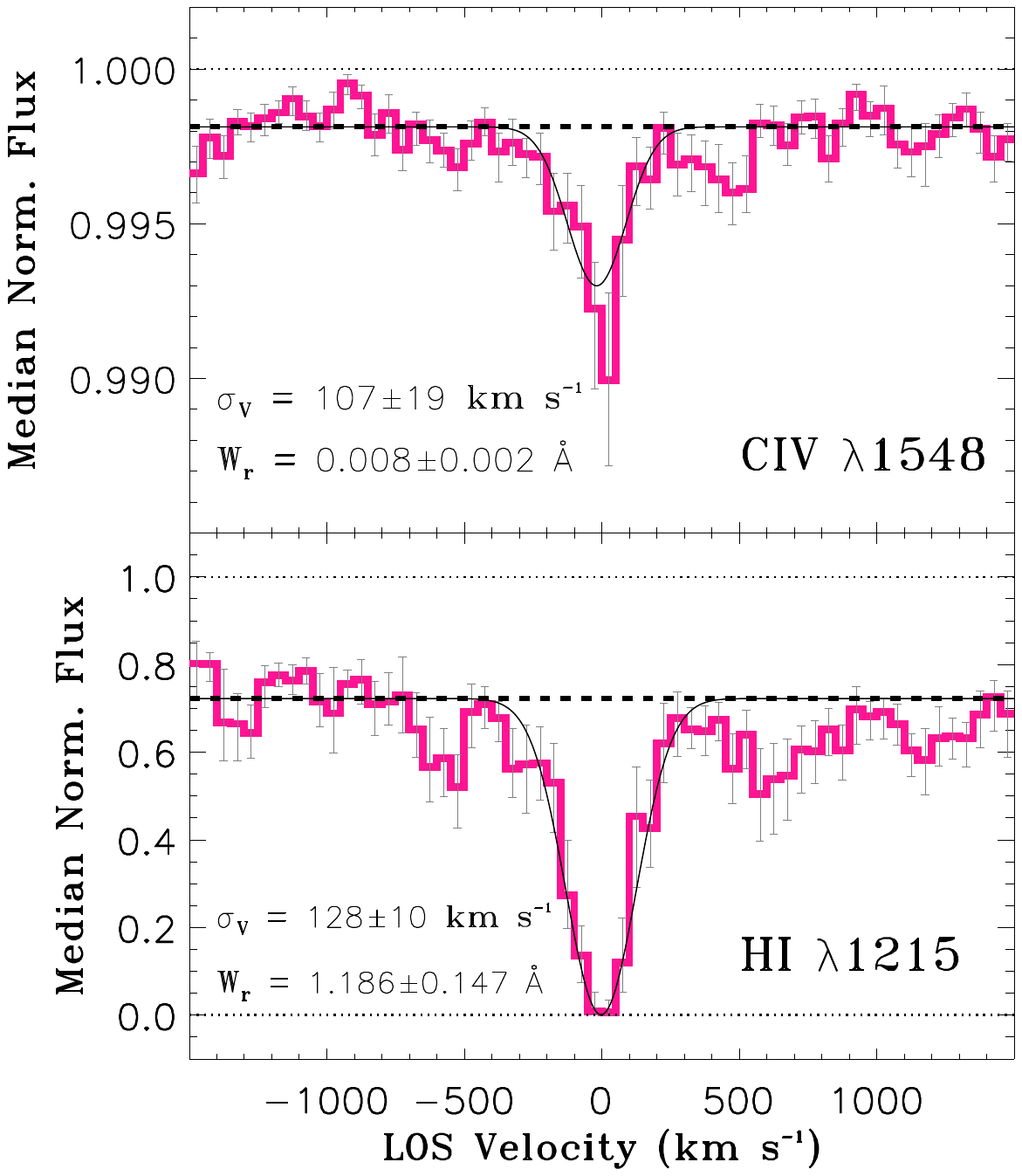} 
\includegraphics[trim=0cm 2cm 0cm 2.5cm,width=0.35\textwidth,angle=00]{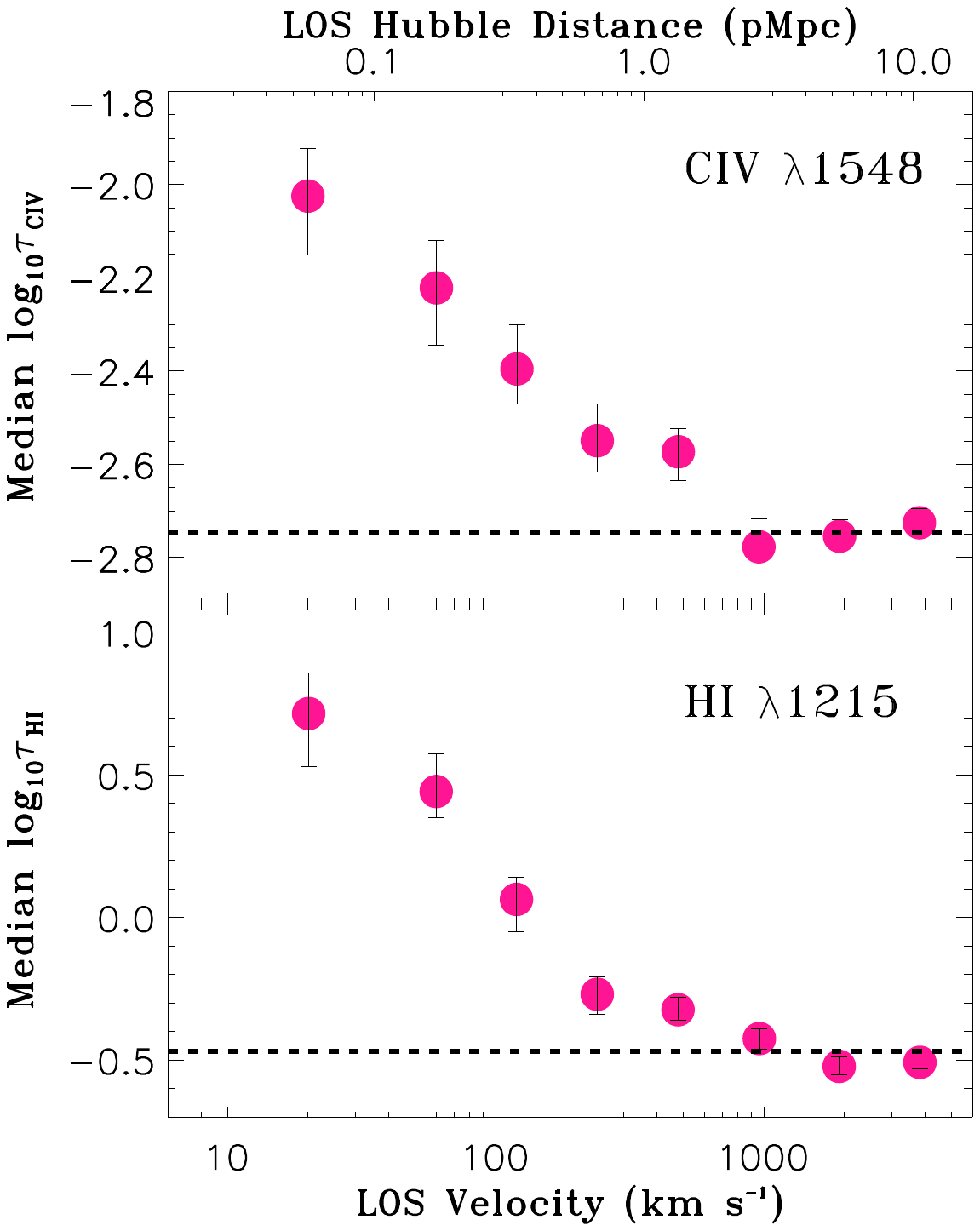} 
}}
}}   
	\caption{{\tt Left:} Mean stacked absorption profiles of \HI\ (bottom) and \CIV\ (top) for the full sample of LAEs. Note the different y--axis ranges for \HI\ and \CIV. There are 77 (92) LAEs contributing to the \HI\ (\CIV) stack. The dotted lines at normalized flux value of 1.0 indicate the actual continua. The dashed lines shows the pseudo continua (i.e., the mean fluxes in random regions) for the stacked profiles, estimated far away from the systemic velocity. The 68\% confidence intervals in all panels are obtained from 1000 bootstrap realizations of the LAE sample. The best--fit single--component Gaussian profiles are shown by the smooth black curves. The line widths, obtained from Gaussian fitting, and $W_r$, determined directly from the stacked profiles (within $\pm300$~\kms\ of the line centroid), are indicated in the corresponding panels. The errors in $W_r$ are estimated from the stacks of the bootstrap samples.       
	{\tt Middle:} The same as the left panels but for median stacked absorption profiles. 
	{\tt Right:} Median stacked optical depth profiles of \HI\ (bottom) and \CIV\ (top) as a function of the absolute line of sight velocity difference. The dashed lines indicate the median optical depth in random regions i.e., far away from the systemic velocity. The top axis indicates the proper line of sight distance at $z=3.3$ (the median redshift) corresponding to the velocity plotted along the bottom axis, assuming pure Hubble flow.}       
\label{fig:fullsample}    
\end{figure*} 

\subsection{Pixel flux distributions} 
\label{subsec:pixelflux}

The dense \lya\ forest absorption seen in the spectra of high--$z$ quasars arises from both virialized halos (i.e., the CGM of foreground galaxies) and from cosmologically evolving density fields, in the form of sheets and filaments in the intergalactic medium (IGM). In Fig.~\ref{fig:CGMvsIGM}, we compare the pixel flux distributions of the \lya\ forest regions in all 8 quasars to that of the CGM of 77 LAEs in our sample. For each quasar, we used fluxes in all the {\sc PODPy} recovered raw spectral pixels (i.e., without binning) lying between the \lyb\ emission and 3000~\kms\ bluewards of the \lya\ emission to determine the IGM+CGM flux distribution. For the CGM pixel flux distributions, we used $\pm100$~\kms\ (bottom) and $\pm1000$~\kms\ (top) velocity windows around \zcorr.

The bottom panel of Fig.~\ref{fig:CGMvsIGM} shows that the CGM pixel flux distribution is significantly different from the distribution of the overall \lya\ forest fluxes. A two-sided Kolmogorov--Smirnov (KS) test returns $D_{\rm KS}= 0.39, P_{\rm KS}=10^{-10}$. The CGM pixel flux distribution shows a strong peak in the lowest flux bin, suggesting significantly stronger absorption (higher optical depth) than seen in the \lya\ forest regions at similar redshifts. The CGM flux distribution has a $+ve$ skew (right--skewed) with a median flux of $0.05$ which  is much lower than the mean value of $0.29$. The \lya\ forest flux distribution, on the other hand, is left--skewed with a median value ($0.74$) higher than the mean ($0.61$). Note that the mean and median values of the two distributions are also significantly different. As expected, the difference between the CGM and the \lya\ forest pixel flux distributions is somewhat reduced ($D_{\rm KS}=0.13 ~\&~ P_{\rm KS}=10^{-10}$) for the $\pm1000$~\kms\ velocity window shown in the top panel. The mean ($0.51$) and median ($0.59$) values of the CGM pixel flux distribution for the $\pm1000$~\kms\ velocity window are somewhat closer to the corresponding values for the \lya\ forest pixel flux distribution. In the next section we show the CGM absorption as a function of line of sight velocity, and indeed confirm that significant excess absorption is seen around galaxy redshifts.

\subsection{Flux and optical depth profiles for the full sample} 
\label{subsec:fullsample}

The mean and median stacked absorption profiles of \HI\ and \CIV\ for the full sample are shown in the left and middle panels of Fig.~\ref{fig:fullsample}, respectively. For a given species (\HI\ or \CIV), the stacked absorption profile is generated by shifting the quasar spectra to the rest-frame of each LAE using the \zcorr\ values, and subsequently calculating the mean/median fluxes in bins of 50~\kms. We verified that our results are insensitive to the bin size. However, we preferred a bin size of 50~\kms\ in order to minimize the effects of correlated noise .  Note that each quasar-galaxy pair provides an independent probe of the CGM. Thus, all quasar spectra were treated equally without putting any weight based on the $S/N$ of the spectra around the wavelength ranges of interests. Since the typical $S/N$ of the spectra is $\gtrsim30$ in the \lya\ forest and $\gtrsim60$ redward of the QSO's \lya\ emission, the scatter in the stacked spectra is likely dominated by the stochastic nature of the gaseous environments around galaxies, and not by the $S/N$ of the individual spectra.

We would like to point out two features seen in the \lya\ stacks shown in the left--bottom and middle--bottom panels of Fig.~\ref{fig:fullsample}: (i) The effective continua (i.e., the pseudo-continua) determined at large velocities ($>3000$~\kms) from the line center are considerably lower than the actual continuum level where the normalized flux is 1.0. This is due to the stochastic absorption in the \lya\ forest. Indeed, the pseudo-continua levels for the mean ($0.60$) and median ($0.72$) stacks are very similar to the mean and median values we obtained for the \lya\ forest pixel flux distribution (IGM+CGM) in the previous section (see \S\ref{subsec:pixelflux}). (ii) The median stacked \lya\ absorption is significantly stronger than the mean stacked absorption. This is owing to the fact that the CGM pixel flux distribution is strongly right--skewed with median value much lower than the mean (see lower panel of Fig.~\ref{fig:CGMvsIGM}). The pseudo-continuum for the median stacked profile is however somewhat higher compared to the mean stacked profile. Consequently, the \lya\ rest-frame equivalent width ($W_r$) measured for the median profile is considerably higher than that of the mean absorption profile as indicated in the figure.

The mean \CIV\ absorption (top--left panel in Fig.~\ref{fig:fullsample}), on the contrary, is significantly stronger than the median \CIV\ absorption (top--middle panel), because the \CIV\ pixel flux distribution is left--skewed. The pseudo-continuum for the median stack is higher compared to the mean stack. This is consistent with what is seen for \lya, however the pseudo-continua values are significantly higher for \CIV\ compared to \HI. This is expected since \CIV\ falls in the red-part of the quasar spectra where the line density for unrelated absorption is significantly lower than in the \lya\ forest.

The 68\% confidence intervals (Fig.~\ref{fig:fullsample}) and standard deviations (not shown) of the stacked spectra are calculated from 1000 bootstrap realizations of the LAE sample. However, we do not use these asymmetric 68\% confidence intervals in any measurements. Instead, we use the standard deviation of flux to obtain the errors on line centroid ($V_0$) and line width ($\sigma_v$) by fitting a single component Gaussian to each stacked absorption profile. We do not use the bootstrap realizations of the stacks to derive uncertainties on $V_0$ and $\sigma_v$, since the Gaussian fits did not converge for many realizations, particularly for the subsamples.

The line centroids of the mean and median stacked \lya\ profiles are $-8\pm11$~\kms\ and $-2\pm10$~\kms, respectively, which are fully consistent with the systemic velocity (i.e., 0~\kms). Thus, the use of \zcorr, instead of \zlae, naturally produces stacked profiles with line centroids consistent with the systemic velocity.  This, however, is not true for every subsample that we explore here. In some cases, particularly for the weaker \CIV\ stacks, we do notice residual offsets owing to sample variance and/or the large intrinsic scatter in the empirical relation used to correct the \zlae\ values. In passing, we note that the uncertainties in the line centroids (and $\sigma_v$) we quoted here should be treated as lower limits, since the fitting errors do not take into account the fact that the noise in nearby spectral pixels is correlated for the adopted bin size of $50$~\kms.

The rest-frame equivalent width, $W_r$, for each absorption stack is derived from direct integration of the observed profiles within $\pm300$~\kms\ (and $\pm500$~\kms) of the line centroid. Such velocity windows encompass most of the absorption (see Fig.~\ref{fig:fullsample}), and are commonly used in the literature \citep[e.g.,][]{Rudie12a,Werk13}. The uncertainties in the equivalent widths are determined from the stacks of the 1000 bootstrap samples. Finally, we calculated column densities using the $W_r$ measured from the mean stacked profiles assuming that the line falls on the linear part of the curve of growth. We caution that the column densities quoted for \HI\ stacks should be considered conservative lower limits, since the \lya\ stacks are heavily saturated. Tables~\ref{tab:HIsummary} and \ref{tab:CIVsummary} summarize all the measurements performed on the \HI\ and \CIV\ profiles, respectively, for all the stacks generated for this study.

The measured \lya\ equivalent width for the median stack of $1.186\pm0.147$~\AA\ ($0.709\pm0.085$~\AA\ for the mean stack) indicates that the line is heavily saturated, and detected at a $>8\sigma$ significance. The $W_r$ of $0.008\pm0.002$~\AA\ for \CIV\ ($0.066\pm0.017$~\AA\ for the mean stack), on the other hand, suggests that the line is very weak but detected at a $\approx4\sigma$ significance. As mentioned earlier, \CIV\ is the only metal line that is detected at a $>3\sigma$ significance. The widths (i.e., velocity dispersions) of the \HI\ and \CIV\ profiles of $128\pm10$~\kms\ ($141\pm11$~\kms\ for the mean stack) and $107\pm19$~\kms\ ($134\pm23$~\kms\ for the mean stack), respectively, are consistent with each other. These line widths are much larger than expected from thermal broadening of $\approx10^{4}$~K gas. \footnote{The thermal line widths ($\sigma$) corresponding to a gas temperature of $10^{4}$~K is only $9.1$ and $2.6$~\kms\ for \HI\ and \CIV, respectively.} The observed line width can be approximated as: $\sigma_{v}^2 = \sigma_{\rm int}^2 + \sigma_{\rm eff}^2$, where $\sigma_{\rm int}$ is the intrinsic line width and $\sigma_{\rm eff}$ is the effective spectral resolution. The $\sigma_{\rm eff}$ will be dominated by the scatter in the relation used to obtain \zcorr, and, to a lesser extent, by the bin size used ($50$~\kms). The $\sigma_{\rm int}$ will have contributions from gravitational motions, dynamical processes such as inflow/outflow, and the Hubble flow.

The median optical depth profiles of \HI\ and \CIV\ are shown in the right panel of Fig.~\ref{fig:fullsample}. Instead of using the flux, here we used the optical depths for stacking. The median optical depth values are calculated in constant logarithmic LOS velocity bins of 0.301 dex. In order to increase the $S/N$, the optical depth profiles are made one-sided by using the absolute of LOS velocity difference. The errors in the optical depths (68\% confidence intervals) are calculated from 1000 bootstrap realizations of the LAE sample. The optical depth signals for both \HI\ and \CIV\ are significantly enhanced for LOS velocities $\le500$~\kms, compared to the values expected in random regions in the universe as indicated by the dashed horizontal lines in the plot. For LOS velocities $\le30$~\kms, the optical depths for both \HI\ and \CIV\ are enhanced by almost an order of magnitude. Owing to the large dynamic range of optical depth, such profiles are particularly useful to compare stacks constructed using different samples of galaxies.


\subsection{Environmental dependence}  
\label{subsec:environment}

\begin{figure}
\includegraphics[trim=2.0cm 8.2cm 1.0cm 10.0cm,width=0.50\textwidth,angle=00]{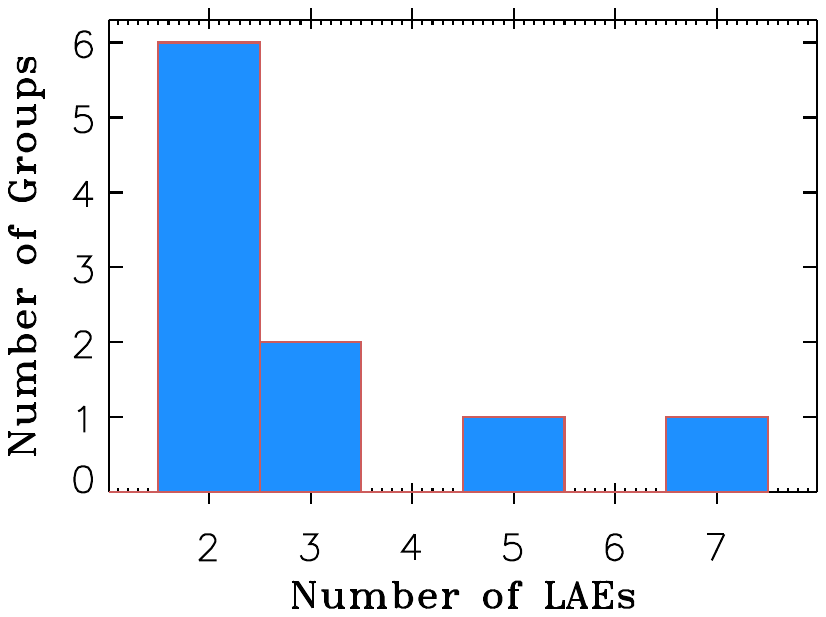} 
\caption{Histogram of the number of galaxies per group in the `group' subsample. Most of the `groups' in our sample consist of 2 detected LAEs.}        
\label{fig:group_ngal}      
\end{figure} 

\begin{figure*}
\centerline{\vbox{
\centerline{\hbox{ 
\includegraphics[trim=0cm 1cm 0cm 3.0cm,clip,width=0.45\textwidth,angle=00]{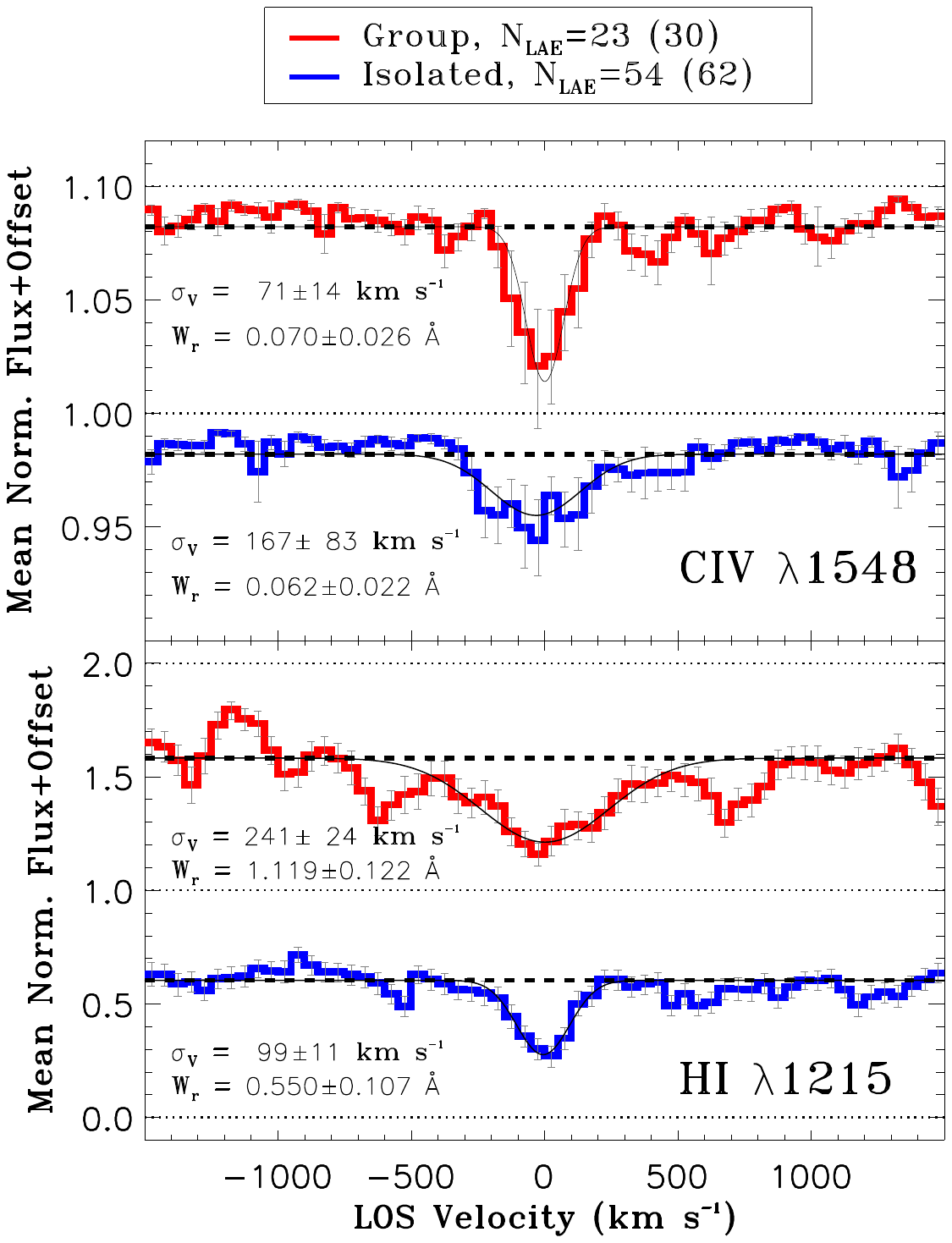} 
\includegraphics[trim=0cm 1cm 0cm 3.0cm,clip,width=0.45\textwidth,angle=00]{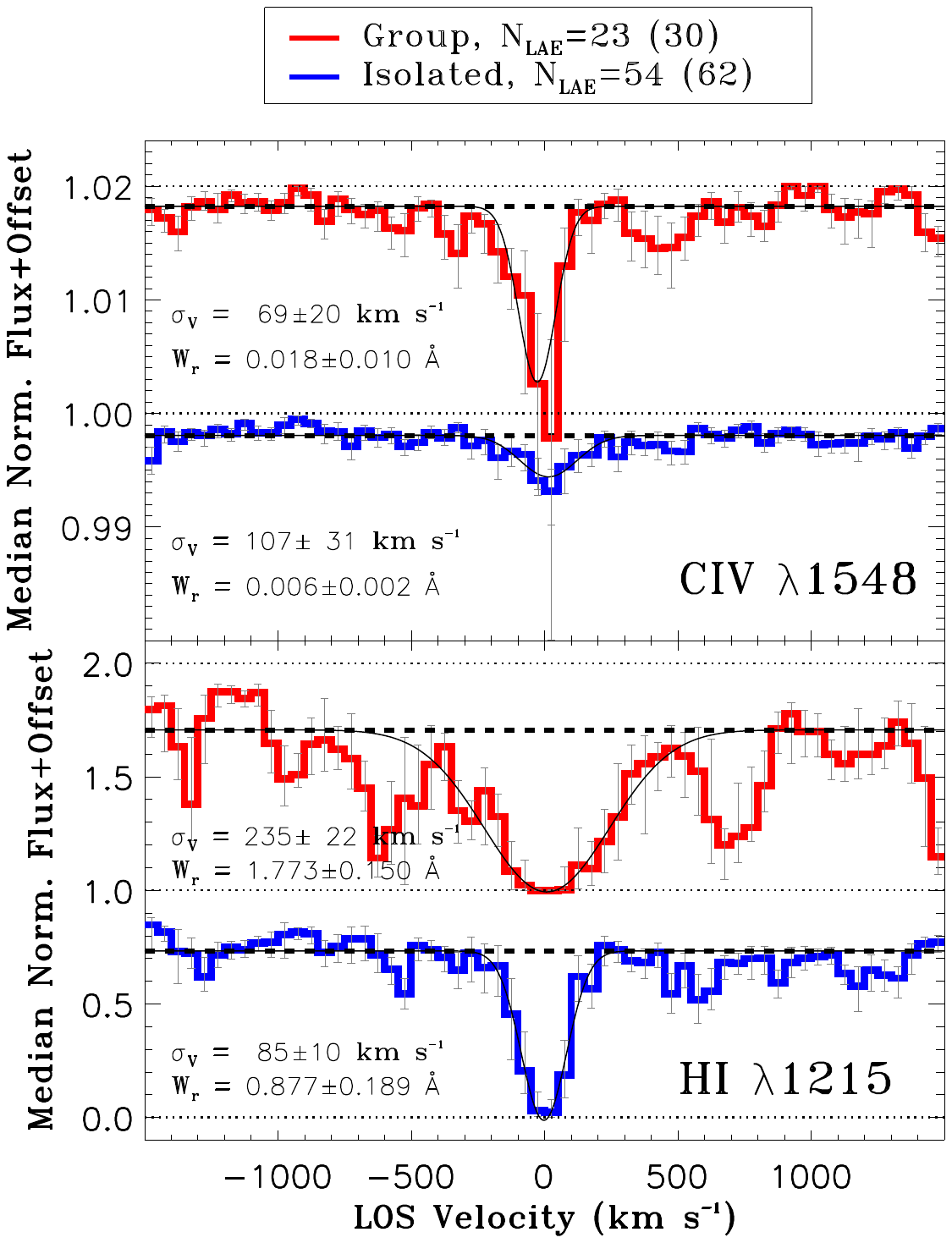} 
}}
}}   
\caption{{\tt Left:} Mean stacked \HI\ (bottom) and \CIV\ (top) profiles for the isolated and `group' subsamples as marked by the legends on the top. The number of LAEs contributing to the \HI\ stack is indicated. The corresponding number for the \CIV\ stack is given in parentheses. The stacked profiles for the `group' subsample are shifted vertically by 1.0 for \HI\ and by 0.10 for \CIV\ for clarity. All other features are the same as described in Fig.~\ref{fig:fullsample}.    
{\tt Right:} The same as the left panels but for median stacked profiles. The \CIV\ profile for the `group' subsample is offset by 0.02 for clarity. A strong (marginal) environmental dependence is seen for the \lya\ (\CIV) absorption.}   
\label{fig:flux_environment}      
\end{figure*} 

\begin{figure}
\centerline{\vbox{
\centerline{\hbox{ 
\includegraphics[trim=0cm 1cm 0cm 2.5cm,clip,width=0.45\textwidth,angle=00]{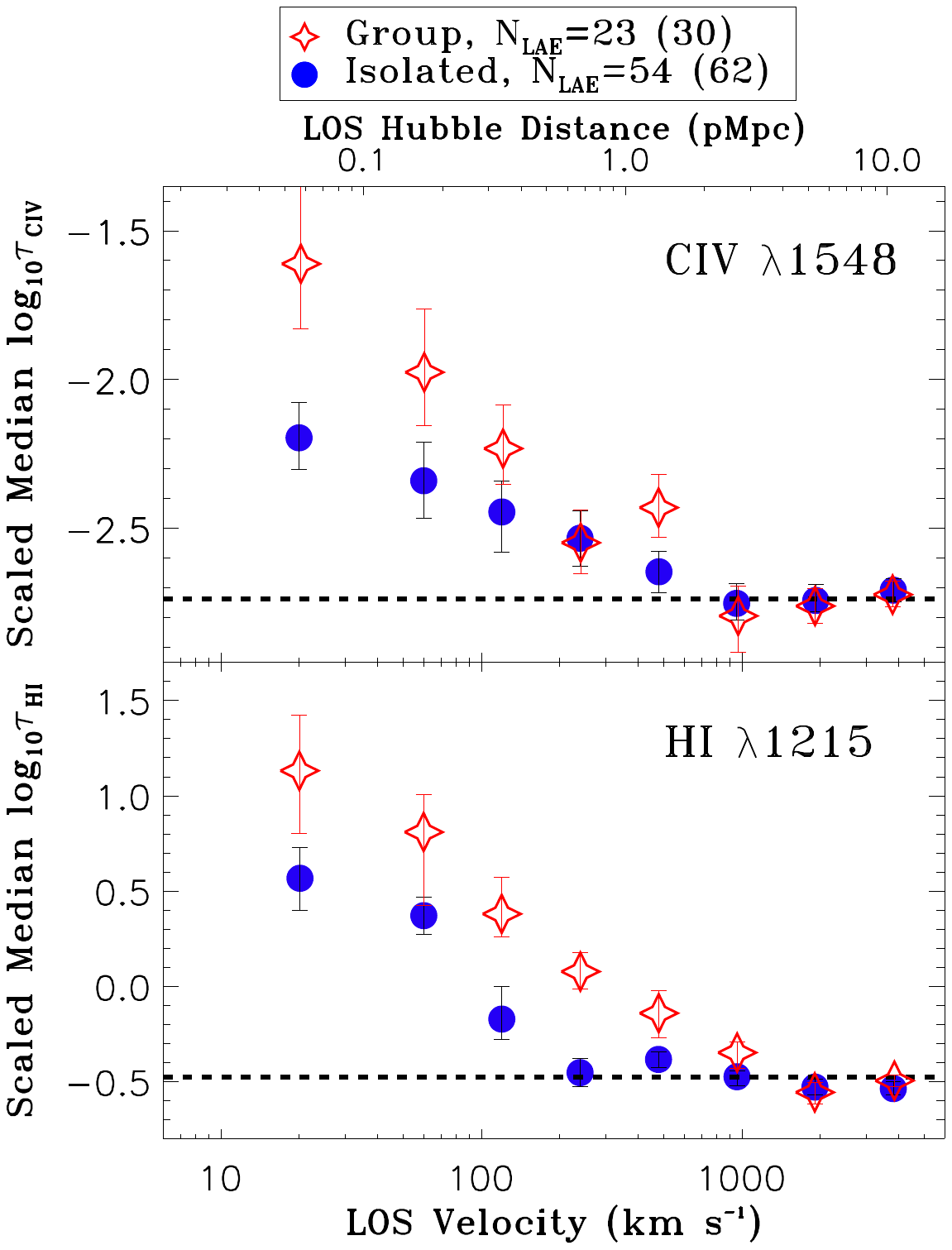} 
}}
}}   
	\caption{The median optical depth profiles of \HI\ (bottom) and \CIV\ (top) for the isolated (blue filled circles) and `group' (red star symbols) subsamples. The top axis indicates the proper line of sight distance at $z=3.3$ corresponding to the velocity plotted along the bottom axis, assuming pure Hubble flow. To facilitate a proper comparison, the profiles of isolated and `group' subsamples are taken to the same common baseline by linearly shifting one of the profiles by the difference in the background levels measured for the two profiles. A strong environmental dependence is seen for both \HI\ and \CIV\ with larger excess absorption around LAEs in `groups'.}   
\label{fig:tau_environment}      
\end{figure} 

The properties of the CGM of a galaxy may vary with the large-scale environment in which it resides. To examine the effects of the environment on the CGM absorption, we  divided our sample into `isolated' and `group' subsamples. A galaxy (LAE) is called isolated if there is one and only one LAE detected within the MUSE FoV and within a linking velocity, $v_{\rm link}$, of $\pm500$~\kms\ of its redshift (`all\_environment\_iso' in Table~\ref{tab:sample_summary}). A total of 66/96 LAEs satisfy these criteria. For the remaining 30 `group' LAEs, there is at least one companion LAE with a LOS velocity less than $v_{\rm link}$  (`all\_environment\_group' in Table~\ref{tab:sample_summary}). We emphasize here that by `group' we {\em do not} necessarily refer to a virialized structure. It is rather a measure of overdensity. Fig.~\ref{fig:group_ngal} shows the distribution of the number of LAEs in the `group' subsample. There are a total of 10 `groups', 60\% (20\%) of them are consist of two (three) member LAEs, and 20\% of them have 5 or more LAEs. Note that we found 96 LAEs over a total redshift path-length of $\approx6.8$, leading to $d\mathcal{N}_{\rm LAE}/dz$ of $\approx14$ and $d\mathcal{N}_{\rm LAE}/dv$ of $\approx0.2$ for $\delta v = 1000$~\kms. Thus, detecting 2 LAEs within $\pm500$~\kms\ corresponds to an overdensity of $\approx10$.


The left panels of Fig.~\ref{fig:flux_environment} shows the mean stacked \HI\ (bottom) and \CIV\ (top) absorption profiles arising from the isolated (blue) and `group' (red) subsamples. The corresponding median stacked profiles are shown in the right panels. A strong environmental dependence, particularly for \HI, is evident. The CGM absorption is significantly stronger for the `group' galaxies compared to the isolated ones. The rest equivalent width measured from the mean \HI\ profile, within a $\pm500$~\kms\ velocity window, for the `group' galaxies is $2.279\pm0.305$~\AA\ ($1.467\pm0.210$~\AA\ for the mean stack), which is $\approx2.5$ times larger than that measured for the isolated galaxies (see Table~\ref{tab:HIsummary}). In addition, the width of the \HI\ absorption is $235\pm22$~\kms\ ($241\pm24$~\kms\ for the mean stack) for the `group' subsample which is $\approx2.8$ times larger than for the isolated subsample. The equivalent widths of both the median and mean stacked \CIV\ profiles for the `group' galaxies are somewhat larger than for the isolated galaxies (see Table~\ref{tab:CIVsummary}). However, owing to the intrinsic weakness of the \CIV\ profiles, the difference in equivalent width is only marginally significant.

The differences in strengths and widths of the \HI\ absorption for the two subsamples are also evident from the optical depth profiles shown in the bottom panel of Fig.~\ref{fig:tau_environment}. The optical depths for the `group' subsample (red star symbols) are consistently higher compared to the isolated subsample (blue filled circles) out to $\approx500$~\kms. The optical depth profile for the isolated subsample shows a sharp decline at LOS velocity of $\approx100$~\kms, and quickly become consistent with the value measured at random regions (dashed line). The decline in the `group' optical depth profile, on the contrary, is more gradual with an excess optical depth seen out to $\approx500$~\kms. The difference in \CIV\ absorption is also conspicuous as shown in the top panel, at least at small LOS velocities.

In order to investigate whether the observed strong environmental dependence is driven by any underlying properties of the LAEs in our sample, we performed several two-sided KS-tests. The median values of different properties of the isolated and `group' subsamples and the KS-test results are summarized in Table~\ref{tab:environment2}. The distributions of these properties are not statistically different between the isolated and `group' subsamples. Although the difference is not statistically significant, the median impact parameter of the isolated subsample is somewhat smaller compared to the `group' subsample. However, this difference would produce the opposite effect (i.e., stronger CGM absorption for the isolated subsample) than what we find here. Besides, the median impact parameter of the closest (to the background quasar) members of the `group' galaxies is 152~pkpc, which is very similar to the isolated subsample (Table~\ref{tab:environment2}). A two-sided KS-test is consistent with the hypothesis that the two distributions are drawn from the same parent populations ($D_{KS}=0.16$, $P_{KS}=0.97$). We therefore rule out the possibility that the observed difference is driven by differences in the impact parameter distributions between the isolated and `group' subsamples.

Owing to the limited FoV of MUSE, it is not possible to determine if an isolated galaxy at the edge of the field is truly isolated, since there might be companion LAEs just outside the FoV. If at all, this would produce stronger absorption for the isolated sample. However, we verified that the \HI\ absorption for the isolated subsample does not change appreciably even if we exclude the LAEs at the edge of the FoV by selecting the isolated LAEs with $\rho <200$~pkpc only. Next, we used all 23/30 (30/30) `group' LAEs satisfying our redshift bounds (see Table~\ref{tab:sample_summary}) for \HI\ (\CIV) stacking. We find that the environmental dependence remains significant even if we use only one LAE from each `group' for stacking.\footnote{Eight (ten) galaxies from the 10 `groups' satisfy the redshift bounds for the \HI\ (\CIV) stacking.} Finally, we find that the stronger absorption for the `group' subsample is not entirely driven by the 2 `groups' with the highest numbers of LAEs (see Fig.~\ref{fig:group_ngal}). The difference remains significant when we exclude the 12 LAEs (5+7) arising from the two `groups'. We therefore conclude that the observed environmental dependence of the CGM absorption is real and robust. It is quite remarkable that such a simple indicator of environment produces such a strong difference in \HI\ absorption.

For the results presented above, we assumed a $v_{\rm link}$ of $\pm500$~\kms\ for defining the isolated/`group' galaxies. We verified that the strong environmental dependence persists for $v_{\rm link}=\pm250$~\kms. However, the difference is reduced significantly for $v_{\rm link}=\pm1000$~\kms. A velocity difference of 1000~\kms\ at $z\approx3.3$ corresponds to a length-scale of $\approx2.8$~pMpc assuming a pure Hubble flow. This length-scale is considerably larger than the scale of a typical galaxy group at similar redshifts. Therefore, the `group' subsample defined with $v_{\rm link}=1000$~\kms\ includes LAEs that are isolated (i.e., not related dynamically). Consequently, the CGM absorption of the corresponding `group' sample is weakened, suppressing the strong environmental dependence otherwise seen for smaller $v_{\rm link}$ values.

Besides being wider and stronger, the \lya\ stack for the `group' subsample seems to show marginally significant secondary peaks both blue- ($\rm \approx [-900, -400]$~\kms) and red- ($\rm \approx [+500, +1000]$~\kms) ward of the primary peak at $\approx0$~\kms\ (Fig.~\ref{fig:flux_environment}). Peaks at particular velocities other than zero are not expected a priori. Indeed, the peaks at negative and positive velocities appear at somewhat different absolute velocity separations. A measurement of their true significance would thus have to account for the `look elsewhere effect', i.e. the fact that it is more likely to find a noise spike somewhere when we search over a larger velocity interval, as well as for the correlations between neighboring pixels. These corrections would further decrease the significance of the secondary peaks. Such an analysis is best done using comparisons with forward models with mock/simulation data, which is beyond the scope of the present study and in our opinion unwarranted given the already marginal significance of the enhancement of the optical depth compared to random regions (Fig.~\ref{fig:tau_environment}).

\begin{table} 
\begin{center}   
\begin{threeparttable}[b] 
\caption{Statistics of the isolated and `group' subsamples}     
\begin{tabular}{lrrc}    
\hline  
Properties      &  ~~~~~Isolated     &   ~~~~~Group  &  ~~~~~~~~KS--test results  \\  
(1)             &  (2)           &  (3)           &  (4)               \\  
\hline  
$\rho$ (pkpc)            	        &  $153.1$     &   $194.7$   &  $D=0.22$, $P=0.26$  \\    
$z$                    			&  $3.35$      &   $3.32$    &  $D=0.21$, $P=0.31$  \\  
$\log_{\rm 10} L$(\lya)/erg s$^{-1}$    &  $41.99$     &   $41.97$   &  $D=0.10$, $P=0.98$  \\  
FWHM (\kms)              		&  $248.9$     &   $231.6$   &  $D=0.14$, $P=0.76$  \\ 
$\log_{\rm 10} \rm SFR/M_{\odot}~yr^{-1}$$^{a}$  &  $-0.13$     &   $-0.02$   &  $D=0.19$, $P=0.44$  \\   
$EW_0$ (\AA)$^{a}$                  &  $48.3$      &   $52.3$    &  $D=0.22$, $P=0.29$  \\ 
\hline  
\end{tabular} 
\label{tab:environment2}     
	Notes-- (1) Properties; (2) Median value of the property in \#1 for the isolated subsample; (3) Median value of the property in \#1 for the `group' subsample; (4) Results of the two-sided KS-test performed on the isolated and `group' subsamples for the property in \#1. Here, $D$ denotes the maximum difference between the cumulative distributions, and $P$ denotes the probability of obtaining the value $D$ by chance. $^{a}$Excluding the 15 LAEs blended with low-$z$ continuum emitters and assuming limits for the continuum undetected LAEs as measurements.    
\end{threeparttable}  
\end{center}   
\end{table} 

\begin{figure*}
\centerline{\vbox{
\centerline{\hbox{ 
\includegraphics[trim=0.0cm 0.0cm 0.0cm 0.5cm,width=0.45\textwidth,angle=00]{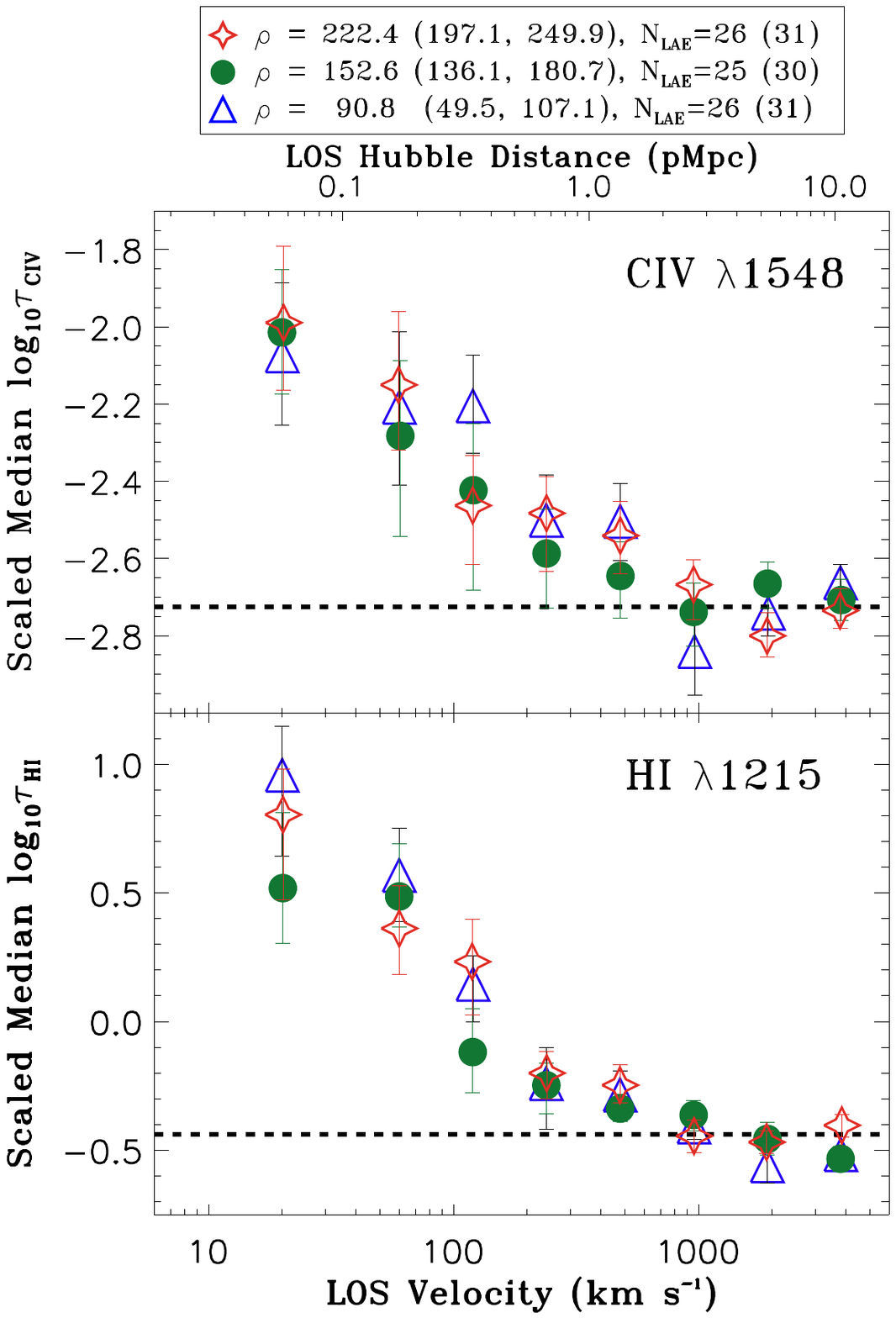} 
\includegraphics[trim=0.0cm 0.0cm 0.0cm 0.5cm,width=0.45\textwidth,angle=00]{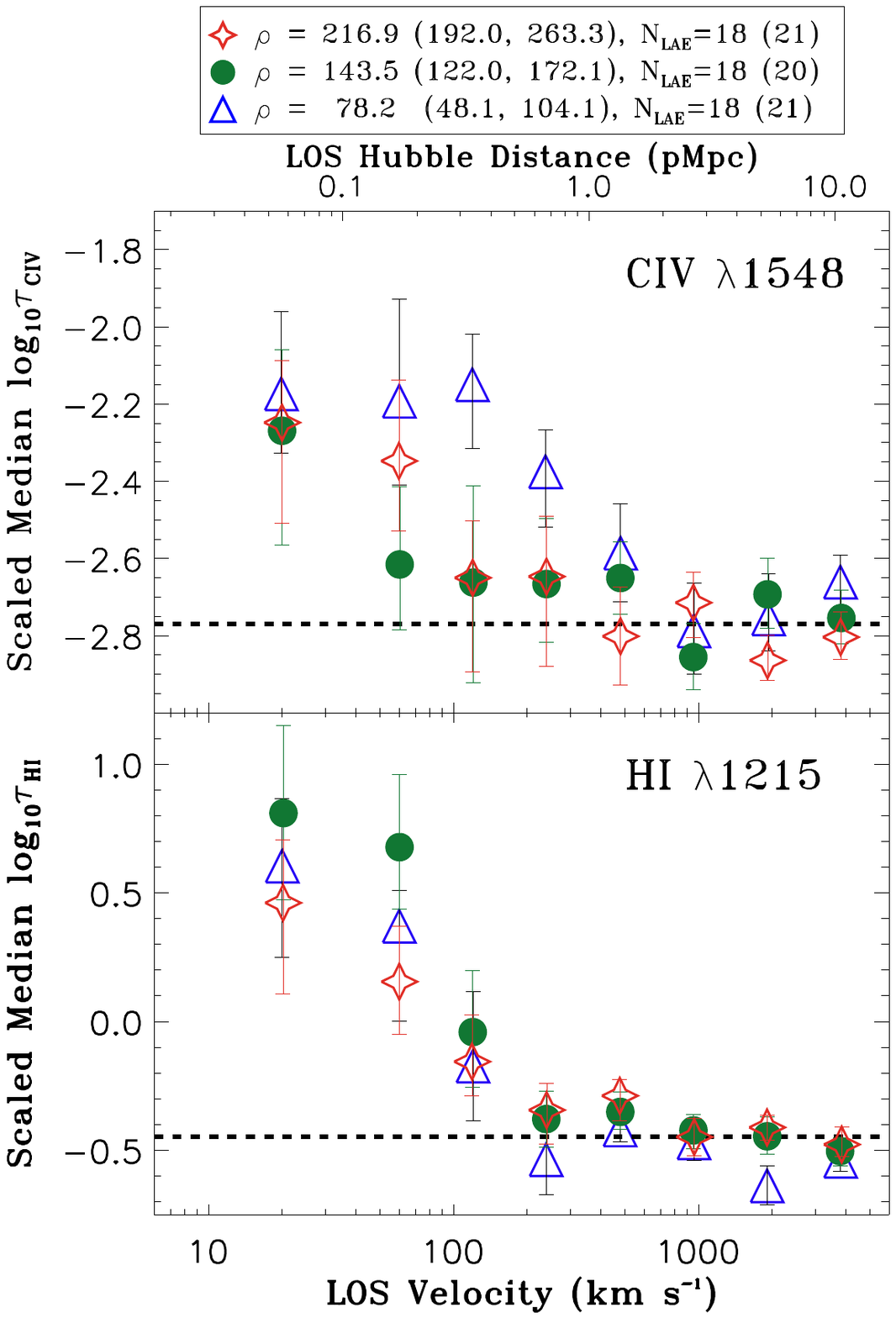} 
}}
}}   
\caption{{\tt Left:} Similar to Fig.~\ref{fig:tau_environment} but for the three impact parameter bins: all\_$\rho$\_low, all\_$\rho$\_mid, all\_$\rho$\_high, as defined in Table~\ref{tab:sample_summary}. The median values and 68 percentile ranges in the parentheses in the legends are for the LAEs contributing to the \HI\ stack. The corresponding values for the \CIV\ stack can be found in Table~\ref{tab:sample_summary}. {\tt Right:} The same as the left panels but for the iso\_$\rho$\_low, iso\_$\rho$\_mid, and iso\_$\rho$\_high subsamples (see Table~\ref{tab:sample_summary}). Neither \HI\ nor \CIV\ absorption show any significant, systematic dependence on impact parameter.}  
\label{fig:rhodep}     
\end{figure*} 

\subsection{Impact parameter dependence}    
\label{subsec:radialprofile}

Determining how gas and metals are distributed around our galaxies is one of our primary goals. We thus generated stacks of \HI\ and \CIV\ lines for the three different impact parameter bins (`all\_$\rho$\_low', `all\_$\rho$\_mid', and `all\_$\rho$\_high' in Table~\ref{tab:sample_summary}) corresponding to the three tertiles of the $\rho$-distribution. Since a strong environmental dependence was seen in the previous section, we further generated stacks for three impact parameter bins (`iso\_$\rho$\_low', `iso\_$\rho$\_mid', and `iso\_$\rho$\_high' in Table~\ref{tab:sample_summary}) using only the isolated LAEs (`isolated only').

The median optical depth profiles of \HI\ and \CIV\ for the three impact parameter bins constructed from the full sample are shown in the left panel Fig.~\ref{fig:rhodep}. The right panel of the figure shows the corresponding profiles for the isolated LAEs. Owing to the smaller sample sizes, the profiles in the right hand panel are noisier. We do not find any significant impact parameter dependence of \HI\ or \CIV\ absorption. The optical depth profiles are consistent with each other within the allowed uncertainties across the whole LOS velocity range.

\begin{figure}
\includegraphics[trim=1.0cm 0.8cm 0.3cm 2cm,width=0.45\textwidth,angle=00]{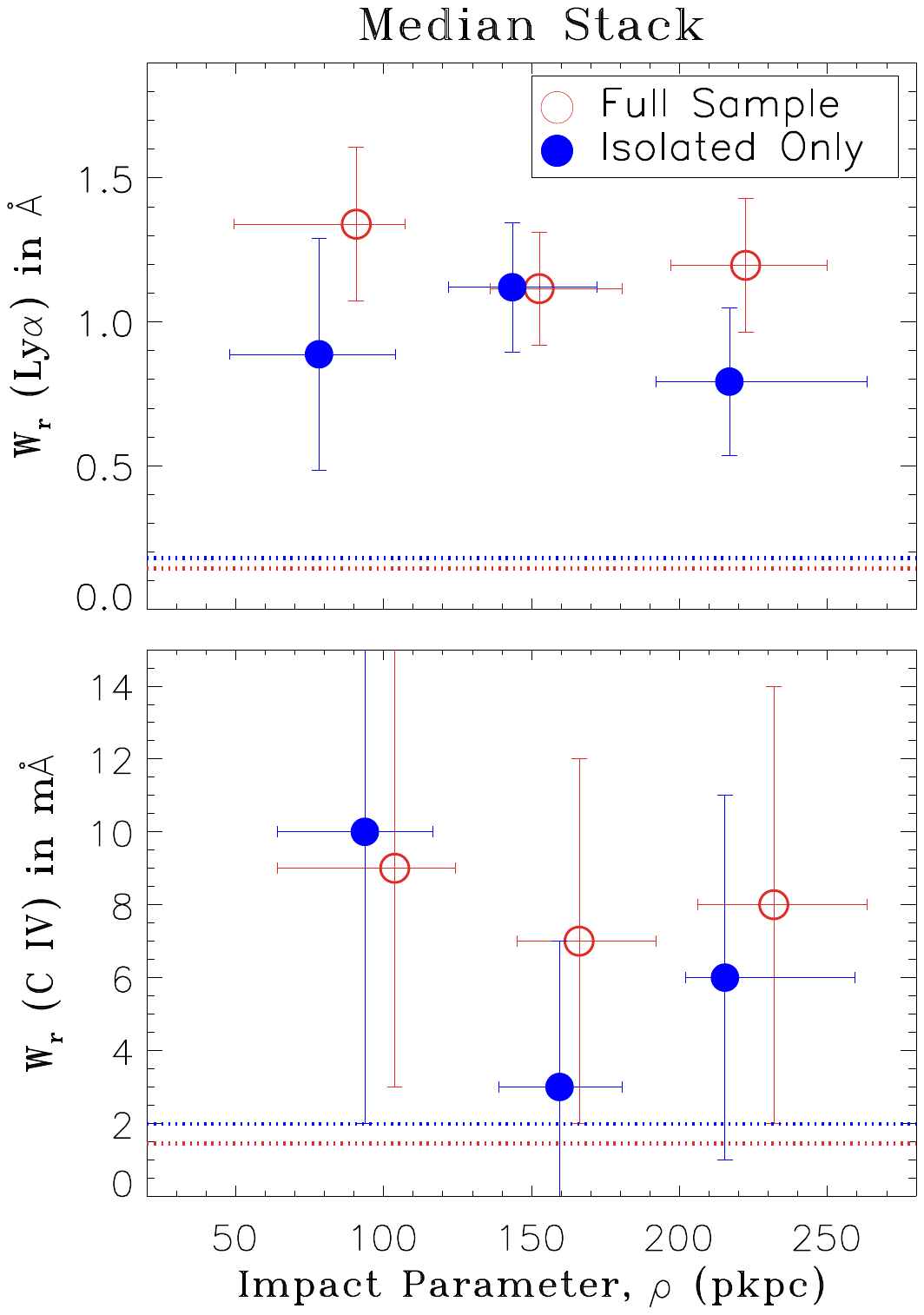} 
\caption{Rest equivalent widths measured within $\pm300$~\kms\ of the line centroids of the median stacked absorption profiles of \lya\ (top) and \CIV\ (bottom) as a function of impact parameter. The $W_r$-profiles for the full sample are shown by the open red circles, whereas the filled blue circles are for the `isolated only' subsample. The error bars along the x--axis indicate 68 percentile ranges. The error bars along y-axis are the $1\sigma$ measurement uncertainties from bootstrap resampling. The horizontal dotted lines represent the $3\sigma$ limiting $W_r$ values. The limiting $W_r$ values are calculated from the continuum $S/N$ of the stacked spectra and assuming a line width, $\sigma_v$, of 100~\kms. We do not find any significant trend between CGM absorption and $\rho$ for the impact parameter range probed by our survey.  
}     
\label{fig:radialprofile}    
\end{figure}

While the profiles of absorption as a function of LOS velocity are important, redshift space distortions due to peculiar motions complicate their interpretation. The dependence of the $W_r$ and column density of different species (neutral hydrogen and metal ions) on impact parameter provides complementary information on the overall distribution of circumgalactic gas and metals. Fig.~\ref{fig:radialprofile} shows the median equivalent widths of \HI\ and \CIV\ absorption as a function of impact parameter.  The corresponding flux profiles are shown in Figs.~\ref{fig:rhodep_full} \& \ref{fig:rhodep_isolated}. The \HI\ and \CIV\ equivalent widths measured from these profiles are summarized in Tables~\ref{tab:HIsummary} \& \ref{tab:CIVsummary} respectively. Interestingly, the \HI\ $W_r$-profile for the full sample (open red circles) is flat with a slope $dW_r/d\rho\approx0$. This is true for both the median and mean (not shown) stacked profiles. The \HI\ $W_r$-profiles corresponding to the `isolated only' sample (filled blue circles) are noisier owing to the smaller sample sizes. However, no significant trend with impact parameter is seen even for the `isolated only' sample either. We note here that for galaxies near the edge of the FoV we can only search for neighbours on one side, whereas for galaxies in the center we can search for group members on all sides. This may lead us to misclassify group galaxies near the edge of the field as isolated objects. Because the absorption is stronger around group galaxies, this `edge effect' may bias us to overestimate the absorption for larger impact parameters for the `isolated only' sample, thus obscuring a possible trend of decreasing absorption with increasing galactocentric distance.

Due to the intrinsic weakness of the \CIV\ absorption, the corresponding $W_r$-profiles are significantly more noisier. Given the large uncertainties in individual measurements, we cannot draw any robust conclusion on the \CIV\ $W_r$-profile. However, a tentative negative trend is seen between \CIV\ equivalent width and impact parameter when we split the `isolated only' sample into two $\rho$ bins.

\subsection{Redshift dependence}  
\label{subsec:redshiftevolution}    

The LAEs in our sample span the redshift range 2.92--3.82, covering $\approx0.6$~Gyr of cosmic time. In this section, we inspect whether there is any redshift evolution of the CGM absorption. Note that the redshift range probed by our sample is too small to expect strong evolution. For this exercise, we first exclude the 30 `group' LAEs in order to eliminate any possible effects due to the environment. The remaining 66 LAEs are divided into two subsamples iso\_\zlae\_low and iso\_\zlae\_high (see Table~\ref{tab:sample_summary} for details) bifurcated at the median redshift.

\begin{figure}
\centerline{\vbox{
\centerline{\hbox{ 
\includegraphics[trim=0.0cm 1.0cm 0.5cm 2.5cm,width=0.45\textwidth,angle=00]{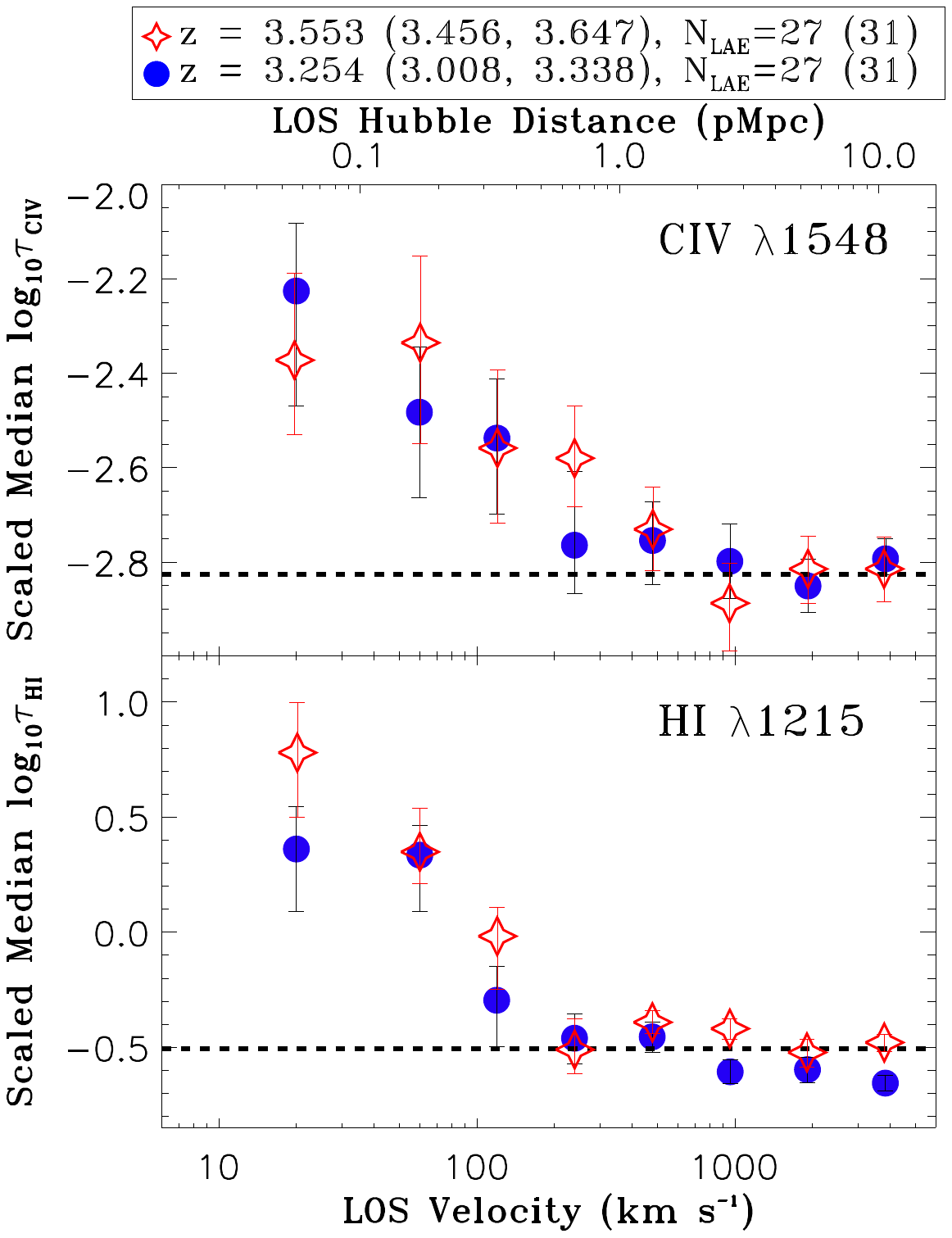} 
}}
}}   
\caption{Similar to Fig.~\ref{fig:tau_environment} but for two different redshift bins (iso\_\zlae\_low and iso\_\zlae\_high). The median redshifts and 68 percentile ranges in the parentheses in the legends are for the LAEs contributing to the \HI\ stack. The corresponding values for the \CIV\ stack can be found in Table~\ref{tab:sample_summary}. No significant dependence on redshift is seen for either \HI\ or \CIV\ absorption. 
}  
\label{fig:redshiftevolution}    
\end{figure}

The optical depth profiles of \HI\ and \CIV\ corresponding to these two subsamples are shown in Fig.~\ref{fig:redshiftevolution}. They do not show any dependence on redshift. The rest-frame equivalent widths measured from the stacked profiles (see Fig.~\ref{fig:redshiftevolution_appendix}) also confirm the lack of any significant redshift evolution in the CGM absorption in our sample (see Tables~\ref{tab:HIsummary} \& \ref{tab:CIVsummary}).

\subsection{\lya\ luminosity dependence}  
\label{subsec:lumdependence}   

Next, we investigate the dependence of CGM absorption on the \lya\ line luminosity, $L$(\lya). In order to avoid any underlying environmental dependence as seen in Section~\ref{subsec:environment}, we will use only the isolated LAEs for this analysis. The $L$(\lya) of the isolated LAEs span the range of $10^{41.3} - 10^{42.9}~\rm erg~s^{-1}$, with a median value of $10^{42.0}~\rm erg~s^{-1}$. To investigate the $L$(\lya) dependence, we split the isolated LAEs into two subsamples (`iso\_$L$(\lya)\_low' and `iso\_$L$(\lya)\_high') about the median value (see Table~\ref{tab:sample_summary} for details).

\begin{figure}
\centerline{\vbox{
\centerline{\hbox{ 
\includegraphics[trim=0.0cm 1.0cm 0.5cm 2.5cm,width=0.45\textwidth,angle=00]{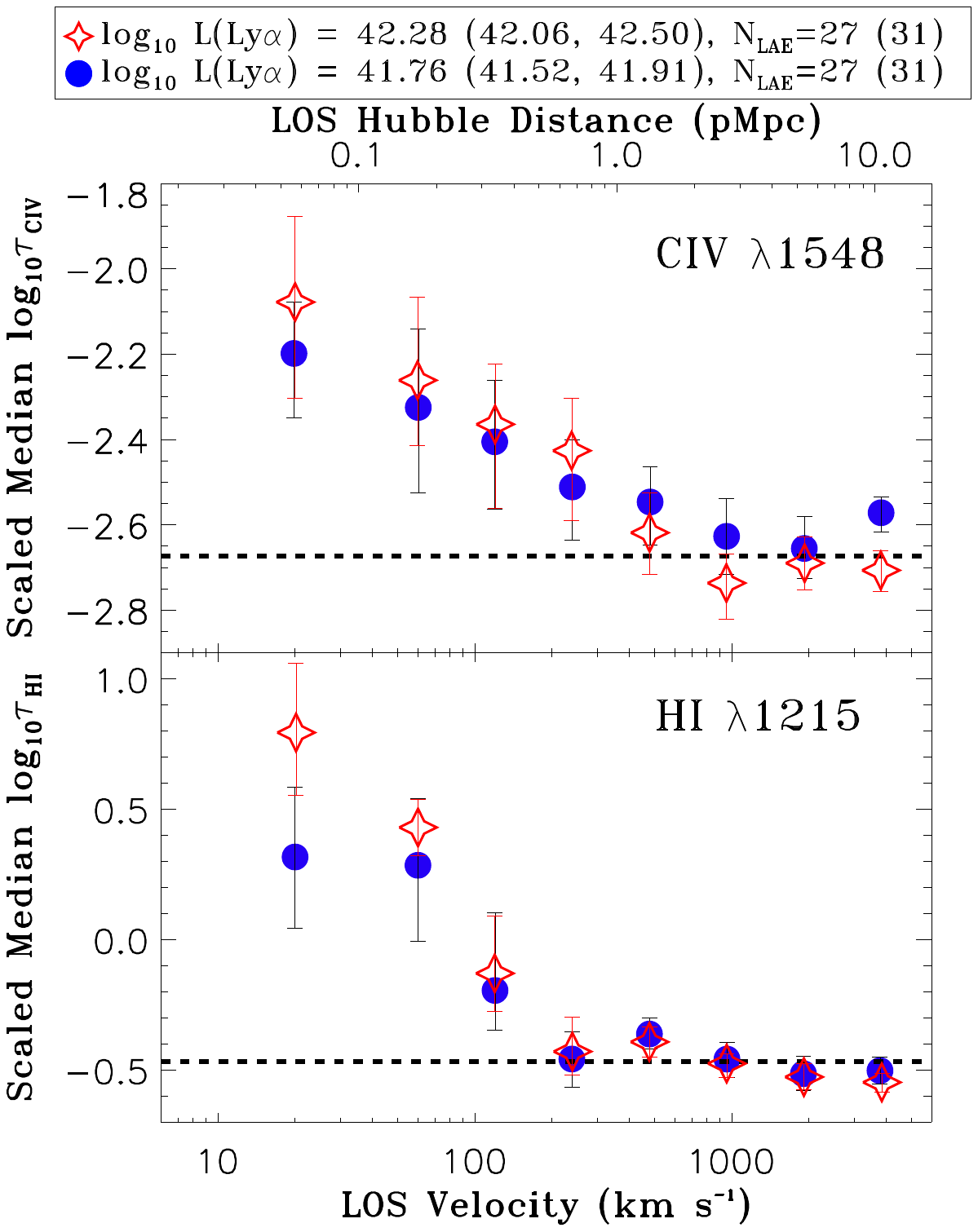} 
}}
}}   
\caption{Similar to Fig.~\ref{fig:tau_environment} but for two different luminosity bins (iso\_$L\rm(Ly\alpha)$\_low and iso\_$L\rm(Ly\alpha)$\_high). The median luminosities and 68 percentile ranges in the parentheses in the legends are for the LAEs contributing to the \HI\ stack. The corresponding values for the \CIV\ stack can be found in Table~\ref{tab:sample_summary}. No significant dependence on \lya\ line luminosity is seen for either \HI\ or \CIV\ absorption.}  
\label{fig:lumdependence}    
\end{figure}

The optical depth profiles of \HI\ and \CIV\ for these two subsamples are shown in Fig.~\ref{fig:lumdependence}. Neither shows any trend with $L$(\lya). The flux profiles, shown in Fig.~\ref{fig:lumdependence_appendix}, also confirm the lack of any strong $L$(\lya) dependence. Nonetheless, we note that the equivalent widths of both \HI\ and \CIV\ are somewhat higher for the iso\_$L$(\lya)\_high subsample compared to the iso\_$L$(\lya)\_low subsample, but consistent with each other within the $1\sigma$ allowed ranges (see Tables~\ref{tab:HIsummary} \& \ref{tab:CIVsummary}). We point out that the median $L$(\lya) of the iso\_$L$(\lya)\_high sample is only a factor $\approx3.3$ higher compared to the iso\_$L$(\lya)\_low sample. Samples with larger dynamic ranges in $L$(\lya) are required to confirm this tentative trend.

\subsection{FWHM dependence}  
\label{subsec:fwhmdependence}   

The FWHM of an emission line is expected to be correlated with its luminosity since both are related to the area under the line. We found a strong ($4.4\sigma$) correlation between $L$(\lya) and FWHM in our sample (see Fig.~\ref{fig:cornerplot}). Nonetheless, there is a significant scatter in the relation. We thus investigate here whether the CGM absorption depends on the FWHM of the \lya\ emission line. Note that, the FWHM of the \lya\ lines were directly calculated from the MUSE spectra without any modeling and without correcting for the instrumental broadening. The median FWHM of the 66 isolated LAEs is 248.9~\kms. We generated two subsamples `iso\_FWHM\_low' and `iso\_FWHM\_high' by bifurcating the isolated LAEs at the median FWHM value. The median FWHM and 68 percentile ranges of each subsamples are summarized in Table~\ref{tab:sample_summary}.

\begin{figure}
\centerline{\vbox{
\centerline{\hbox{ 
\includegraphics[trim=0.5cm 1.0cm 0.5cm 2.5cm,width=0.45\textwidth,angle=00]{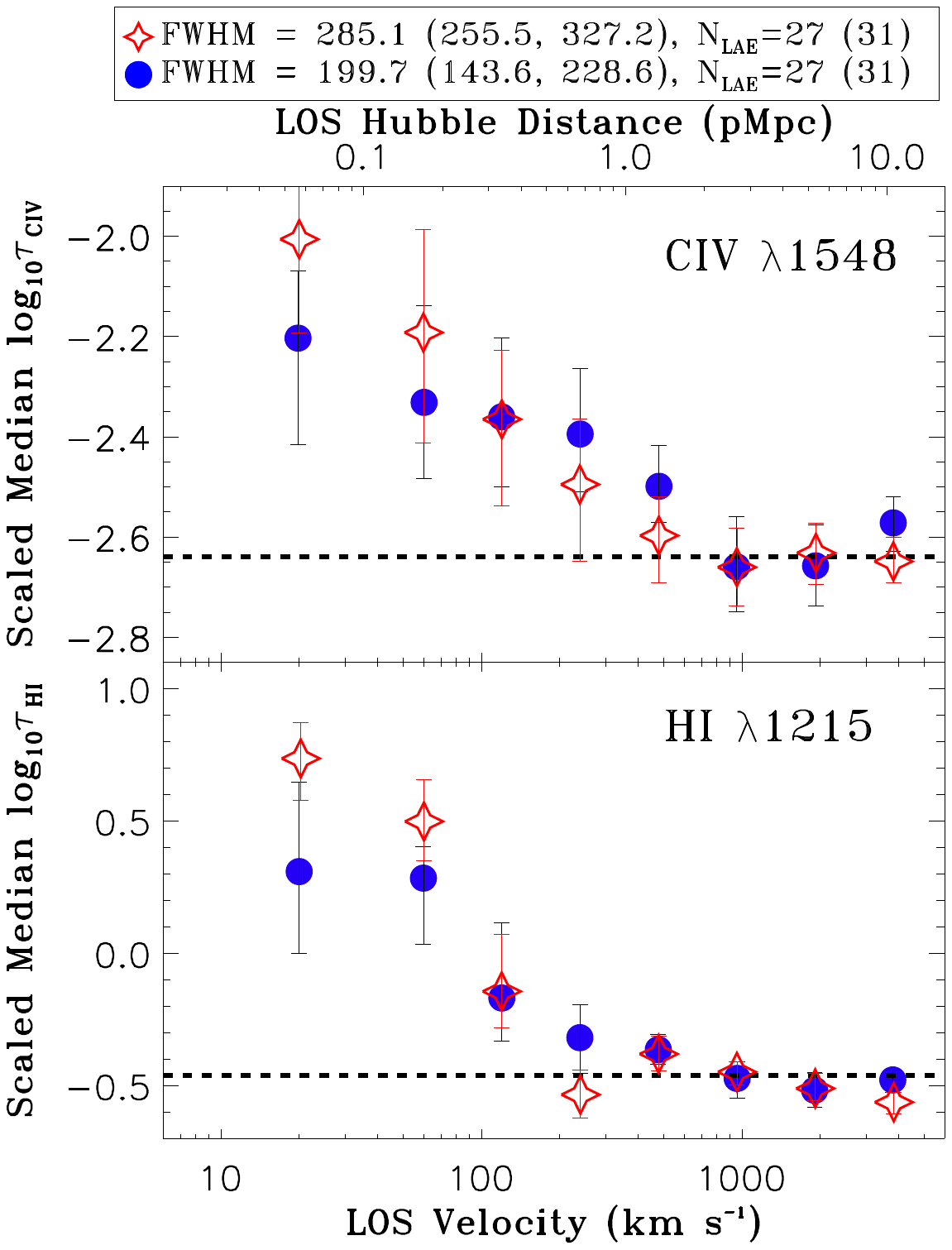} 
}}
}}   
\caption{Similar to Fig.~\ref{fig:tau_environment} but for two different FWHM bins (iso\_FWHM\_low and iso\_FWHM\_high). The median FWHM values and 68 percentile ranges in the parentheses in the legends are for the LAEs contributing to the \HI\ stack. The corresponding values for the \CIV\ stack can be found in Table~\ref{tab:sample_summary}. No significant dependence on FWHM of \lya\ emission is seen for either \HI\ or \CIV\ absorption.}  
\label{fig:fwhmdependence}     
\end{figure}

Fig.~\ref{fig:fwhmdependence} shows the stacked \HI\ and \CIV\ optical depth profiles for the two FWHM bins. Both the \HI\ and \CIV\ optical depth measurements for the two bins are consistent with each other within the $1\sigma$ uncertainties, suggesting a lack of a strong FWHM dependence. The equivalent widths measured from the flux profiles (see Fig.~\ref{fig:fwhmdependence_appendix}) for the two FWHM bins are also consistent with each other (see Table~\ref{tab:HIsummary} \& Table~\ref{tab:CIVsummary}). We however note that the \HI\ line widths measured for the iso\_FWHM\_low sample is somewhat larger than that of the iso\_FWHM\_high sample. This is likely due to the larger scatter in the FWHM distribution for the iso\_FWHM\_low sample (Table~\ref{tab:sample_summary}) leading to larger redshift errors. Recall that we used the empirical FWHM -- \voff\ relation to correct the \lya\ redshifts.

According to models of \lya\ radiative transfer \citep[e.g.,][]{Verhamme15}, the FWHM of \lya\ emission is expected to correlate with the neutral hydrogen column density ($N(\HI)$) of a galaxy. Thus, the lack of FWHM-dependence may suggests a lack of a strong between the \HI\ content of galaxies and the neutral gas in their environments.

\subsection{SFR dependence}  
\label{subsec:sfrdependence}   

Star formation-driven large-scale winds are ubiquitous in high-redshift galaxies. The SFR is thus thought to be one of the key parameters that shapes the distributions of gas and metals surrounding galaxies. The dust-uncorrected SFRs of the LAEs in our sample are estimated from the UV continuum luminosity. Out of the 77 (92) LAEs satisfying the redshift limits for the \HI\ (\CIV) stack, 30 (38) are detected in the UV continuum emission. 17/30 (18/38 for \CIV) LAEs are from the isolated subsample. Because of the small sample size, we used all 30 (38) UV-continuum-detected LAEs to investigate the SFR dependence of \HI\ (\CIV) absorption irrespective of their environments. The details of the two SFR bins (`all\_SFR\_low' and `all\_SFR\_high') are given in Table~\ref{tab:sample_summary}. We created a third SFR bin (`all\_SFR\_non-det') containing all the UV-continuum-undetected objects.

\begin{figure}
\centerline{\vbox{
\centerline{\hbox{ 
\includegraphics[trim=0.5cm 0.0cm 0.5cm 1cm,width=0.5\textwidth,angle=00]{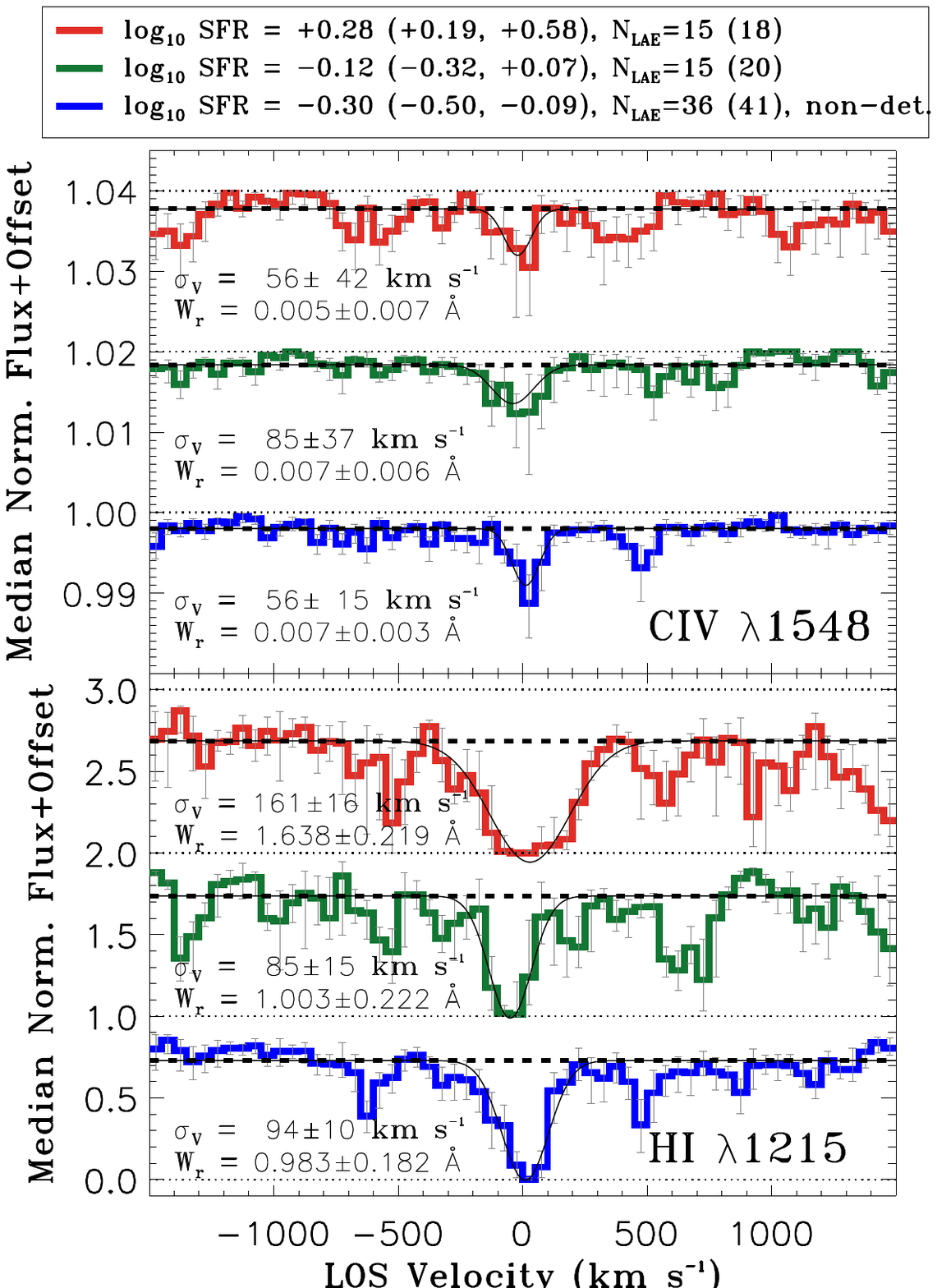} 
}}
}}   
\caption{Similar to the right panel of Fig.~\ref{fig:flux_environment} but for three different SFR bins (all\_SFR\_non-det (blue), all\_SFR\_low (green), and all\_SFR\_high (red)). The profiles in blue represent the subsample of the UV-continuum-undetected objects. The median SFR values and 68 percentile ranges in the parentheses in the legends are for the LAEs contributing to the \HI\ stack. The corresponding values for \CIV\ stack can be found in Table~\ref{tab:sample_summary}. The \HI\ absorption shows a strong dependence on the SFR. The noisier \CIV\ absorption profiles do not show any significant differences.}        
\label{fig:flux_sfrdependence}    
\end{figure}

\begin{figure}
\centerline{\vbox{
\centerline{\hbox{ 
\includegraphics[trim=0.0cm 0.0cm 0.5cm 1cm,width=0.45\textwidth,angle=00]{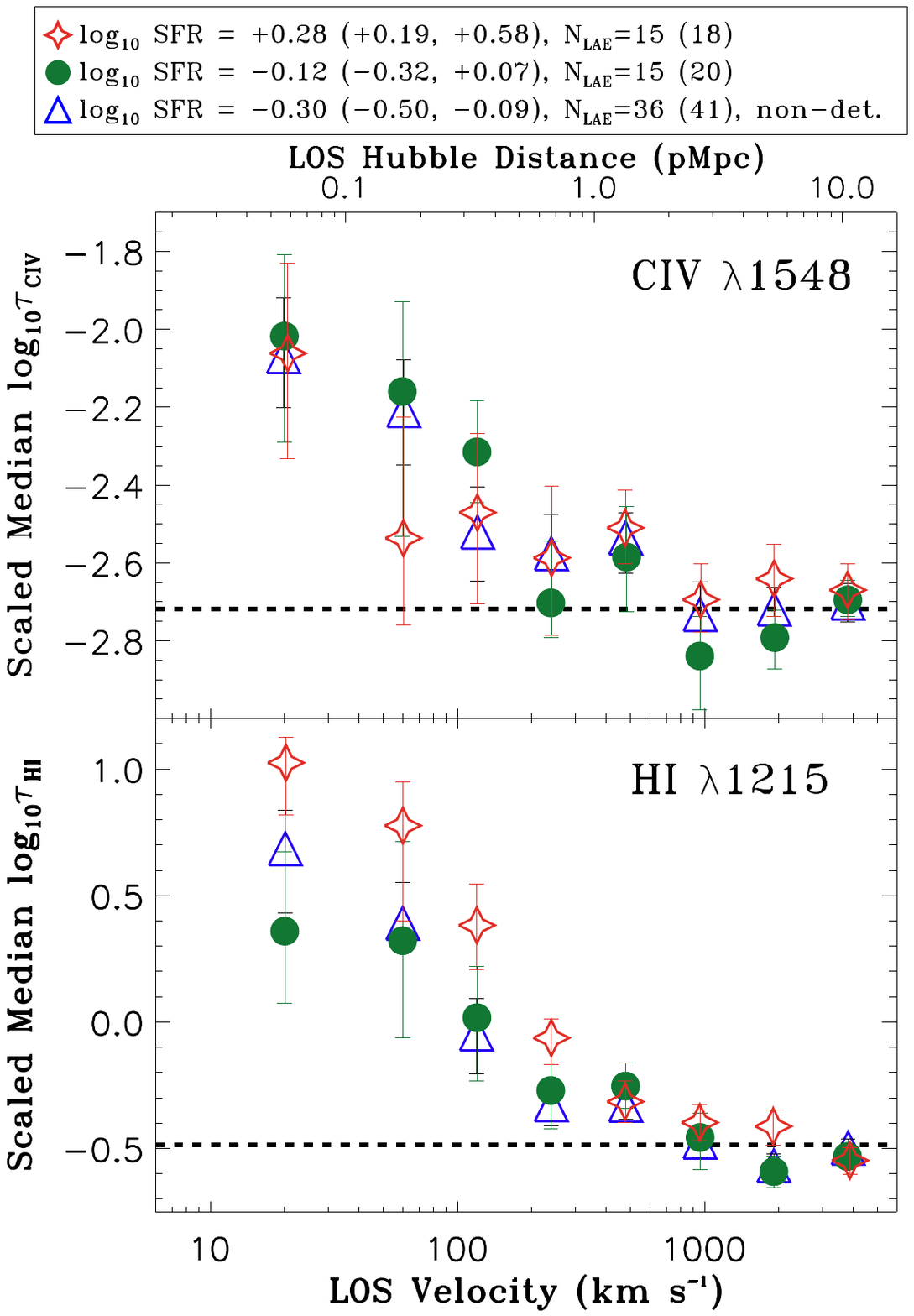} 
}}
}}   
\caption{Similar to Fig.~\ref{fig:tau_environment} but for three different SFR bins (all\_SFR\_non-det (blue), all\_SFR\_low (green), and all\_SFR\_high (red)). The blue triangles represent the subsample of the UV-continuum-undetected objects. The median SFR values and 68 percentile ranges in the parentheses in the legends are for the LAEs contributing to the \HI\ stack. The corresponding values for \CIV\ stack can be found in Table~\ref{tab:sample_summary}. The \HI\ profile is stronger and wider for the highest SFR bin, suggesting a strong SFR dependence. The noisier \CIV\ profiles do not show any SFR dependence.}     
\label{fig:tau_sfrdependence}    
\end{figure}

The median stacked \HI\ and \CIV\ absorption spectra corresponding to the three subsamples are shown in Fig.~\ref{fig:flux_sfrdependence} and the measurements performed on the stacked profiles are summarized in Tables~\ref{tab:HIsummary} \& \ref{tab:CIVsummary}. The median \HI\ equivalent width ($1.638\pm0.219$~\AA) measured for the all\_SFR\_high bin is $\approx1.8$ times ($\approx1.6$ times for the mean stack) larger than that of the all\_SFR\_low bin. The difference is significant at the $>3\sigma$ level. Besides being stronger, the \HI\ absorption is $\approx2$ times wider ($>3\sigma$) for the all\_SFR\_high bin with $\sigma_v = 161\pm16$~\kms ($154\pm19$~\kms\ for the mean stack) as compared to the all\_SFR\_low bin. The larger $W_r$ and $\sigma_v$ of the stacked \HI\ profile for the all\_SFR\_high bin are also prominent in the optical depth profile shown in the bottom panel of Fig.~\ref{fig:tau_sfrdependence}. Therefore, we find that the circumgalactic \lya\ absorption in our sample depends strongly on the SFR of the host LAEs. The detection significance for \CIV\ is $<3\sigma$ for all of the stacks (see top panel of Fig.~\ref{fig:flux_sfrdependence}). No obvious trend is seen from the noisy optical depth profiles shown in the top panel of Fig.~\ref{fig:tau_sfrdependence}.

Is the strong SFR dependence seen for \HI\ driven by the environmental dependence we noticed in Section~\ref{subsec:environment}? We find that 7/15 LAEs contributing to the \HI\ stack for the all\_SFR\_low subsample are part of `groups'. For the all\_SFR\_high subsample, 6/15 of LAEs contributing to the \HI\ stack are part of `groups'. Thus, the number of `group' LAEs contributing to the two competing SFR bins are very similar. We also note here that, depending on the observed SFR, the member galaxies of a given `group' will contribute to to different SFR bins. \footnote{For example, 5 of the member galaxies from the `group' with the highest number (7) of galaxies (Fig.~\ref{fig:group_ngal}) are detected in the UV continuum. 2/5 LAEs contribute to the all\_SFR\_low bin and the remaining 3 LAEs contribute to the all\_SFR\_high bin.} We thus believe that the observed trend with SFR cannot be fully attributed to the environmental dependence.

In order to investigate further, we generated \HI\ stacks (not shown) using the 54 isolated galaxies (see Table~\ref{tab:sample_summary}). Only 17/54 are detected in UV continuum emission, allowing us to measure the SFRs (27 non-detections, 10 LAEs are blended with low-$z$ objects). No significant difference is seen in the stacked \HI\ absorption spectra for the 13/17 UV-continuum-detected LAEs with $\rm \log~ SFR/M_{\odot}~ yr^{-1} > -0.25$ (median $\rm \log~ SFR/M_{\odot}~ yr^{-1}  = 0.15$) and 15/27 UV-continuum-undetected LAEs with $\rm \log~ SFR/M_{\odot}~ yr^{-1} < -0.25$. The same is true for the \CIV\ stack. However, we point out here that the median SFR of the 13 isolated LAEs in this exercise is lower than the median SFR of the all\_SFR\_high bin, and only 8 LAEs have $\rm \log~ SFR/M_{\odot}~ yr^{-1} > 0.1$ whereas all the 15 LAEs in the $\rm all\_SFR\_high$ bin show $\rm \log~ SFR/M_{\odot}~ yr^{-1} > 0.1$. Thus the lack of trend with SFR for the isolated LAEs may owe to the smaller number of isolated high-SFR LAEs reducing the dynamic range in SFR. Nevertheless, this does suggest the possibility that the trend seen in the full sample may be partly driven by the underlying environmental dependence.

The impact parameter distributions of the all\_SFR\_low and all\_SFR\_high subsamples are not statistically different ($D_{\rm KS}=0.2$, $P_{\rm KS}=0.90$), but the median $\rho$ for the all\_SFR\_low sample (164.5~pkpc) is somewhat larger than that of the all\_SFR\_high sample (136.1~pkpc). This is unlikely to drive the observed SFR dependence since we do not find a significant trend between \HI\ absorption and impact parameter (Section~\ref{subsec:radialprofile}).

Finally, we point out that the widths and strengths of the mean and median \lya\ absorption for the UV-continuum-undetected objects (all\_SFR\_non-det) are consistent with those of the all\_SFR\_low subsample. This is not surprising since the bulk of the SFR upper limits are similar to the SFRs measured for the all\_SFR\_low subsample.

\subsection{$EW_0$ dependence}  
\label{subsec:ewdependence}

The rest-frame equivalent width of \lya\ emission ($EW_0$) depends on both line and continuum fluxes ($EW_0=$~Line Flux$/$Continuum Flux Density~$\propto L(\rm Ly\alpha)/\rm SFR$). We noticed a strong positive trend between \HI\ absorption and SFR (Section~\ref{subsec:sfrdependence}), and only a marginal trend with $L(\rm Ly\alpha)$ (Section~\ref{subsec:lumdependence}). Here we investigate whether the CGM absorption shows any trend with $EW_0$.

\begin{figure}
\centerline{\vbox{
\centerline{\hbox{ 
\includegraphics[trim=0.0cm 0.7cm 0.5cm 2.5cm,width=0.45\textwidth,angle=00]{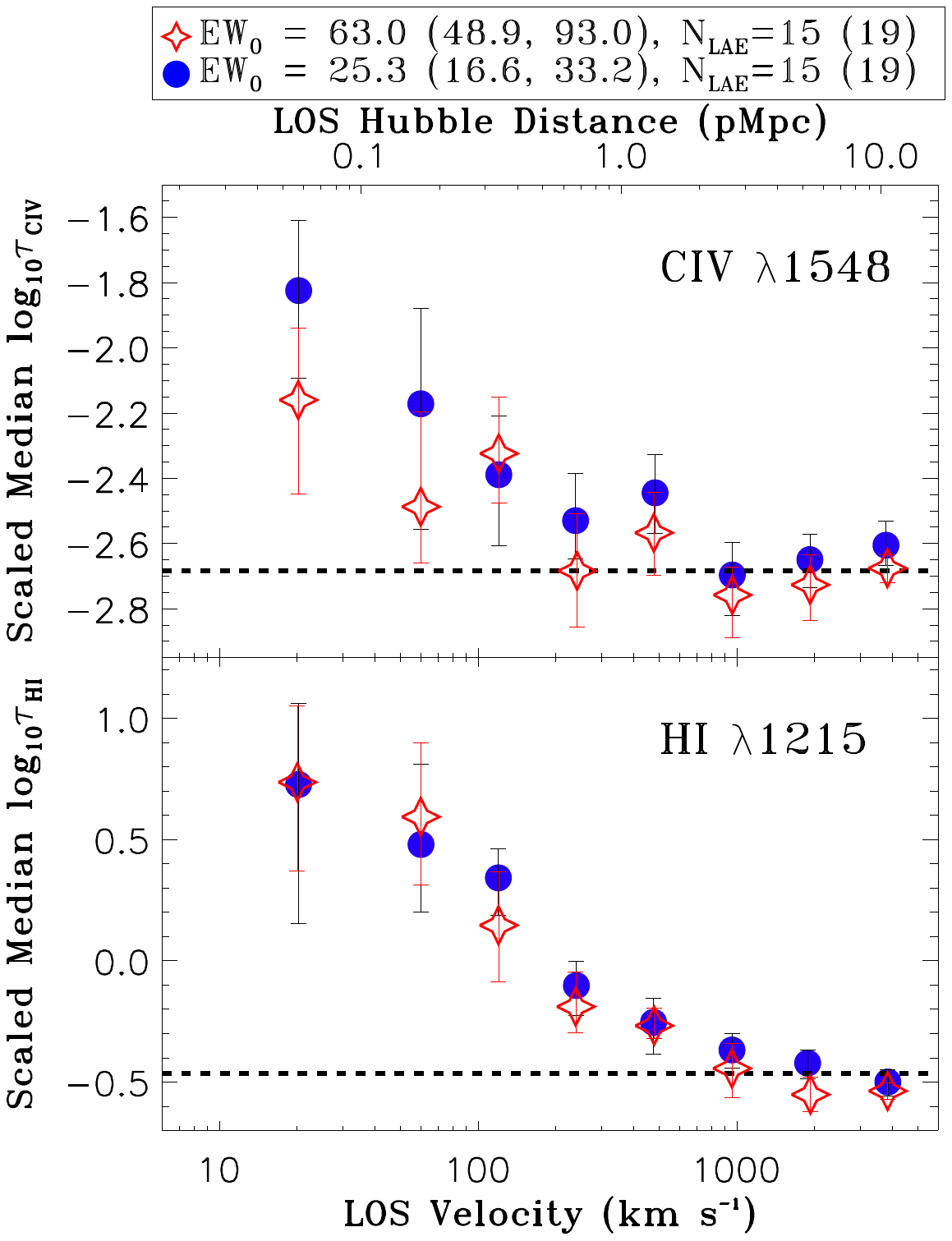} 
}}
}}   
\caption{Similar to Fig.~\ref{fig:tau_environment} but for two different $EW_0$ bins (all\_$EW_0$\_low and all\_$EW_0$\_high). The median $EW_0$ values and 68 percentile ranges in the parentheses in the legends are for the LAEs contributing to the \HI\ stack. The corresponding values for \CIV\ stack can be found in Table~\ref{tab:sample_summary}. No significant dependence on the $EW_0$ of the \lya\ emission is seen in either \HI\ or \CIV\ absorption.}  
\label{fig:ewdependence}    
\end{figure}

Following the same strategy as in the previous section, we split the 30 (38 for \CIV) UV-continuum-detected LAEs into all\_$EW_0$\_low and all\_$EW_0$\_high subsamples at the median $EW_0$ of 41.5~\AA\ (48.7~\AA\ for \CIV) irrespective of their environments. The details for these two subsamples are given in Table~\ref{tab:sample_summary}. The stacked optical depth profiles corresponding to these two subsamples are shown in Fig.~\ref{fig:ewdependence}, and the measurements performed on the stacked flux profiles (see Fig.~\ref{fig:ewdependence_appendix}) are summarized in Tables~\ref{tab:HIsummary} \& \ref{tab:CIVsummary}. There is no significant difference in the optical depths of the two subsamples across the whole range of LOS velocities, suggesting a lack of a strong dependence on $EW_0$ for both \HI\ and \CIV.

\begin{figure*}
\centerline{\vbox{
\centerline{\hbox{ 
\includegraphics[trim=1.0cm 4.2cm 0.5cm 6cm,width=0.52\textwidth,angle=00]{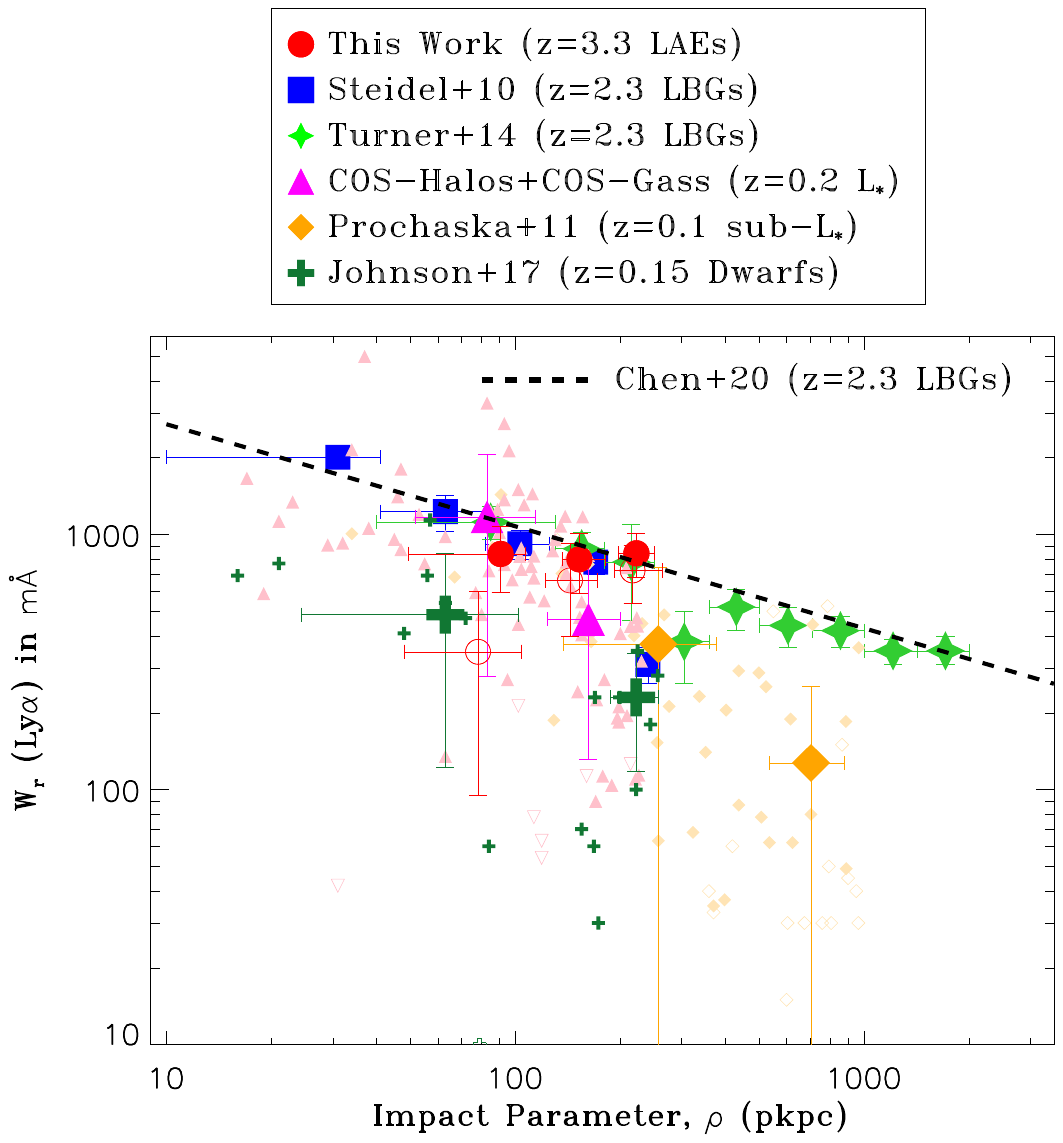} 
\includegraphics[trim=1.0cm 4.2cm 0.5cm 6cm,width=0.52\textwidth,angle=00]{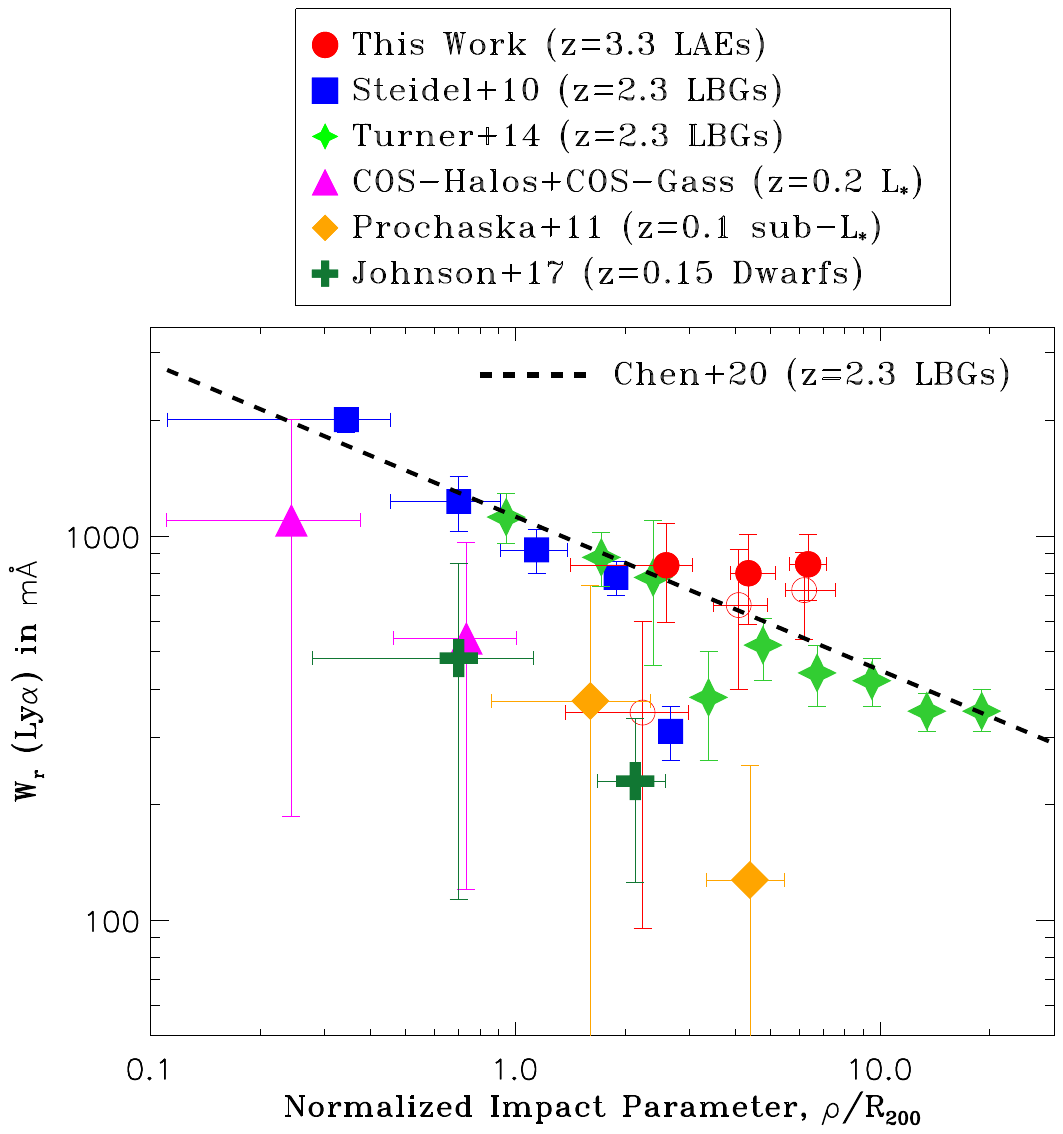} 
}}
}}   
\caption{{\tt Left:} Compilation of \lya\ $W_r$-profiles. The red filled (open) circles mark the measurements of our full (`isolated only') sample. The equivalent widths are calculated from the {\em mean stacked} absorption profiles for our sample within a $\pm500$~\kms\ velocity window (see Table~\ref{tab:HIsummary}). The small filled (open) triangles represent measurements (upper limits) for $\sim L_*$ galaxies from the COS-Halos \citep[]{Tumlinson13} and COS-GASS \citep[]{Borthakur15} surveys at $z\approx 0.2$. The big magenta triangles provide the mean $W_r$ of these data points split into two $\rho$-bins. The small filled (open) diamonds represent measurements (upper limits) for sub-$L_*$ galaxies from \citet{Prochaska11} at $z\approx0.1$. The big diamonds indicate the mean $W_r$ of these data points for two $\rho$-bins. The small plus symbols represent the $z\approx0.15$ dwarf galaxy sample of \citet{Johnson17}. The mean $W_r$ values of the two $\rho$-bins are shown by the big plus symbols. The error bars on the triangles, diamonds, and plus symbols indicate the standard deviations. Survival statistics is used to take into account the censored data points (upper limits) while calculating the mean and standard deviations. The blue squares represent measurements for $z\approx2.3$ LBGs \citep[][]{Steidel10}. The green star symbols are from \citet[][]{Turner14}, and the dashed straight line represents the best-fitting power-law relation from \citet{YChen20} obtained from a significantly larger sample size from the same survey.  The \lya\ equivalent widths measured for the LAEs in our sample are comparable to those measured for $z\approx2.3$ LBGs and $z\approx0.2$ $L_*$ galaxies, but considerably higher than the measurements for sub-$L*$ and dwarf galaxies at low-$z$. {\tt Right:} The same as the left but the $W_r$ values are plotted against the impact parameters normalized by the corresponding median $R_{200}$ values (see text). The environments of LAEs are significantly more rich in \HI\ gas compared to LBGs at $z\approx2.3$ and sub-$L*$ and dwarf galaxies at low-$z$ at a given normalized impact parameter.   
}          
\label{fig:compareHI}    
\end{figure*} 

\begin{figure*}
\centerline{\vbox{
\centerline{\hbox{ 
\includegraphics[trim=1.0cm 4.2cm 0.5cm 6cm,width=0.52\textwidth,angle=00]{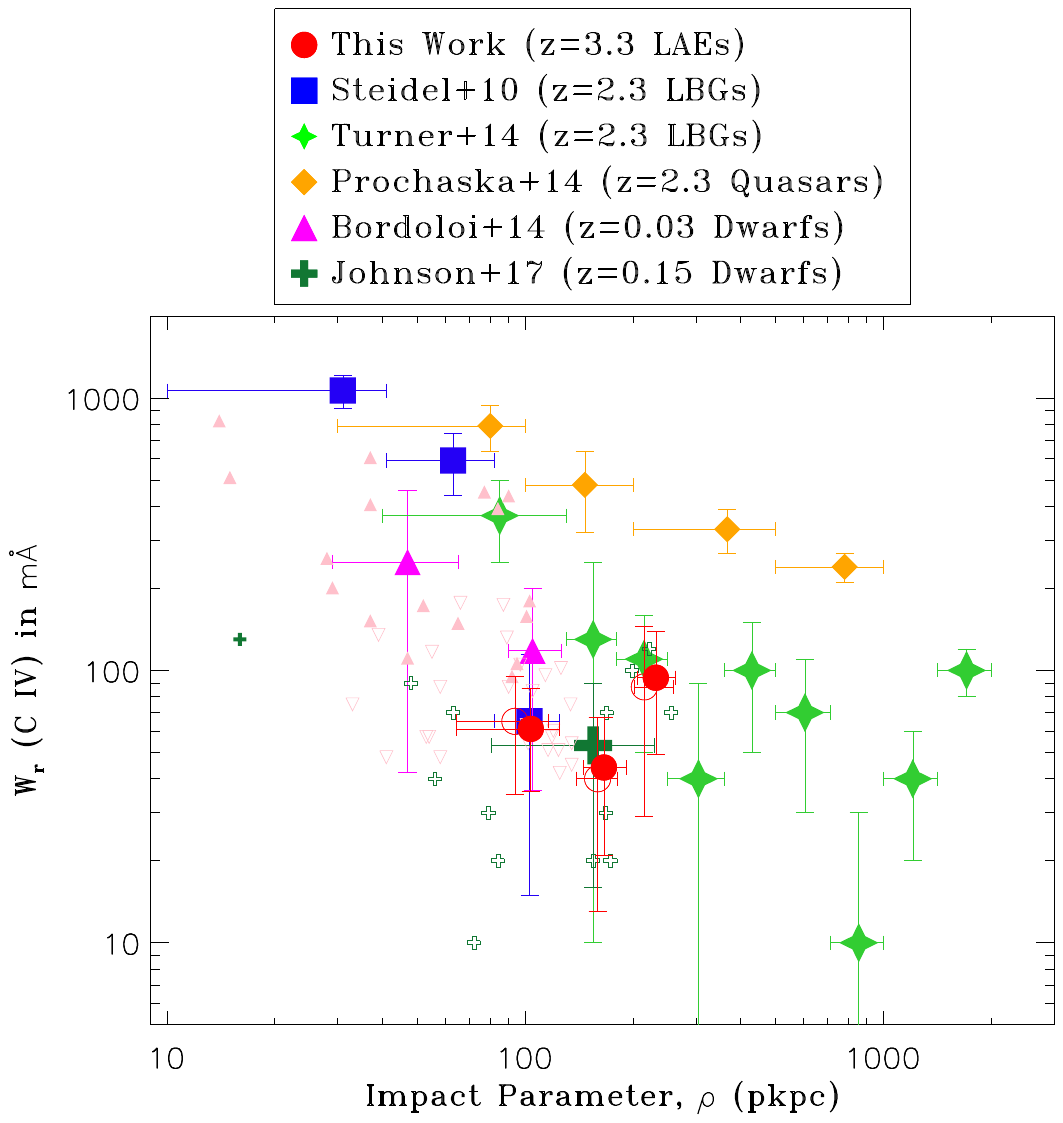} 
\includegraphics[trim=1.0cm 4.2cm 0.5cm 6cm,width=0.52\textwidth,angle=00]{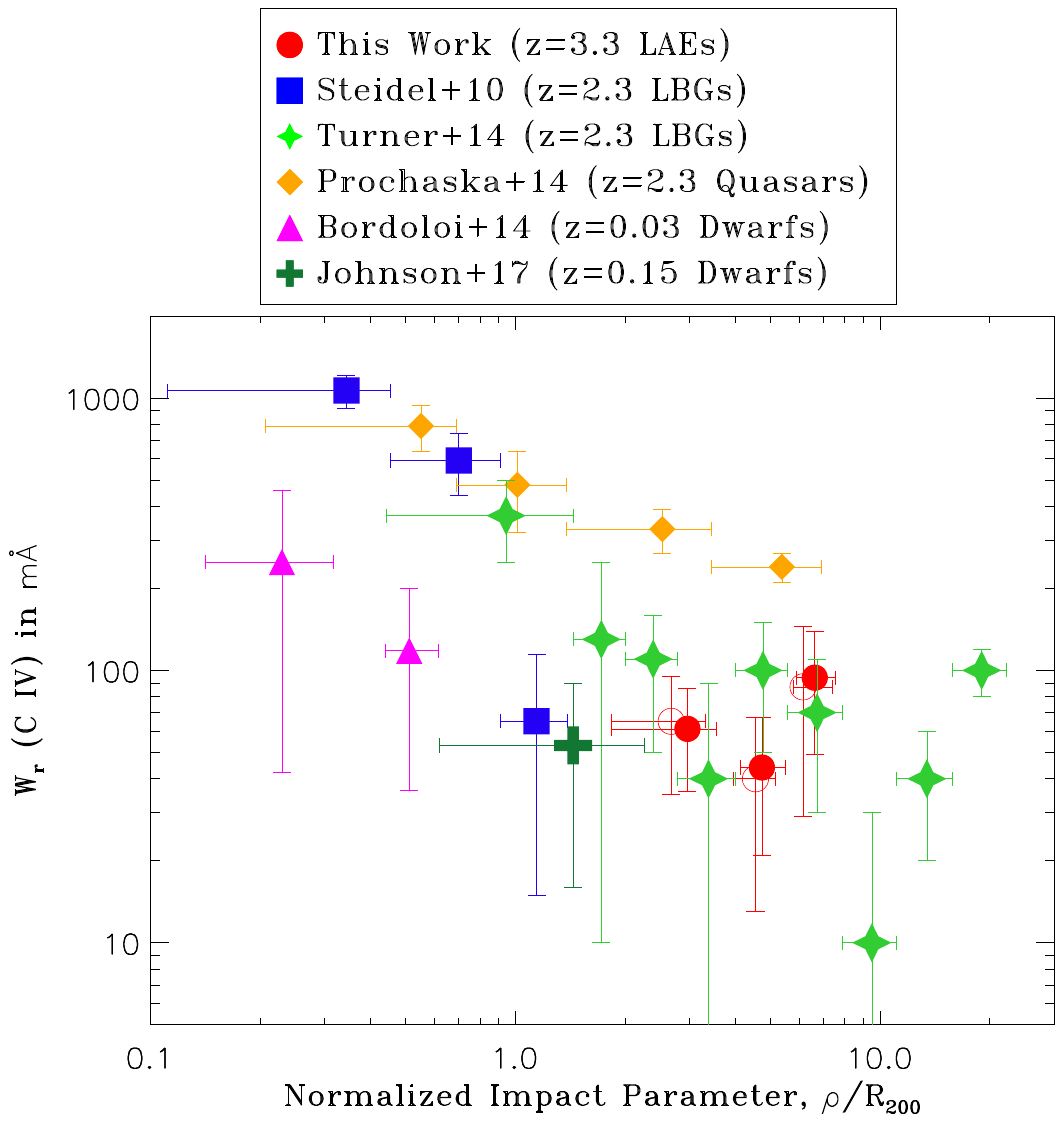} 
}}
}}   
	\caption{{\tt Left:} Compilation of \CIV\ $W_r$-profiles. The equivalent widths for our sample are calculated from the {\em mean stacked} absorption profiles within a $\pm500$~\kms\ velocity window (see Table~\ref{tab:CIVsummary}). The filled and open red circles are for the full and `isolated only' samples respectively. The big magenta triangles indicate the mean $W_r$ of the COS-Dwarf sample \citep[][]{Bordoloi14b} divided into two impact parameter bins. The individual detections (upper limits) are shown by the small solid (open) triangles. The small plus symbols represent the $z\approx0.15$ dwarf galaxy sample of \citet{Johnson17}, with open symbols indicating upper limits. The big plus symbol indicates the mean $W_r$. Survival analysis is used to take into account the censored data points (upper limits) while calculating the mean and standard deviations (error bar along y-axis) for both the dwarf galaxy samples. The blue squares and green star symbols represent measurements of $z\approx2.3$ LBGs from \citet{Steidel10} and \citet{Turner14}, respectively. The orange diamonds indicate observations of $z\approx2$ quasar host-galaxies from \citet{Prochaska14}. The error bars along x-axis for the blue squares, green stars, and orange diamonds represent the full range, and 68 percentiles for the red circles, magenta triangles, and the plus symbol. {\tt Right:} The same as the left but the $W_r$ values are plotted against the impact parameters normalized by the corresponding median $R_{200}$ values (see text). For a fixed normalized impact parameter, our $W_r$(\CIV) measurements are comparable to those of the $z\approx2.3$ LBGs and low-$z$ dwarf galaxies, but considerably lower compared to the $z\approx2.3$ quasar host-galaxies.     
}       
\label{fig:compareCIV}    
\end{figure*} 

\section{Discussion}    
\label{sec:discussion}

In this section we discuss the main results of our survey in the context of other CGM surveys. The CGM surveys in the literature are for more massive galaxies and mostly at low $z$. Before we move on, we recall that the median halo (stellar) mass for our sample is estimated to be $\sim10^{11.1}$~\Msun\ ($\sim10^{8.6}$~\Msun), which corresponds to a virial radius ($R_{200}$) of only $\approx35$~pkpc \citep[see Section~5 of][]{Muzahid20}. This suggests that they are Small-Magellanic-Cloud (SMC)-type objects. Almost all of the existing CGM surveys deal with galaxies that are significantly more massive (by an order of magnitude in stellar mass) than the objects in our sample.

\subsection{Reservoirs of gas and metals around LAEs} 
\label{subsec:ewdiss} 

We detected \lya\ and \CIV~$\lambda1548$ absorption in the composite spectra with $\gtrsim4\sigma$ significance (Fig.~\ref{fig:fullsample}). Given the median impact parameter (165~pkpc) and the median virial radius ($\approx35$~pkpc) of our sample, such detections reveal the presence of diffuse gas and metal reservoirs out to at least $5R_{\rm vir}$ surrounding the \lya\ emitting galaxies.

In Fig.~\ref{fig:compareHI} we compare the $W_r$-profile of \HI~$\lambda1215$ of our sample (red circles) with the ones from the literature. The blue squares represent observations from \citet[][see their Table~4]{Steidel10}, who used stacking of background galaxy spectra to probe the CGM of 512 LBGs out to 125~pkpc. An extension of this study was recently presented by \citet{YChen20} with 200,000 background-foreground galaxy pairs. The dashed line represents the empirical relation from \citet{YChen20}. The median halo mass of these LBGs is $\approx10^{12}$~\Msun, corresponding to a virial radius of $R_{200}$ $\approx90$~pkpc at the median redshift ($z\approx2.3$) of their sample. The median SFR of the LBG sample is $\approx25$~\Msun~$\rm yr^{-1}$. Note that the median halo mass and SFR of the LBG sample are an order of magnitude higher than for our LAE sample. The green star symbols represent observations from \citet[][see their Table~8]{Turner14} who used background quasars to probe the CGM of $z\approx2.3$ LBGs drawn from the KBSS \citep[][]{Steidel14}, which is essentially the same set of LBGs presented in \citet{Steidel10} but probed at larger impact parameters. The small triangles in Fig.~\ref{fig:compareHI} are from the COS-Halos \citep[]{Tumlinson13} and COS-GASS \citep[]{Borthakur15} surveys. Both these surveys characterize the CGM of $\sim L*$ galaxies at $z<0.4$ with stellar masses in the range $M_*\sim10^{9.5-11.5}$~\Msun. The small orange diamonds represent the measurements for sub-$L_*$ ($0.1<L/L_*<1.0$) galaxies at $z\approx0.1$ from \citet[][see their Table~8]{Prochaska11}. The small plus symbols are from the $z\approx0.15$ dwarf galaxy (median $M_*\approx10^{8.3}$~\Msun, $L/L_* < 0.1$) sample of \citet[][see their Table~1]{Johnson17}. For all three samples the open symbols represent upper limits. The individual measurements for each sample are split into two impact parameter bins. The two big triangles, diamonds, and plus symbols represent the corresponding mean $W_r$ at the median $\rho$ of the two bins. The error bars on them are standard deviations. The censored data points (upper limits) are taken into account in calculating the mean $W_r$ and standard deviations in each bins using survival analysis.\footnote{We used the {\em `cenken'} function under the {\tt NADA} package in {\tt R} (http://www.r-project.org/).}



The left panel of Fig.~\ref{fig:compareHI} shows that the \lya\ equivalent widths measured for the LAEs in our sample are comparable to those measured for $z\approx2.3$ LBGs \citep[]{Steidel10,Turner14,YChen20} and $\approx L_*$ galaxies at low $z$ \citep[]{Tumlinson13,Borthakur15}. The $W_r$(\lya) for the low-$z$ sub-$L_*$ and dwarf galaxies are, however, considerably lower compared to our sample. In the right panel of Fig.~\ref{fig:compareHI}, we show the mean $W_r$(\lya) of the same samples as a function of normalized impact parameters, $\rho/R_{200}$. Such an x-axis minimizes the effects of different galaxy masses and redshifts for the different samples provided that the structure/properties of the CGM are self-similar. For the COS-Halos, COS-GASS, and the dwarf galaxy samples, the $R_{200}$ values for the individual galaxies are available. For the sub-$L_*$ galaxies we used the median $R_{200}\approx160$~pkpc as suggested by \citet{Prochaska11}. For the LBG samples we used $R_{200}\approx90$~pkpc \citep[]{Turner14}. The right hand panel of Fig.~\ref{fig:compareHI} conveys a couple of important messages: (1) For a fixed normalized impact parameter, the environment of $z\approx2.3$ LBGs is more enriched in \HI\ compared to the dwarf ($<0.1L_*$), sub-$L_*$ ($<L*$), and $L_*$ galaxies at $z\approx0.2$. (2) The environments of the LAEs in our sample are significantly more \HI-rich compared to those of the LBGs at $z\approx2.3$ for a fixed normalized impact parameter.   

A compilation of \CIV\ $W_r$-profiles is shown in the left panel of Fig.~\ref{fig:compareCIV}. The red circles represent our $W_r$ measurements from the mean stacked spectra (open red circles are for the `isolated only' subsample). The blue squares and green stars symbols are for the $z\approx2.3$ LBGs from \citet[][Table~4]{Steidel10} and \citet[][Table~8]{Turner14}, respectively. The orange diamonds are for $z\approx2$ quasar host-galaxy sample of \citet[][see their Table~3]{Prochaska14}. \footnote{The average $W_r$ presented in Table~3 of \citet{Prochaska14} are calculated taking the non-detections at their measured values.} The small magenta triangles indicate the measurements from the COS-dwarf survey \citep[$M_*\sim10^{9.5}$~\Msun\ galaxies at $z\approx0.03$][]{Bordoloi14b} with open triangles indicating upper limits. We divided the sample into two impact parameter bins. The bigger triangles mark the mean $W_r$(\CIV) at the median $\rho$ of the two bins. As in Fig.~\ref{fig:compareHI}, survival analysis is used to take into account the upper limits while calculating the means and the standard deviations (shown by the error bars). The small plus symbols denote the $z\approx0.15$ dwarf galaxy sample of \citet[][Table~1]{Johnson17}. All except one are upper limits. The big plus symbol represents the mean $W_r$ derived from survival analysis. The mean $W_r$(\CIV) values measured for the low-$z$ dwarf-galaxy samples are broadly consistent with our measurements given the scatter in individual $W_r$ values. The \CIV\ $W_r$-profile for the LBGs shows a sharp decline at $\approx100$~pkpc which makes it consistent with our measurements. The $W_r$(\CIV) measurements of \citet{Turner14} are somewhat higher compared to our measurements. Finally, at any given impact parameter probed by our sample, the $W_r$(\CIV) measured for the quasar hosts are at least an order of magnitude higher compared to our measurements. The quasar-hosts also show significantly higher $W_r$(\CIV) compared to the COS-Dwarf sample, as noted previously by \citet{Prochaska14}, and the LBG sample of \citet{Turner14}.      

The five different samples plotted in Fig.~\ref{fig:compareCIV} have halo masses ranging from $\sim 10^{10.8} - 10^{12.5}$~\Msun. The galaxies in our sample have masses comparable to the dwarf galaxy sample of \citet{Johnson17}. The COS-Dwarf survey presents a sample of 43 dwarf galaxies at $z\approx0.03$ with a median $M_* \approx10^{9.5}$~\Msun\ ($M_{h}\sim10^{11.5}$~\Msun) and a median $R_{200} \approx205$~pkpc. As mentioned earlier, the LBGs in the samples of \citet{Steidel10} and \citet{Turner14} have a median $M_{h} \sim 10^{12.0}$~\Msun\ (median $R_{200}\approx90$~pkpc). The quasar host-galaxies studied by \cite{Prochaska14} have the highest halo masses, $M_{h} \sim10^{12.5}$~\Msun, corresponding to a median $R_{200} \approx$ 145~pkpc at $z\approx2$. The right panel of Fig.~\ref{fig:compareCIV} shows the $W_r$(\CIV) as a function of normalized impact parameter, which minimizes the effects of different masses and redshifts for the different samples. Overall, the right panel confirms the conclusions we derived from the left panel. The $W_r$(\CIV) measured for the LAEs in our sample are now fully consistent with the LBG measurements of \citet{Turner14}.

\subsection{Direct comparison with LBG optical depth profiles}   
\label{subsec:LBGvsLAE}  

In Section~\ref{subsec:ewdiss} we showed that at a given impact parameter, the mean $W_r$(\lya) for sightlines near LBGs are comparable to the sightlines near the LAEs in our sample, and marginally higher at $\rho\lesssim100$~pkpc (see the left panel of Fig.~\ref{fig:compareHI}). The LAEs, however, shows a significantly higher $W_r$(\lya) for $\approx 3-8~ \rho/R_{200}$ (see the right panel of Fig.~\ref{fig:compareHI}). The $W_r$(\CIV) measured for LBGs and LAEs are consistent with each other at a given $\rho/R_{200}$ (see the right panel of Fig.~\ref{fig:compareCIV}). Here we directly compare the \HI\ and \CIV\ optical depth profiles for $z\approx2.3$ LBGs \citep{Turner14} and our LAEs.

\begin{figure}
\centerline{\vbox{
\centerline{\hbox{ 
\includegraphics[trim=0.5cm 2.0cm 0.5cm 4cm,width=0.45\textwidth,angle=00]{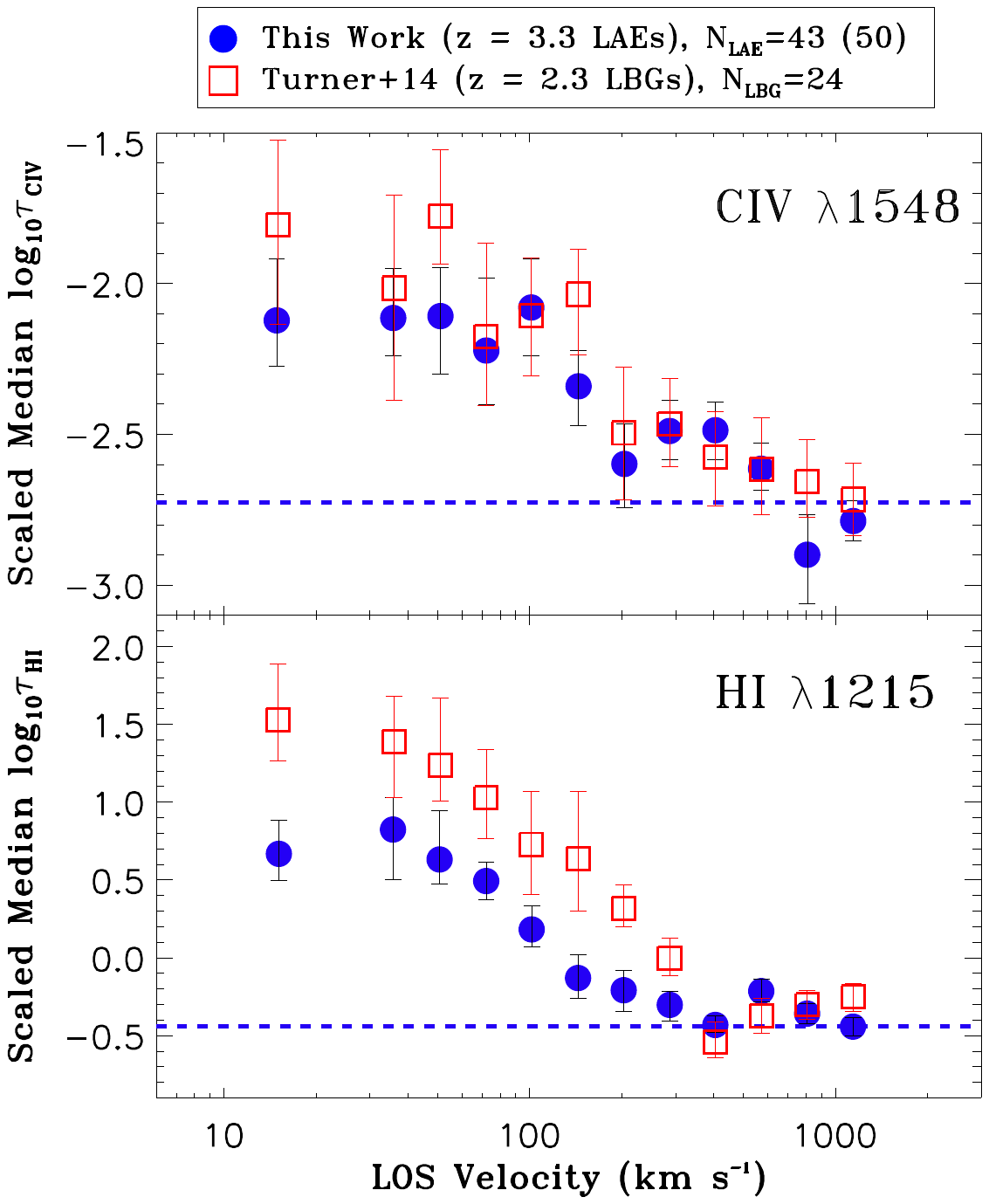} 
}}
}}   
\caption{Comparison of the median stacked optical depth profiles of our LAEs and $z\approx2.3$ LBGs for impact parameters of 40--180~pkpc. {\tt Bottom:} The median \HI~$\lambda1215$ optical depth profile for our LAE sample is shown by the blue circles. The corresponding  median optical depth in random regions is shown by the blue dashed line. The median \lya\ optical depth profile for LBGs \citep[]{Turner14} is shown by the red squares, after linearly scaling to the same background level as indicated by the dashed line. In both cases the $1\sigma$ confidence intervals are calculated from 1000 bootstrap realizations of the corresponding galaxy samples. The number of galaxies contributing to the plots are indicated in the legends. {\tt Top:} Same as the bottom but for the \CIV~$\lambda1548$ line. The number in parentheses in the legend is the number of LAEs contributing to the \CIV\ stack. For a fixed impact parameter range (i.e., 40--180~pkpc), the CGM of LBGs appears to be more \HI-rich. \CIV, on the other hand, shows only a marginal difference.    
}        
\label{fig:LBGvsLAE}    
\end{figure}

As mentioned earlier, \citet{Turner14} studied the CGM of 854 LBGs drawn from the KBSS survey using background quasars (see Section~\ref{subsec:ewdiss}). In Fig.~\ref{fig:LBGvsLAE} we reproduced the \HI~$\lambda1215$ and \CIV~$\lambda1548$ optical depth profiles for the LBG sample using Table~5 of \citet[][]{Turner14}. We also show the corresponding profiles obtained for our LAE sample, using the same logarithmic LOS velocity bins as used in their study. Moreover, for consistency, we used the same impact parameter range (i.e., 40--180~pkpc) as used by \citet[][see their Fig.~6]{Turner14}. There are 43 (50) LAEs contributing to the \HI\ (\CIV) composites for our sample, which is $\approx2$ times larger than the number of LBGs (24) that contribute to the stacks.

It is evident from Fig.~\ref{fig:LBGvsLAE} that the \HI\ absorption is significantly stronger and wider for the LBG sample compared to our LAE sample. \CIV\ absorption, on the other hand, is only marginally stronger for LBGs. These findings are consistent with the conclusions we have drawn in Section~\ref{subsec:ewdiss} (see Fig.~\ref{fig:compareHI}). The average $W_r$(\lya) for the first two impact parameter bins (40--180~pkpc) of \citet{Turner14} is $1000\pm220$~m\AA\ (see their Table~8), which is larger than the corresponding average $W_r$(\lya) of $810\pm138$~m\AA\ measured for the LAEs in our sample.

Three possible reasons that could cause such a difference in the CGM absorption are: (i) difference in mass (ii) difference in SFR and (ii) difference in redshift between the two samples. The galaxies from \citet{Turner14} are UV-continuum selected ($m_{R} \leqslant 25.5$), drawn from the KBSS survey \citep[]{Steidel10}. As pointed out earlier, both the median mass and SFR of the LBG sample are at least an order of magnitude higher compared to our LAE sample. Moreover, the median redshift of our sample is $3.3$, whereas the median redshift of the LBG sample is $2.3$. Given the lack of evidence for redshift evolution in our sample (Section~\ref{subsec:redshiftevolution}), the difference is unlikely due to the redshift difference. This conclusion however assumes that the lack of redshift evolution extends down to $z \approx 2.3$, which need not be true.

Owing to the lack of rest-frame optical data, we could not measure the stellar masses of the individual LAEs in our sample. We, however, have seen a strong SFR dependence for \HI\ absorption in Section~\ref{subsec:sfrdependence}, which in turn can be driven by an underlying mass dependence if the LAEs in our sample follow a star-forming main sequence relation as seen for low-$z$ galaxies. The higher halo mass (and the lower redshift) of the LBGs also imply that the fixed impact parameter range (40--180~pkpc) corresponds to smaller normalized impact parameters ($\rho/R_{200}$) for the LBGs owing to larger virial radii. This can cause the observed differences as well.

%

\subsection{Strong environment dependence} 
\label{subsec:envdiss} 

The strongest trend that we observed in this study is due to the environment (Section~\ref{subsec:environment}; Figs. \ref{fig:flux_environment} \& \ref{fig:tau_environment}). The CGM absorption (both for \HI\ and \CIV) is significantly stronger for the `group' galaxies compared to the isolated ones. The $W_r$ measured for the stacked \HI\ profile of the `group' subsample is $>2$ times stronger than the isolated subsample with a $>4\sigma$ significance. The $\sigma_v$ of the median stacked \HI\ profile for the `group' subsample is $235\pm22$~\kms\ ($241\pm24$~\kms\ for the mean stack), which is $\approx2.7$ times larger (with a $>5\sigma$ significance) than that of the isolated subsample. Under pure Hubble flow, the $\sigma_v$ of the `group' subsample corresponds to a physical scale of $\approx700$~pkpc at $z=3.3$, which is much larger than the typical group scale at these redshifts and exceeds the impact parameter range that we probe. On the other hand, the $\sigma_v$ of $\approx240$~\kms\ at $z=3.3$ would correspond to a halo mass of $\sim10^{12}$~\Msun, provided it is dominated by dynamical motions.

\citet{Bordoloi11} studied the effect of environment on \MgII\ absorbing gas at $z\approx0.7$ using stacking of background galaxy spectra. They found that the $W_r$-profile of group galaxies is considerably flatter compared to that of the isolated galaxies, so that at any given impact parameter at $\rho>50$~pkpc, the $W_r$(\MgII) is significantly higher for the group galaxies \citep[see also][]{Nielsen18}. More recently, using a sample of 228 MUSE-detected galaxies at $z\approx0.8-1.5$, \citet{Dutta20} reported five times stronger absorption and three times higher covering fraction of \MgII\ for group galaxies (two or more galaxies within 500~kpc and 500~\kms). A simple model that assumes that the \MgII\ absorption of a group can be represented by the superposition of the absorption of individual member galaxies successfully reproduces the $W_r$-profile for the group galaxies. \citet{Nielsen18} showed that while such a simple model predicts the right $W_r$ for groups, it overpredicts the absorption kinematics.

Recently, using a handful of galaxy groups (more than two galaxies within 400~kpc and 400~\kms), \citet{Fossati19} showed that the \MgII\ absorption in groups is stronger than what is typically seen for more isolated galaxies at comparable impact parameters leading to a rather flat $W_r$-profile \citep[see also][]{Chen10b,Dutta20}. They argued that stripping of metal-rich gas from the individual halos due to gravitational interactions between the member galaxies in a group gives rise to the large cross-section for \MgII. Here we emphasize that in most of these \MgII\ absorber studies discussed above, `group' does not necessarily mean a virialized structure. It is rather a measure of overdensity.

The strong environmental dependence we observe for \HI\ and \CIV\ is consistent with what various authors have seen for \MgII. However, we point out that for the majority of our LAEs, the impact parameters (median $\rho \approx$165~pkpc) are larger than the virial radii, which are only a few tens of kpc. It is likely that the LAEs in `group' subsample reside in relatively more massive halos. The hierarchical nature of cosmic structures naturally facilitates more gas in such environments. The metallicity of the absorbing gas, which will be presented in future work, will provide crucial insights into the nature of the absorbers.

Using a \CIV-absorption-selected sample, \citet{Burchett16} found that the detection rate of \CIV\ is lower for group ($M_h>10^{12.5}$~\Msun) galaxies at $z<0.015$. \HI, on the other hand, did not show any dependence on mass or environment. Indeed, at such high halo masses, a decline in \MgII\ equivalent width has been reported by \citet{Bouche06a} from the cross-correlation of \MgII\ absorbers with luminous red galaxies \citep[see also][]{Lundgren09}. However, we point out that $10^{12.5}$~\Msun\ is considerably larger than the halo masses of the LAEs in our sample ($\sim10^{11}$~\Msun).

\citet{Johnson15} studied environmental effects for \OVI\ somewhat beyond the virial radii of individual galaxies at low redshift ($z\approx0.2$). They found a relatively enhanced detection rate for \OVI\ absorption around galaxies with nearby neighbors, and suggested that galaxy interactions in the form of ram-pressure and tidal stripping can distribute metal enriched gas out to distances well beyond individual halos. Finally, \citet{Pointon17} studied the effect of group environment on circumgalactic \OVI\ absorption at $z\approx0.2$, using a sample of 18 galaxy groups (two or more galaxies with a LOS velocity difference smaller than 1000~\kms\ and located within 350~pkpc of the background quasars). They found a $3\sigma$ result that the average $W_r$(\OVI) for the group galaxies are smaller than for the isolated ones. But, they find that the covering fractions of \OVI\ are comparable between the two samples. They suggested that \OVI\ traces warm/hot gas in the halo, which is too highly ionized owing to the higher virial temperature for the groups. The \HI\ and \CIV\ as studied here are tracers of cool, photoionized gas \citep[see e.g.,][]{Muzahid12a} which are not particularly sensitive to the virial temperature.

\subsection{Strong SFR dependence} 
\label{subsec:sfrdiss} 

SF-driven outflows are thought to play an important role in determining the chemical, kinematical, and physical structure of the CGM. We noticed a strong SFR dependence for the \HI\ absorption in Section~\ref{subsec:sfrdependence} (see Figs.~\ref{fig:flux_sfrdependence} \& \ref{fig:tau_sfrdependence}), with the highest SFR bin showing significantly stronger and wider absorption compared to the lower SFR bins. No significant SFR dependence is seen when we limited our sample to `isolated only' LAEs. However, in that case we do lose a significant number (about half) of the high SFR LAEs with $\rm \log~ SFR/M_{\odot}~yr^{-1} > 0.1$, which reduces the dynamic range and the median SFR of the high-SFR bin. We thus believe that the strong SFR dependence seen for the full sample cannot be entirely due to the underlying environmental dependence.

The SFRs calculated from UV continua has a characteristic time scale of $\approx100$~Myr \citep[]{Kennicutt98}. If the observed trend of enhanced \HI\ in the CGM of high SFR galaxies is owing to the wind materials entrained by SFR-driven outflows, then the wind velocity required to travel a distance of 165~pkpc (median $\rho$) is $\approx165$~pkpc$/100$~Myr~$\approx1600$~\kms. In models of momentum-driven outflows, the wind speed scales as the halo circular velocity \citep[]{Murray05,Heckman15}. The expected circular velocity for the low-mass galaxies in our sample is only $\approx120$~\kms, which is an order of magnitude lower than the velocity required for a causal connection between wind driven by the observed star-formation and CGM absorption. While wind velocities much greater than the circular velocity have been observed on ISM scales \citep[e.g.,][]{Steidel10}, it is unclear that such high velocities are maintained out to distances comparable to or larger than the halo virial radius, particularly since the mass in the CGM is expected to exceed that in the ISM by a large factor.

In other words, the distance travelled by an outflow in 100~Myr with a speed of $\approx120$~\kms\ is only $\approx12$~pkpc, which is an order of magnitude lower than the median impact parameter of our sample. Unless the SFR has been high for time scales $\gg100$~Myr, the observed trend cannot be explained by outflows associated with the star formation episode that we observe. This is also consistent with the lack of a trend seen for \CIV. 

We instead speculate that the higher SFR is the consequence of the availability of more cool, neutral gas in the CGM. In particular, the observed SFR dependence can be explained by an underlying mass dependence provided LAEs at these redshifts follow a star-forming main sequence relation like the normal low-$z$ galaxy population. Alternatively, galaxies of a fixed mass whose environment is more \HI-rich, may also tend to be more gas rich themselves and hence have higher star formation rates.

\section{Summary}    
\label{sec:summary}  

We presented the first results of the MUSEQuBES survey on the CGM of 96 LAEs with redshifts in the range $z=2.9-3.8$ (median $z=3.3$). This is the first study characterizing the gaseous environment of LAEs using background quasars on relatively small scales (impact parameters $\rho = 16 - 315$~pkpc; median $\rho = 165$~pkpc) with a statistically significant sample. The LAEs are detected within $1'' \times 1''$ fields centered on bright $z=3.6-3.9$ quasars with the MUSE integral field spectrograph on the VLT. Our MUSE exposures are deep, on average 6 hours per field, yielding a relatively faint sample of LAEs with \lya\ luminosities $\sim 10^{42}\,\text{erg}\,\text{s}^{-1}$. The star formation rates are estimated from the rest-frame UV continua, which we detect with $>5\sigma$ confidence for about half the sample. Neglecting dust corrections, we obtain SFR~$\sim 1~\text{M}_\odot\,\text{yr}^{-1}$, which for main sequence galaxies corresponds to stellar masses of $\sim 10^{8.6}$~\Msun. Subhalo abundance matching results suggest that this stellar mass corresponds to a halo mass of only $\sim 10^{11}$~\Msun. High-quality quasar spectra were obtained using over 250 hours of (mostly archival) observations with the UVES echelle spectrograph on the VLT, as well as some Keck/HIRES observations to fill in gaps in the UVES coverage.

We used mean and median spectral stacking to investigate possible trends between LAE properties and the CGM absorption. We focused on the \HI\ and \CIV\ absorption lines, since these are the only two lines detected at $>3\sigma$ significance in the composite spectrum. In order to minimize contamination from unrelated absorption and to correct for possible line saturation, we used pixel optical depths recovered using the Python module {\tt PODPy} \citep{Turner14}, though this only affects our quantitative conclusions. Since the \lya\ emission peaks do not provide the systemic redshifts owing to the resonant nature of the \lya\ line, we used the empirical relation between velocity offset (with respect to the systemic redshift) and FWHM as obtained in \citet{Muzahid20} to correct the \lya\ redshifts.

Our main findings are:

\begin{itemize} 
\item There is significant excess \HI\ and \CIV\ absorption near the LAEs out to 500~$\text{km}\,\text{s}^{-1}$ (Fig.~\ref{fig:fullsample}) and at least $\approx 250$ pkpc (corresponding to $\approx 7$ virial radii) (Figs.~\ref{fig:rhodep} and \ref{fig:radialprofile}). At $\lesssim 30~\text{km}\,\text{s}^{-1}$ from the galaxies the median \HI\ and \CIV\ optical depths are enhanced by an order of magnitude (Fig.~\ref{fig:fullsample}). 

\item A strong environmental dependence is seen for both \HI\ and \CIV\ (Figs.~\ref{fig:flux_environment} and \ref{fig:tau_environment}). The absorption is significantly stronger near the $\approx 1/3$ of our LAEs that are part of a `group' (multiple LAEs in the MUSE FoV within $\pm500$~\kms) than near isolated LAEs. We suggest that the large-scale structure around more massive halos, hosting multiple LAEs, is the cause of the enhanced CGM absorption for `group' galaxies.

\item We searched for a dependence on impact parameter of the optical depth profile as a function of velocity (Fig.~\ref{fig:rhodep}) and of the equivalent widths in fixed velocity windows (Fig.~\ref{fig:radialprofile}). We do not detect any significant, systematic dependence of either the \HI\ or \CIV\ absorption on transverse distance (over the range $\approx 50-250$~pkpc). This is true for both the full sample and the isolated LAEs. We argued that the lack of a trend with impact parameter for the isolated LAEs could in part be due to fact that galaxies near the edge of the FoV may belong to groups whose members reside outside it.

\item  Comparing subsamples of isolated LAEs with median redshifts differing by $\Delta z\approx 0.4$, we do not find any evidence for redshift evolution (Fig.~\ref{fig:redshiftevolution}).

\item Focusing on the isolated LAEs, we do not detect any significant dependence of the absorption on the luminosity (Fig.~\ref{fig:lumdependence}), full width at half maximum (Fig.~\ref{fig:fwhmdependence}), or equivalent width (Fig.~\ref{fig:ewdependence}) of the \lya\ emission line when we compare subsamples that differ by factors of $\approx 3$, 1.5, and 2 in $L$(\lya), FWHM, and $EW_0$, respectively.

\item The \HI\ absorption at $\lesssim 100\,\text{km}\,\text{s}^{-1}$ from the LAE increases with the SFR, but the \CIV\ absorption does not show any trend with SFR (Figs.~\ref{fig:flux_sfrdependence} and \ref{fig:tau_sfrdependence}). Using simple velocity/time-scale arguments we conclude that the enhanced \HI\ absorption seen for the high-SFR subsample cannot be due to the material entrained by SFR-driven winds. Instead, we favor a scenario in which galaxies surrounded by more cool neutral gas, either because the galaxies are more massive or because the \HI\ fraction is relatively high in their environments, tend to have higher SFRs owing to the readily available fuel for star formation. We do not see any significant trend with SFR for the isolated LAEs, but this may owe to the smaller number of isolated high-SFR LAEs reducing the dynamic range in SFR. Nevertheless, this does suggest the possibility that the trend seen in the full sample may be partly driven by the underlying environmental dependence.

\item Direct comparison of the excess optical depth profiles as a function of velocity for impact parameters of 40--180 pkpc with those of $z\approx 2.3$ LBGs shows that the environments of LBGs are more enhanced in \HI\ than is the case for our LAEs (Fig.~\ref{fig:LBGvsLAE}). We argue that the order of magnitude higher SFRs and halo masses of the LBGs are the most likely cause for the enhanced \HI\ absorption. However, no significant differences are found for \CIV.

\item At a fixed impact parameter, the equivalent widths of the \HI\ absorption for our sample are larger than observed for low-$z$ dwarf/sub-$L_*$ galaxies, but comparable to those of $\approx L_*$ galaxies at low $z$ and $z\approx2.3$ LBGs. However, we find that at a given normalized impact parameter ($\rho/R_{200}$) the LAEs show stronger \lya\ absorption compared to LBGs (Fig.~\ref{fig:compareHI}).

\item A comparison with \CIV\ equivalent width profiles from the literature indicates that quasar-host halos are significantly richer in \CIV\ compared to LAEs (and LBGs). The \CIV\ equivalent widths measured near the LAEs are however comparable to those for low-$z$ dwarf galaxies and $z\approx2.3$ LBGs (Fig.~\ref{fig:compareCIV}).  

\end{itemize}  

This is the second paper reporting results from the high-$z$ MUSEQuBES survey and the first focusing on the properties of the CGM of LAEs. Our results extend studies of the gas around high-$z$ galaxies to much lower galaxy masses than were probed by previous surveys, thus extending observational constraints from rare objects such as quasars and the progenitors of massive galaxies to the ubiquitous but faint progenitors of the galaxies that dominate the cosmic stellar mass density. In future work we plan to extend our analysis from flux statistics to measurements of column densities and line widths. We also intend to study some individual systems in more detail.

\section{Data Availability} 

The data underlying this article are available in ESO (http://archive.eso.org/cms.html) and Keck (https://www2.keck.hawaii.edu/koa/public/koa.php) public archives.

\noindent  
{\it Acknowledgements:} 
We would like to thank the anonymous referee for the useful comments which improved the clarity of the paper. This work is partly funded by Vici grant 639.043.409 from the Dutch Research Council (NWO). SM thanks the Alexander von Humboldt-Stiftung (Germany). SM also thanks Monica Turner, Lorrie Straka, Marijke Segers, and Peter Mitchell for useful discussion. SC gratefully acknowledges support from Swiss National Science Foundation grant PP00P2\_190092 and from the European Research Council (ERC) under the European Union's Horizon 2020 research and innovation programme grant agreement No 864361. NB acknowledges  support from the ANR 3DGasFlows (ANR-17-CE31-0017).

\def\aj{AJ}%
\def\actaa{Acta Astron.}%
\def\araa{ARA\&A}%
\def\apj{ApJ}%
\def\apjl{ApJ}%
\def\apjs{ApJS}%
\def\ao{Appl.~Opt.}%
\def\apss{Ap\&SS}%
\def\aap{A\&A}%
\def\aapr{A\&A~Rev.}%
\def\aaps{A\&AS}%
\def\azh{AZh}%
\def\baas{BAAS}%
\def\bac{Bull. astr. Inst. Czechosl.}%
\def\caa{Chinese Astron. Astrophys.}%
\def\cjaa{Chinese J. Astron. Astrophys.}%
\def\icarus{Icarus}%
\def\jcap{J. Cosmology Astropart. Phys.}%
\def\jrasc{JRASC}%
\def\mnras{MNRAS}%
\def\memras{MmRAS}%
\def\na{New A}%
\def\nar{New A Rev.}%
\def\pasa{PASA}%
\def\pra{Phys.~Rev.~A}%
\def\prb{Phys.~Rev.~B}%
\def\prc{Phys.~Rev.~C}%
\def\prd{Phys.~Rev.~D}%
\def\pre{Phys.~Rev.~E}%
\def\prl{Phys.~Rev.~Lett.}%
\def\pasp{PASP}%
\def\pasj{PASJ}%
\def\qjras{QJRAS}%
\def\rmxaa{Rev. Mexicana Astron. Astrofis.}%
\def\skytel{S\&T}%
\def\solphys{Sol.~Phys.}%
\def\sovast{Soviet~Ast.}%
\def\ssr{Space~Sci.~Rev.}%
\def\zap{ZAp}%
\def\nat{Nature}%
\def\iaucirc{IAU~Circ.}%
\def\aplett{Astrophys.~Lett.}%
\def\apspr{Astrophys.~Space~Phys.~Res.}%
\def\bain{Bull.~Astron.~Inst.~Netherlands}%
\def\fcp{Fund.~Cosmic~Phys.}%
\def\gca{Geochim.~Cosmochim.~Acta}%
\def\grl{Geophys.~Res.~Lett.}%
\def\jcp{J.~Chem.~Phys.}%
\def\jgr{J.~Geophys.~Res.}%
\def\jqsrt{J.~Quant.~Spec.~Radiat.~Transf.}%
\def\memsai{Mem.~Soc.~Astron.~Italiana}%
\def\nphysa{Nucl.~Phys.~A}%
\def\physrep{Phys.~Rep.}%
\def\physscr{Phys.~Scr}%
\def\planss{Planet.~Space~Sci.}%
\def\procspie{Proc.~SPIE}%
\let\astap=\aap
\let\apjlett=\apjl
\let\apjsupp=\apjs
\let\applopt=\ao
\bibliographystyle{mn}
\bibliography{mybib}

\begin{thebibliography}{}
\makeatletter
\relax
\def\mn@urlcharsother{\let\do\@makeother \do\$\do\&\do\#\do\^\do\_\do\%\do\~}
\def\mn@doi{\begingroup\mn@urlcharsother \@ifnextchar [ {\mn@doi@}
  {\mn@doi@[]}}
\def\mn@doi@[#1]#2{\def\@tempa{#1}\ifx\@tempa\@empty \href
  {http://dx.doi.org/#2} {doi:#2}\else \href {http://dx.doi.org/#2} {#1}\fi
  \endgroup}
\def\mn@eprint#1#2{\mn@eprint@#1:#2::\@nil}
\def\mn@eprint@arXiv#1{\href {http://arxiv.org/abs/#1} {{\tt arXiv:#1}}}
\def\mn@eprint@dblp#1{\href {http://dblp.uni-trier.de/rec/bibtex/#1.xml}
  {dblp:#1}}
\def\mn@eprint@#1:#2:#3:#4\@nil{\def\@tempa {#1}\def\@tempb {#2}\def\@tempc
  {#3}\ifx \@tempc \@empty \let \@tempc \@tempb \let \@tempb \@tempa \fi \ifx
  \@tempb \@empty \def\@tempb {arXiv}\fi \@ifundefined
  {mn@eprint@\@tempb}{\@tempb:\@tempc}{\expandafter \expandafter \csname
  mn@eprint@\@tempb\endcsname \expandafter{\@tempc}}}

\bibitem[\protect\citeauthoryear{{Adelberger}, {Shapley}, {Steidel}, {Pettini},
  {Erb}  \& {Reddy}}{{Adelberger} et~al.}{2005}]{Adelberger05}
{Adelberger} K.~L.,  {Shapley} A.~E.,  {Steidel} C.~C.,  {Pettini} M.,  {Erb}
  D.~K.,   {Reddy} N.~A.,  2005, \mn@doi [\apj] {10.1086/431753}, \href
  {http://adsabs.harvard.edu/abs/2005ApJ...629..636A} {629, 636}

\bibitem[\protect\citeauthoryear{{Aguirre}, {Dow-Hygelund}, {Schaye}  \&
  {Theuns}}{{Aguirre} et~al.}{2008}]{Aguirre08}
{Aguirre} A.,  {Dow-Hygelund} C.,  {Schaye} J.,   {Theuns} T.,  2008, \mn@doi
  [\apj] {10.1086/592554}, \href
  {http://ads.iucaa.ernet.in/abs/2008ApJ...689..851A} {689, 851}

\bibitem[\protect\citeauthoryear{{Bacon} et~al.,}{{Bacon}
  et~al.}{2010}]{Bacon10}
{Bacon} R.,  et~al., 2010, in Ground-based and Airborne Instrumentation for
  Astronomy III. p. 773508, \mn@doi{10.1117/12.856027}

\bibitem[\protect\citeauthoryear{{Bacon} et~al.,}{{Bacon}
  et~al.}{2015}]{Bacon15}
{Bacon} R.,  et~al., 2015, \mn@doi [\aap] {10.1051/0004-6361/201425419}, \href
  {http://esoads.eso.org/abs/2015A%26A...575A..75B} {575, A75}

\bibitem[\protect\citeauthoryear{{Bacon} et~al.,}{{Bacon}
  et~al.}{2021}]{Bacon21}
{Bacon} R.,  et~al., 2021, \mn@doi [\aap] {10.1051/0004-6361/202039887}, \href
  {https://ui.adsabs.harvard.edu/abs/2021A&A...647A.107B} {647, A107}

\bibitem[\protect\citeauthoryear{{Behroozi}, {Wechsler}, {Hearin}  \&
  {Conroy}}{{Behroozi} et~al.}{2019}]{Behroozi19}
{Behroozi} P.,  {Wechsler} R.~H.,  {Hearin} A.~P.,   {Conroy} C.,  2019,
  \mn@doi [\mnras] {10.1093/mnras/stz1182}, \href
  {https://ui.adsabs.harvard.edu/abs/2019MNRAS.tmp.1134B} {p.~1134}

\bibitem[\protect\citeauthoryear{{Bielby} et~al.,}{{Bielby}
  et~al.}{2017}]{Bielby17}
{Bielby} R.~M.,  et~al., 2017, \mn@doi [\mnras] {10.1093/mnras/stx1772}, \href
  {https://ui.adsabs.harvard.edu/abs/2017MNRAS.471.2174B} {471, 2174}

\bibitem[\protect\citeauthoryear{{Bielby} et~al.,}{{Bielby}
  et~al.}{2020}]{Bielby20}
{Bielby} R.~M.,  et~al., 2020, \mn@doi [\mnras] {10.1093/mnras/staa546}, \href
  {https://ui.adsabs.harvard.edu/abs/2020MNRAS.493.5336B} {493, 5336}

\bibitem[\protect\citeauthoryear{{Bordoloi} et~al.,}{{Bordoloi}
  et~al.}{2011}]{Bordoloi11}
{Bordoloi} R.,  et~al., 2011, \mn@doi [\apj] {10.1088/0004-637X/743/1/10},
  \href {http://adsabs.harvard.edu/abs/2011ApJ...743...10B} {743, 10}

\bibitem[\protect\citeauthoryear{{Bordoloi} et~al.,}{{Bordoloi}
  et~al.}{2014}]{Bordoloi14b}
{Bordoloi} R.,  et~al., 2014, \mn@doi [\apj] {10.1088/0004-637X/796/2/136},
  \href {http://adsabs.harvard.edu/abs/2014ApJ...796..136B} {796, 136}

\bibitem[\protect\citeauthoryear{{Borisova} et~al.,}{{Borisova}
  et~al.}{2016}]{Borisova16}
{Borisova} E.,  et~al., 2016, \mn@doi [\apj] {10.3847/0004-637X/831/1/39},
  \href {http://esoads.eso.org/abs/2016ApJ...831...39B} {831, 39}

\bibitem[\protect\citeauthoryear{{Borthakur} et~al.,}{{Borthakur}
  et~al.}{2015}]{Borthakur15}
{Borthakur} S.,  et~al., 2015, \mn@doi [\apj] {10.1088/0004-637X/813/1/46},
  \href {http://esoads.eso.org/abs/2015ApJ...813...46B} {813, 46}

\bibitem[\protect\citeauthoryear{{Bouch{\'e}}, {Murphy}, {P{\'e}roux}, {Csabai}
   \& {Wild}}{{Bouch{\'e}} et~al.}{2006}]{Bouche06a}
{Bouch{\'e}} N.,  {Murphy} M.~T.,  {P{\'e}roux} C.,  {Csabai} I.,   {Wild} V.,
  2006, \mn@doi [\mnras] {10.1111/j.1365-2966.2006.10685.x}, \href
  {https://ui.adsabs.harvard.edu/abs/2006MNRAS.371..495B} {371, 495}

\bibitem[\protect\citeauthoryear{{Burchett} et~al.,}{{Burchett}
  et~al.}{2016}]{Burchett16}
{Burchett} J.~N.,  et~al., 2016, \mn@doi [\apj] {10.3847/0004-637X/832/2/124},
  \href {https://ui.adsabs.harvard.edu/abs/2016ApJ...832..124B} {832, 124}

\bibitem[\protect\citeauthoryear{{Cai} et~al.,}{{Cai} et~al.}{2019}]{Cai19}
{Cai} Z.,  et~al., 2019, \mn@doi [\apjs] {10.3847/1538-4365/ab4796}, \href
  {https://ui.adsabs.harvard.edu/abs/2019ApJS..245...23C} {245, 23}

\bibitem[\protect\citeauthoryear{{Cantalupo}, {Arrigoni-Battaia}, {Prochaska},
  {Hennawi}  \& {Madau}}{{Cantalupo} et~al.}{2014}]{Cantalupo14}
{Cantalupo} S.,  {Arrigoni-Battaia} F.,  {Prochaska} J.~X.,  {Hennawi} J.~F.,
  {Madau} P.,  2014, \mn@doi [\nat] {10.1038/nature12898}, \href
  {https://ui.adsabs.harvard.edu/abs/2014Natur.506...63C} {506, 63}

\bibitem[\protect\citeauthoryear{{Cantalupo} et~al.,}{{Cantalupo}
  et~al.}{2019}]{Cantalupo19}
{Cantalupo} S.,  et~al., 2019, \mn@doi [\mnras] {10.1093/mnras/sty3481}, \href
  {https://ui.adsabs.harvard.edu/abs/2019MNRAS.483.5188C} {483, 5188}

\bibitem[\protect\citeauthoryear{{Chabrier}}{{Chabrier}}{2003}]{Chabrier03}
{Chabrier} G.,  2003, \mn@doi [\pasp] {10.1086/376392}, \href
  {https://ui.adsabs.harvard.edu/abs/2003PASP..115..763C} {115, 763}

\bibitem[\protect\citeauthoryear{{Chen}, {Helsby}, {Gauthier}, {Shectman},
  {Thompson}  \& {Tinker}}{{Chen} et~al.}{2010}]{Chen10b}
{Chen} H.-W.,  {Helsby} J.~E.,  {Gauthier} J.-R.,  {Shectman} S.~A.,
  {Thompson} I.~B.,   {Tinker} J.~L.,  2010, \mn@doi [\apj]
  {10.1088/0004-637X/714/2/1521}, \href
  {http://adsabs.harvard.edu/abs/2010ApJ...714.1521C} {714, 1521}

\bibitem[\protect\citeauthoryear{{Chen} et~al.,}{{Chen} et~al.}{2020}]{YChen20}
{Chen} Y.,  et~al., 2020, \mn@doi [\mnras] {10.1093/mnras/staa2808}, \href
  {https://ui.adsabs.harvard.edu/abs/2020MNRAS.tmp.2723C} {}

\bibitem[\protect\citeauthoryear{{Cowie} \& {Songaila}}{{Cowie} \&
  {Songaila}}{1998}]{Cowie98}
{Cowie} L.~L.,  {Songaila} A.,  1998, \mn@doi [\nat] {10.1038/27845}, \href
  {http://adsabs.harvard.edu/abs/1998Natur.394...44C} {394, 44}

\bibitem[\protect\citeauthoryear{{Crighton} et~al.,}{{Crighton}
  et~al.}{2011}]{Crighton11}
{Crighton} N.~H.~M.,  et~al., 2011, \mn@doi [\mnras]
  {10.1111/j.1365-2966.2011.17247.x}, \href
  {http://adsabs.harvard.edu/abs/2011MNRAS.414...28C} {414, 28}

\bibitem[\protect\citeauthoryear{{D{\'{\i}}az}, {Ryan-Weber}, {Cooke}, {Koyama}
   \& {Ouchi}}{{D{\'{\i}}az} et~al.}{2015}]{Diaz15}
{D{\'{\i}}az} C.~G.,  {Ryan-Weber} E.~V.,  {Cooke} J.,  {Koyama} Y.,   {Ouchi}
  M.,  2015, \mn@doi [\mnras] {10.1093/mnras/stu2738}, \href
  {http://esoads.eso.org/abs/2015MNRAS.448.1240D} {448, 1240}

\bibitem[\protect\citeauthoryear{{D{\'\i}az}, {Ryan-Weber}, {Karman}, {Caputi},
  {Salvadori}, {Crighton}, {Ouchi}  \& {Vanzella}}{{D{\'\i}az}
  et~al.}{2021}]{Diaz21}
{D{\'\i}az} C.~G.,  {Ryan-Weber} E.~V.,  {Karman} W.,  {Caputi} K.~I.,
  {Salvadori} S.,  {Crighton} N.~H.,  {Ouchi} M.,   {Vanzella} E.,  2021,
  \mn@doi [\mnras] {10.1093/mnras/staa3129}, \href
  {https://ui.adsabs.harvard.edu/abs/2021MNRAS.502.2645D} {502, 2645}

\bibitem[\protect\citeauthoryear{{Dutta} et~al.,}{{Dutta}
  et~al.}{2020}]{Dutta20}
{Dutta} R.,  et~al., 2020, \mn@doi [\mnras] {10.1093/mnras/staa3147}, \href
  {https://ui.adsabs.harvard.edu/abs/2020MNRAS.499.5022D} {499, 5022}

\bibitem[\protect\citeauthoryear{{Fossati} et~al.,}{{Fossati}
  et~al.}{2019}]{Fossati19}
{Fossati} M.,  et~al., 2019, \mn@doi [\mnras] {10.1093/mnras/stz2693}, \href
  {https://ui.adsabs.harvard.edu/abs/2019MNRAS.490.1451F} {490, 1451}

\bibitem[\protect\citeauthoryear{{Fumagalli}, {Cantalupo}, {Dekel}, {Morris},
  {O'Meara}, {Prochaska}  \& {Theuns}}{{Fumagalli} et~al.}{2016}]{Fumagalli16}
{Fumagalli} M.,  {Cantalupo} S.,  {Dekel} A.,  {Morris} S.~L.,  {O'Meara}
  J.~M.,  {Prochaska} J.~X.,   {Theuns} T.,  2016, \mn@doi [\mnras]
  {10.1093/mnras/stw1782}, \href
  {https://ui.adsabs.harvard.edu/abs/2016MNRAS.462.1978F} {462, 1978}

\bibitem[\protect\citeauthoryear{{Heckman}, {Alexandroff}, {Borthakur},
  {Overzier}  \& {Leitherer}}{{Heckman} et~al.}{2015}]{Heckman15}
{Heckman} T.~M.,  {Alexandroff} R.~M.,  {Borthakur} S.,  {Overzier} R.,
  {Leitherer} C.,  2015, \mn@doi [\apj] {10.1088/0004-637X/809/2/147}, \href
  {https://ui.adsabs.harvard.edu/abs/2015ApJ...809..147H} {809, 147}

\bibitem[\protect\citeauthoryear{{Inami} et~al.,}{{Inami}
  et~al.}{2017}]{Inami17}
{Inami} H.,  et~al., 2017, \mn@doi [\aap] {10.1051/0004-6361/201731195}, \href
  {http://esoads.eso.org/abs/2017A%26A...608A...2I} {608, A2}

\bibitem[\protect\citeauthoryear{{Johnson}, {Chen}  \& {Mulchaey}}{{Johnson}
  et~al.}{2015}]{Johnson15}
{Johnson} S.~D.,  {Chen} H.-W.,   {Mulchaey} J.~S.,  2015, \mn@doi [\mnras]
  {10.1093/mnras/stv553}, \href {http://esoads.eso.org/abs/2015MNRAS.449.3263J}
  {449, 3263}

\bibitem[\protect\citeauthoryear{{Johnson}, {Chen}, {Mulchaey}, {Schaye}  \&
  {Straka}}{{Johnson} et~al.}{2017}]{Johnson17}
{Johnson} S.~D.,  {Chen} H.-W.,  {Mulchaey} J.~S.,  {Schaye} J.,   {Straka}
  L.~A.,  2017, \mn@doi [\apjl] {10.3847/2041-8213/aa9370}, \href
  {https://ui.adsabs.harvard.edu/abs/2017ApJ...850L..10J} {850, L10}

\bibitem[\protect\citeauthoryear{{Keeney} et~al.,}{{Keeney}
  et~al.}{2017}]{Keeney17}
{Keeney} B.~A.,  et~al., 2017, \mn@doi [\apjs] {10.3847/1538-4365/aa6b59},
  \href {http://esoads.eso.org/abs/2017ApJS..230....6K} {230, 6}

\bibitem[\protect\citeauthoryear{{Kennicutt}}{{Kennicutt}}{1998}]{Kennicutt98}
{Kennicutt} Jr. R.~C.,  1998, \mn@doi [\araa] {10.1146/annurev.astro.36.1.189},
  \href {http://adsabs.harvard.edu/abs/1998ARA%26A..36..189K} {36, 189}

\bibitem[\protect\citeauthoryear{{Leclercq} et~al.,}{{Leclercq}
  et~al.}{2017}]{Leclercq17}
{Leclercq} F.,  et~al., 2017, \mn@doi [\aap] {10.1051/0004-6361/201731480},
  \href {http://esoads.eso.org/abs/2017A%26A...608A...8L} {608, A8}

\bibitem[\protect\citeauthoryear{{Lofthouse} et~al.,}{{Lofthouse}
  et~al.}{2020}]{Lofthouse20}
{Lofthouse} E.~K.,  et~al., 2020, \mn@doi [\mnras] {10.1093/mnras/stz3066},
  \href {https://ui.adsabs.harvard.edu/abs/2020MNRAS.491.2057L} {491, 2057}

\bibitem[\protect\citeauthoryear{{Lundgren} et~al.,}{{Lundgren}
  et~al.}{2009}]{Lundgren09}
{Lundgren} B.~F.,  et~al., 2009, \mn@doi [\apj] {10.1088/0004-637X/698/1/819},
  \href {https://ui.adsabs.harvard.edu/abs/2009ApJ...698..819L} {698, 819}

\bibitem[\protect\citeauthoryear{{Mackenzie} et~al.,}{{Mackenzie}
  et~al.}{2019}]{Mackenzie19}
{Mackenzie} R.,  et~al., 2019, \mn@doi [\mnras] {10.1093/mnras/stz1501}, \href
  {https://ui.adsabs.harvard.edu/abs/2019MNRAS.487.5070M} {487, 5070}

\bibitem[\protect\citeauthoryear{{Marino} et~al.,}{{Marino}
  et~al.}{2018}]{Marino18}
{Marino} R.~A.,  et~al., 2018, \mn@doi [\apj] {10.3847/1538-4357/aab6aa}, \href
  {http://esoads.eso.org/abs/2018ApJ...859...53M} {859, 53}

\bibitem[\protect\citeauthoryear{{Momose} et~al.,}{{Momose}
  et~al.}{2021}]{Momose21b}
{Momose} R.,  et~al., 2021, \mn@doi [\apj] {10.3847/1538-4357/abd2af}, \href
  {https://ui.adsabs.harvard.edu/abs/2021ApJ...909..117M} {909, 117}

\bibitem[\protect\citeauthoryear{{Moster}, {Naab}  \& {White}}{{Moster}
  et~al.}{2013}]{Moster13}
{Moster} B.~P.,  {Naab} T.,   {White} S.~D.~M.,  2013, \mn@doi [\mnras]
  {10.1093/mnras/sts261}, \href {http://esoads.eso.org/abs/2013MNRAS.428.3121M}
  {428, 3121}

\bibitem[\protect\citeauthoryear{{Mukae} et~al.,}{{Mukae}
  et~al.}{2017}]{Mukae17}
{Mukae} S.,  et~al., 2017, \mn@doi [\apj] {10.3847/1538-4357/835/2/281}, \href
  {https://ui.adsabs.harvard.edu/abs/2017ApJ...835..281M} {835, 281}

\bibitem[\protect\citeauthoryear{{Murphy}, {Kacprzak}, {Savorgnan}  \&
  {Carswell}}{{Murphy} et~al.}{2019}]{Murphy19}
{Murphy} M.~T.,  {Kacprzak} G.~G.,  {Savorgnan} G.~A.~D.,   {Carswell} R.~F.,
  2019, \mn@doi [\mnras] {10.1093/mnras/sty2834}, \href
  {http://esoads.eso.org/abs/2019MNRAS.482.3458M} {482, 3458}

\bibitem[\protect\citeauthoryear{{Murray}, {Quataert}  \& {Thompson}}{{Murray}
  et~al.}{2005}]{Murray05}
{Murray} N.,  {Quataert} E.,   {Thompson} T.~A.,  2005, \mn@doi [\apj]
  {10.1086/426067}, \href
  {https://ui.adsabs.harvard.edu/abs/2005ApJ...618..569M} {618, 569}

\bibitem[\protect\citeauthoryear{{Muzahid}, {Srianand}, {Bergeron}  \&
  {Petitjean}}{{Muzahid} et~al.}{2012}]{Muzahid12a}
{Muzahid} S.,  {Srianand} R.,  {Bergeron} J.,   {Petitjean} P.,  2012, \mn@doi
  [\mnras] {10.1111/j.1365-2966.2011.20324.x}, \href
  {http://adsabs.harvard.edu/abs/2012MNRAS.421..446M} {421, 446}

\bibitem[\protect\citeauthoryear{{Muzahid} et~al.,}{{Muzahid}
  et~al.}{2020}]{Muzahid20}
{Muzahid} S.,  et~al., 2020, \mn@doi [\mnras] {10.1093/mnras/staa1347}, \href
  {https://ui.adsabs.harvard.edu/abs/2020MNRAS.496.1013M} {496, 1013}

\bibitem[\protect\citeauthoryear{{Nielsen}, {Kacprzak}, {Pointon}, {Churchill}
  \& {Murphy}}{{Nielsen} et~al.}{2018}]{Nielsen18}
{Nielsen} N.~M.,  {Kacprzak} G.~G.,  {Pointon} S.~K.,  {Churchill} C.~W.,
  {Murphy} M.~T.,  2018, \mn@doi [\apj] {10.3847/1538-4357/aaedbd}, \href
  {https://ui.adsabs.harvard.edu/abs/2018ApJ...869..153N} {869, 153}

\bibitem[\protect\citeauthoryear{{Nielsen}, {Kacprzak}, {Pointon}, {Murphy},
  {Churchill}  \& {Dav{\'e}}}{{Nielsen} et~al.}{2020}]{Nielsen20}
{Nielsen} N.~M.,  {Kacprzak} G.~G.,  {Pointon} S.~K.,  {Murphy} M.~T.,
  {Churchill} C.~W.,   {Dav{\'e}} R.,  2020, \mn@doi [\apj]
  {10.3847/1538-4357/abc561}, \href
  {https://ui.adsabs.harvard.edu/abs/2020ApJ...904..164N} {904, 164}

\bibitem[\protect\citeauthoryear{{O'Meara} et~al.,}{{O'Meara}
  et~al.}{2015}]{OMeara15}
{O'Meara} J.~M.,  et~al., 2015, \mn@doi [\aj] {10.1088/0004-6256/150/4/111},
  \href {http://esoads.eso.org/abs/2015AJ....150..111O} {150, 111}

\bibitem[\protect\citeauthoryear{{Peeples}, {Werk}, {Tumlinson}, {Oppenheimer},
  {Prochaska}, {Katz}  \& {Weinberg}}{{Peeples} et~al.}{2014}]{Peeples14}
{Peeples} M.~S.,  {Werk} J.~K.,  {Tumlinson} J.,  {Oppenheimer} B.~D.,
  {Prochaska} J.~X.,  {Katz} N.,   {Weinberg} D.~H.,  2014, \mn@doi [\apj]
  {10.1088/0004-637X/786/1/54}, \href
  {http://adsabs.harvard.edu/abs/2014ApJ...786...54P} {786, 54}

\bibitem[\protect\citeauthoryear{{Pointon}, {Nielsen}, {Kacprzak}, {Muzahid},
  {Churchill}  \& {Charlton}}{{Pointon} et~al.}{2017}]{Pointon17}
{Pointon} S.~K.,  {Nielsen} N.~M.,  {Kacprzak} G.~G.,  {Muzahid} S.,
  {Churchill} C.~W.,   {Charlton} J.~C.,  2017, \mn@doi [\apj]
  {10.3847/1538-4357/aa7743}, \href
  {http://esoads.eso.org/abs/2017ApJ...844...23P} {844, 23}

\bibitem[\protect\citeauthoryear{{Prochaska}, {Weiner}, {Chen}, {Mulchaey}  \&
  {Cooksey}}{{Prochaska} et~al.}{2011}]{Prochaska11}
{Prochaska} J.~X.,  {Weiner} B.,  {Chen} H.-W.,  {Mulchaey} J.,   {Cooksey} K.,
   2011, \mn@doi [\apj] {10.1088/0004-637X/740/2/91}, \href
  {http://esoads.eso.org/abs/2011ApJ...740...91P} {740, 91}

\bibitem[\protect\citeauthoryear{{Prochaska}, {Lau}  \& {Hennawi}}{{Prochaska}
  et~al.}{2014}]{Prochaska14}
{Prochaska} J.~X.,  {Lau} M.~W.,   {Hennawi} J.~F.,  2014, \mn@doi [\apj]
  {10.1088/0004-637X/796/2/140}, \href
  {https://ui.adsabs.harvard.edu/abs/2014ApJ...796..140P} {796, 140}

\bibitem[\protect\citeauthoryear{{Rakic}, {Schaye}, {Steidel}  \&
  {Rudie}}{{Rakic} et~al.}{2011}]{Rakic11}
{Rakic} O.,  {Schaye} J.,  {Steidel} C.~C.,   {Rudie} G.~C.,  2011, \mn@doi
  [\mnras] {10.1111/j.1365-2966.2011.18624.x}, \href
  {http://esoads.eso.org/abs/2011MNRAS.414.3265R} {414, 3265}

\bibitem[\protect\citeauthoryear{{Rakic}, {Schaye}, {Steidel}  \&
  {Rudie}}{{Rakic} et~al.}{2012}]{Rakic12}
{Rakic} O.,  {Schaye} J.,  {Steidel} C.~C.,   {Rudie} G.~C.,  2012, \mn@doi
  [\apj] {10.1088/0004-637X/751/2/94}, \href
  {http://esoads.eso.org/abs/2012ApJ...751...94R} {751, 94}

\bibitem[\protect\citeauthoryear{{Rakic}, {Schaye}, {Steidel}, {Booth}, {Dalla
  Vecchia}  \& {Rudie}}{{Rakic} et~al.}{2013}]{Rakic13}
{Rakic} O.,  {Schaye} J.,  {Steidel} C.~C.,  {Booth} C.~M.,  {Dalla Vecchia}
  C.,   {Rudie} G.~C.,  2013, \mn@doi [\mnras] {10.1093/mnras/stt950}, \href
  {http://esoads.eso.org/abs/2013MNRAS.433.3103R} {433, 3103}

\bibitem[\protect\citeauthoryear{{Rudie} et~al.,}{{Rudie}
  et~al.}{2012}]{Rudie12a}
{Rudie} G.~C.,  et~al., 2012, \mn@doi [\apj] {10.1088/0004-637X/750/1/67},
  \href {http://esoads.eso.org/abs/2012ApJ...750...67R} {750, 67}

\bibitem[\protect\citeauthoryear{{Rudie}, {Steidel}, {Pettini}, {Trainor},
  {Strom}, {Hummels}, {Reddy}  \& {Shapley}}{{Rudie} et~al.}{2019}]{Rudie19}
{Rudie} G.~C.,  {Steidel} C.~C.,  {Pettini} M.,  {Trainor} R.~F.,  {Strom}
  A.~L.,  {Hummels} C.~B.,  {Reddy} N.~A.,   {Shapley} A.~E.,  2019, \mn@doi
  [\apj] {10.3847/1538-4357/ab4255}, \href
  {https://ui.adsabs.harvard.edu/abs/2019ApJ...885...61R} {885, 61}

\bibitem[\protect\citeauthoryear{{Schaye}, {Aguirre}, {Kim}, {Theuns}, {Rauch}
  \& {Sargent}}{{Schaye} et~al.}{2003}]{Schaye03}
{Schaye} J.,  {Aguirre} A.,  {Kim} T.-S.,  {Theuns} T.,  {Rauch} M.,
  {Sargent} W.~L.~W.,  2003, \mn@doi [\apj] {10.1086/378044}, \href
  {http://adsabs.harvard.edu/abs/2003ApJ...596..768S} {596, 768}

\bibitem[\protect\citeauthoryear{{Schaye} et~al.,}{{Schaye}
  et~al.}{2015}]{Schaye15}
{Schaye} J.,  et~al., 2015, \mn@doi [\mnras] {10.1093/mnras/stu2058}, \href
  {http://adsabs.harvard.edu/abs/2015MNRAS.446..521S} {446, 521}

\bibitem[\protect\citeauthoryear{{Steidel}, {Erb}, {Shapley}, {Pettini},
  {Reddy}, {Bogosavljevi{\'c}}, {Rudie}  \& {Rakic}}{{Steidel}
  et~al.}{2010}]{Steidel10}
{Steidel} C.~C.,  {Erb} D.~K.,  {Shapley} A.~E.,  {Pettini} M.,  {Reddy} N.,
  {Bogosavljevi{\'c}} M.,  {Rudie} G.~C.,   {Rakic} O.,  2010, \mn@doi [\apj]
  {10.1088/0004-637X/717/1/289}, \href
  {http://esoads.eso.org/abs/2010ApJ...717..289S} {717, 289}

\bibitem[\protect\citeauthoryear{{Steidel} et~al.,}{{Steidel}
  et~al.}{2014}]{Steidel14}
{Steidel} C.~C.,  et~al., 2014, \mn@doi [\apj] {10.1088/0004-637X/795/2/165},
  \href {https://ui.adsabs.harvard.edu/abs/2014ApJ...795..165S} {795, 165}

\bibitem[\protect\citeauthoryear{{Trainor} \& {Steidel}}{{Trainor} \&
  {Steidel}}{2012}]{Trainor12}
{Trainor} R.~F.,  {Steidel} C.~C.,  2012, \mn@doi [\apj]
  {10.1088/0004-637X/752/1/39}, \href
  {https://ui.adsabs.harvard.edu/abs/2012ApJ...752...39T} {752, 39}

\bibitem[\protect\citeauthoryear{{Trainor}, {Steidel}, {Strom}  \&
  {Rudie}}{{Trainor} et~al.}{2015}]{Trainor15}
{Trainor} R.~F.,  {Steidel} C.~C.,  {Strom} A.~L.,   {Rudie} G.~C.,  2015,
  \mn@doi [\apj] {10.1088/0004-637X/809/1/89}, \href
  {https://ui.adsabs.harvard.edu/abs/2015ApJ...809...89T} {809, 89}

\bibitem[\protect\citeauthoryear{{Tumlinson} et~al.,}{{Tumlinson}
  et~al.}{2011}]{Tumlinson11}
{Tumlinson} J.,  et~al., 2011, \mn@doi [\apj] {10.1088/0004-637X/733/2/111},
  \href {http://adsabs.harvard.edu/abs/2011ApJ...733..111T} {733, 111}

\bibitem[\protect\citeauthoryear{{Tumlinson} et~al.,}{{Tumlinson}
  et~al.}{2013}]{Tumlinson13}
{Tumlinson} J.,  et~al., 2013, \mn@doi [\apj] {10.1088/0004-637X/777/1/59},
  \href {http://esoads.eso.org/abs/2013ApJ...777...59T} {777, 59}

\bibitem[\protect\citeauthoryear{{Tumlinson}, {Peeples}  \& {Werk}}{{Tumlinson}
  et~al.}{2017}]{Tumlinson17}
{Tumlinson} J.,  {Peeples} M.~S.,   {Werk} J.~K.,  2017, \mn@doi [\araa]
  {10.1146/annurev-astro-091916-055240}, \href
  {http://esoads.eso.org/abs/2017ARA%26A..55..389T} {55, 389}

\bibitem[\protect\citeauthoryear{{Turner}, {Schaye}, {Steidel}, {Rudie}  \&
  {Strom}}{{Turner} et~al.}{2014}]{Turner14}
{Turner} M.~L.,  {Schaye} J.,  {Steidel} C.~C.,  {Rudie} G.~C.,   {Strom}
  A.~L.,  2014, \mn@doi [\mnras] {10.1093/mnras/stu1801}, \href
  {http://esoads.eso.org/abs/2014MNRAS.445..794T} {445, 794}

\bibitem[\protect\citeauthoryear{{Turner}, {Schaye}, {Steidel}, {Rudie}  \&
  {Strom}}{{Turner} et~al.}{2015}]{Turner15}
{Turner} M.~L.,  {Schaye} J.,  {Steidel} C.~C.,  {Rudie} G.~C.,   {Strom}
  A.~L.,  2015, \mn@doi [\mnras] {10.1093/mnras/stv750}, \href
  {https://ui.adsabs.harvard.edu/abs/2015MNRAS.450.2067T} {450, 2067}

\bibitem[\protect\citeauthoryear{{Turner}, {Schaye}, {Crain}, {Rudie},
  {Steidel}, {Strom}  \& {Theuns}}{{Turner} et~al.}{2017}]{Turner17}
{Turner} M.~L.,  {Schaye} J.,  {Crain} R.~A.,  {Rudie} G.,  {Steidel} C.~C.,
  {Strom} A.,   {Theuns} T.,  2017, \mn@doi [\mnras] {10.1093/mnras/stx1616},
  \href {https://ui.adsabs.harvard.edu/abs/2017MNRAS.471..690T} {471, 690}

\bibitem[\protect\citeauthoryear{{Umehata} et~al.,}{{Umehata}
  et~al.}{2019}]{Umehata19}
{Umehata} H.,  et~al., 2019, \mn@doi [Science] {10.1126/science.aaw5949}, \href
  {https://ui.adsabs.harvard.edu/abs/2019Sci...366...97U} {366, 97}

\bibitem[\protect\citeauthoryear{{Verhamme}, {Orlitov{\'a}}, {Schaerer}  \&
  {Hayes}}{{Verhamme} et~al.}{2015}]{Verhamme15}
{Verhamme} A.,  {Orlitov{\'a}} I.,  {Schaerer} D.,   {Hayes} M.,  2015, \mn@doi
  [\aap] {10.1051/0004-6361/201423978}, \href
  {https://ui.adsabs.harvard.edu/abs/2015A&A...578A...7V} {578, A7}

\bibitem[\protect\citeauthoryear{{Weilbacher} et~al.,}{{Weilbacher}
  et~al.}{2020}]{Weilbacher20}
{Weilbacher} P.~M.,  et~al., 2020, \mn@doi [\aap]
  {10.1051/0004-6361/202037855}, \href
  {https://ui.adsabs.harvard.edu/abs/2020A&A...641A..28W} {641, A28}

\bibitem[\protect\citeauthoryear{{Werk}, {Prochaska}, {Thom}, {Tumlinson},
  {Tripp}, {O'Meara}  \& {Peeples}}{{Werk} et~al.}{2013}]{Werk13}
{Werk} J.~K.,  {Prochaska} J.~X.,  {Thom} C.,  {Tumlinson} J.,  {Tripp} T.~M.,
  {O'Meara} J.~M.,   {Peeples} M.~S.,  2013, \mn@doi [\apjs]
  {10.1088/0067-0049/204/2/17}, \href
  {http://adsabs.harvard.edu/abs/2013ApJS..204...17W} {204, 17}

\bibitem[\protect\citeauthoryear{{Werk} et~al.,}{{Werk} et~al.}{2014}]{Werk14}
{Werk} J.~K.,  et~al., 2014, \mn@doi [\apj] {10.1088/0004-637X/792/1/8}, \href
  {http://adsabs.harvard.edu/abs/2014ApJ...792....8W} {792, 8}

\bibitem[\protect\citeauthoryear{{Wisotzki} et~al.,}{{Wisotzki}
  et~al.}{2016}]{Wisotzki16}
{Wisotzki} L.,  et~al., 2016, \mn@doi [\aap] {10.1051/0004-6361/201527384},
  \href {http://esoads.eso.org/abs/2016A%26A...587A..98W} {587, A98}

\bibitem[\protect\citeauthoryear{{Zahedy}, {Rauch}, {Chen}, {Carswell},
  {Stalder}  \& {Stark}}{{Zahedy} et~al.}{2019}]{Zahedy19b}
{Zahedy} F.~S.,  {Rauch} M.,  {Chen} H.-W.,  {Carswell} R.~F.,  {Stalder} B.,
  {Stark} A.~A.,  2019, \mn@doi [\mnras] {10.1093/mnras/stz861}, \href
  {https://ui.adsabs.harvard.edu/abs/2019MNRAS.486.1392Z} {486, 1392}

\makeatother
\end{thebibliography}
\bsp 

\appendix

\section{Measurements performed on the stacked profiles}  

Table~\ref{tab:HIsummary}: presents the summary of measurements performed on the \HI\ stacks.  \\ 
Table~\ref{tab:CIVsummary}: presents the summary of measurements performed on the \CIV\ stacks. \\

\begin{landscape}
\begin{center}
\begin{threeparttable}[b]
\caption{Measurements performed on the \HI\ stacks}
\begin{tabular}{@{\extracolsep{2pt}}lrrrrcccccc@{}}
\hline
	Sample &   \multicolumn{2}{c}{\voff~(\kms)}  &  \multicolumn{2}{c}{$\sigma_v$~(\kms)} & \multicolumn{2}{c}{$W_r$~(\AA)} & \multicolumn{2}{c}{$W_r$~(\AA)} & \multicolumn{2}{c}{$\log_{\rm 10} N(\HI)/\rm cm^{-2}$}     \\ \cline{2-3} \cline{4-5} \cline{6-7} \cline{8-9} \cline{10-11}
          &            Mean &          Median &            Mean &          Median &             Mean  &           Median  &              Mean &            Median &             Mean  &             Mean \\
&                 &                 &                 &                 &   ($\pm300$~\kms) &  ($\pm300$~\kms)  &  ($\pm500$~\kms)  &   ($\pm500$~\kms) &   ($\pm300$~\kms) &  ($\pm500$~\kms) \\  
(1) & (2) & (3) & (4) & (5) & (6) & (7) & (8) & (9) & (10) & (11) \\ \hline
                     all      &    $  -8\pm 11$ &    $  -2\pm 10$ &    $ 141\pm 11$ &    $ 128\pm 10$ &   $0.709\pm0.085$ &   $1.186\pm0.147$ &   $0.823\pm0.120$ &   $1.323\pm0.182$ &   $14.12\pm 0.05$ &  $14.39\pm 0.06$ \\
        all\_environment\_iso &    $  -5\pm 11$ &    $  -2\pm 11$ &    $  99\pm 11$ &    $  85\pm 10$ &   $0.550\pm0.107$ &   $0.877\pm0.189$ &   $0.573\pm0.140$ &   $0.927\pm0.218$ &   $14.00\pm 0.08$ &  $14.23\pm 0.10$ \\
      all\_environment\_group &    $   2\pm 22$ &    $   9\pm 21$ &    $ 241\pm 24$ &    $ 235\pm 22$ &   $1.119\pm0.122$ &   $1.773\pm0.150$ &   $1.467\pm0.210$ &   $2.279\pm0.305$ &   $14.31\pm 0.05$ &  $14.62\pm 0.06$ \\
             all\_$\rho$\_low$^{a}$ &    $  11\pm 18$ &    $  11\pm 17$ &    $ 117\pm 20$ &    $ 127\pm 18$ &   $0.732\pm0.169$ &   $1.339\pm0.267$ &   $0.839\pm0.243$ &   $1.478\pm0.353$ &   $14.13\pm 0.10$ &  $14.43\pm 0.10$ \\
             all\_$\rho$\_mid &    $ -33\pm 13$ &    $ -40\pm 12$ &    $ 105\pm 13$ &    $  95\pm 12$ &   $0.705\pm0.131$ &   $1.115\pm0.195$ &   $0.800\pm0.211$ &   $1.283\pm0.267$ &   $14.11\pm 0.08$ &  $14.37\pm 0.09$ \\
             all\_$\rho$\_high &    $  18\pm 18$ &    $  20\pm 17$ &    $ 145\pm 18$ &    $ 123\pm 14$ &   $0.725\pm0.137$ &   $1.196\pm0.232$ &   $0.844\pm0.165$ &   $1.311\pm0.259$ &   $14.12\pm 0.08$ &  $14.38\pm 0.09$ \\
             iso\_$\rho$\_low &    $  -4\pm 25$ &    $   6\pm 18$ &    $  94\pm 26$ &    $  77\pm 20$ &   $0.469\pm0.220$ &   $0.877\pm0.404$ &   $0.347\pm0.252$ &   $0.869\pm0.442$ &   $13.94\pm 0.20$ &  $14.20\pm 0.22$ \\
             iso\_$\rho$\_mid &    $ -28\pm 12$ &    $ -28\pm 11$ &    $  93\pm 12$ &    $  95\pm 13$ &   $0.721\pm0.164$ &   $1.120\pm0.226$ &   $0.661\pm0.261$ &   $1.140\pm0.327$ &   $14.12\pm 0.10$ &  $14.32\pm 0.12$ \\
             iso\_$\rho$\_high &    $  49\pm 24$ &    $  43\pm 21$ &    $ 131\pm 24$ &    $  96\pm 17$ &   $0.537\pm0.155$ &   $0.792\pm0.256$ &   $0.723\pm0.185$ &   $0.977\pm0.272$ &   $13.99\pm 0.13$ &  $14.25\pm 0.12$ \\
     iso\_$z_{\rm peak}$\_low$^{b}$ &    $ -10\pm 16$ &    $ -17\pm 16$ &    $ 108\pm 16$ &    $  82\pm 14$ &   $0.544\pm0.127$ &   $0.815\pm0.225$ &   $0.672\pm0.175$ &   $0.970\pm0.271$ &   $14.00\pm 0.10$ &  $14.25\pm 0.12$ \\
     iso\_$z_{\rm peak}$\_high &    $   5\pm 13$ &    $   3\pm 12$ &    $  84\pm 14$ &    $  82\pm 13$ &   $0.546\pm0.163$ &   $0.995\pm0.283$ &   $0.446\pm0.202$ &   $0.923\pm0.325$ &   $14.00\pm 0.13$ &  $14.23\pm 0.15$ \\
iso\_$L({\rm Ly\alpha})$\_low &    $ -24\pm 19$ &    $ -29\pm 18$ &    $ 100\pm 18$ &    $  85\pm 15$ &   $0.448\pm0.156$ &   $0.811\pm0.273$ &   $0.440\pm0.216$ &   $0.845\pm0.330$ &   $13.92\pm 0.15$ &  $14.19\pm 0.17$ \\
iso\_$L({\rm Ly\alpha})$\_high &    $   9\pm 12$ &    $   6\pm 11$ &    $  90\pm 12$ &    $  86\pm 12$ &   $0.645\pm0.129$ &   $0.955\pm0.226$ &   $0.697\pm0.153$ &   $1.047\pm0.244$ &   $14.07\pm 0.09$ &  $14.28\pm 0.10$ \\
               iso\_FWHM\_low &    $ -49\pm 21$ &    $ -47\pm 22$ &    $ 113\pm 20$ &    $ 102\pm 18$ &   $0.518\pm0.156$ &   $0.910\pm0.293$ &   $0.518\pm0.223$ &   $0.927\pm0.354$ &   $13.98\pm 0.13$ &  $14.23\pm 0.17$ \\
               iso\_FWHM\_high &    $  19\pm 11$ &    $  11\pm 12$ &    $  83\pm 11$ &    $  83\pm 11$ &   $0.574\pm0.137$ &   $0.900\pm0.200$ &   $0.632\pm0.157$ &   $1.024\pm0.219$ &   $14.02\pm 0.10$ &  $14.27\pm 0.09$ \\
                all\_SFR\_non-det$^{c}$ &    $   2\pm 12$ &    $  11\pm 11$ &    $ 105\pm 12$ &    $  94\pm 10$ &   $0.660\pm0.114$ &   $0.983\pm0.182$ &   $0.788\pm0.176$ &   $1.121\pm0.255$ &   $14.08\pm 0.08$ &  $14.31\pm 0.10$ \\
                all\_SFR\_low$^{d}$ &    $ -47\pm 18$ &    $ -50\pm 17$ &    $  87\pm 18$ &    $  85\pm 15$ &   $0.592\pm0.155$ &   $1.003\pm0.222$ &   $0.754\pm0.248$ &   $1.268\pm0.322$ &   $14.04\pm 0.11$ &  $14.37\pm 0.11$ \\
                all\_SFR\_high &    $  17\pm 19$ &    $  30\pm 15$ &    $ 154\pm 19$ &    $ 161\pm 16$ &   $1.042\pm0.185$ &   $1.638\pm0.219$ &   $1.080\pm0.274$ &   $1.822\pm0.301$ &   $14.28\pm 0.08$ &  $14.52\pm 0.07$ \\
             all\_$EW_0$\_low &    $ -48\pm 19$ &    $ -26\pm 20$ &    $ 126\pm 18$ &    $ 154\pm 18$ &   $0.801\pm0.173$ &   $1.372\pm0.249$ &   $0.912\pm0.292$ &   $1.660\pm0.356$ &   $14.17\pm 0.09$ &  $14.48\pm 0.09$ \\
             all\_$EW_0$\_high &    $  43\pm 22$ &    $  44\pm 14$ &    $ 149\pm 22$ &    $ 132\pm 14$ &   $0.791\pm0.188$ &   $1.314\pm0.259$ &   $0.909\pm0.244$ &   $1.566\pm0.298$ &   $14.16\pm 0.10$ &  $14.46\pm 0.08$ \\
\hline
\end{tabular}
\label{tab:HIsummary}
Notes-- (1) The sample, as defined in Table~\ref{tab:sample_summary}, used for stacking. (2) \& (3) The \voff\ (line centroids) measured for the mean and median stack profiles, respectively. (4) \& (5) The line widths obtained from single--component Gaussian fits to the mean and median stacked profiles, respectively. A spectral range of $\rm [-500, +500]$~\kms\ is used for Gaussian fitting, unless specified otherwise. The uncertainties in columns 2--5 are the fitting errors determined from the error spectra. The error spectra are obtained from the standard deviations of the mean/median flux distributions of 1000 bootstrap samples. (6) \& (7) Rest-frame equivalent widths of the line measured within $\pm300$~\kms\ of the line centroid of the mean and median stacked profiles, respectively. (8) \& (9) Rest-frame equivalent width of the line measured within $\pm500$~\kms\ of the line centroid of the mean and median stacked profiles, respectively. The uncertainties in columns 6--9 are estimated from the stacked absorption corresponding to the bootstrap samples. (10) \& (11) Column density obtained from the $W_r$ listed in (6) \& (8), respectively, assuming that the line falls in the linear part of the curve of growth. \\ 
$^{a}$ The velocity range used for single--component Gaussian fitting to the mean absorption profile is: $[-200.0, +500.0]$~\kms. \\  
$^{b}$ The velocity range used for single--component Gaussian fitting to the mean absorption profile is: $[-300.0, +300.0]$~\kms. \\ 
$^{c}$ The velocity range used for single--component Gaussian fitting to the mean absorption profile is: $[-250.0, +500.0]$~\kms. \\ 
$^{d}$ The velocity range used for single--component Gaussian fitting to the mean absorption profile is: $[-500.0, +200.0]$~\kms. \\  
\end{threeparttable}
\end{center}
\end{landscape}

\begin{landscape}
\begin{center}
\begin{threeparttable}[b]
\caption{Measurements performed on the \CIV\ stacks}
\begin{tabular}{@{\extracolsep{2pt}}lrrrrcccccc@{}}
\hline
	Sample &   \multicolumn{2}{c}{\voff~(\kms)}  &  \multicolumn{2}{c}{$\sigma_v$~(\kms)} & \multicolumn{2}{c}{$W_r$~(\AA)} & \multicolumn{2}{c}{$W_r$~(\AA)} & \multicolumn{2}{c}{$\log_{\rm 10} N(\CIV)/\rm cm^{-2}$}     \\ \cline{2-3} \cline{4-5} \cline{6-7} \cline{8-9} \cline{10-11}
          &            Mean &          Median &            Mean &          Median &             Mean  &           Median  &              Mean &            Median &             Mean  &             Mean \\
&                 &                 &                 &                 &   ($\pm300$~\kms) &  ($\pm300$~\kms)  &  ($\pm500$~\kms)  &   ($\pm500$~\kms) &   ($\pm300$~\kms) &  ($\pm500$~\kms) \\  
(1) & (2) & (3) & (4) & (5) & (6) & (7) & (8) & (9) & (10) & (11)   \\  \hline
                     all            &    $ -30\pm 22$ &    $ -18\pm 21$ &    $ 134\pm 23$ &    $ 107\pm 19$ &   $0.066\pm0.017$ &   $0.008\pm0.002$ &   $0.070\pm0.019$ &   $0.010\pm0.003$ &   $13.21\pm 0.11$ &  $12.39\pm 0.12$ \\
        all\_environment\_iso$^{a}$ &    $ -31\pm 83$ &    $  15\pm 31$ &    $ 167\pm 83$ &    $ 107\pm 31$ &   $0.062\pm0.022$ &   $0.006\pm0.002$ &   $0.064\pm0.025$ &   $0.007\pm0.003$ &   $13.19\pm 0.16$ &  $12.22\pm 0.17$ \\
        all\_environment\_group$^{b}$ &    $   1\pm 17$ &    $ -27\pm 23$ &    $  71\pm 14$ &    $  69\pm 20$ &   $0.070\pm0.026$ &   $0.018\pm0.010$ &   $0.082\pm0.028$ &   $0.020\pm0.011$ &   $13.24\pm 0.16$ &  $12.69\pm 0.23$ \\
             all\_$\rho$\_low$^{c}$ &    $ -43\pm 18$ &    $  32\pm 35$ &    $  69\pm 17$ &    $  90\pm 43$ &   $0.055\pm0.023$ &   $0.009\pm0.006$ &   $0.061\pm0.025$ &   $0.013\pm0.006$ &   $13.13\pm 0.18$ &  $12.49\pm 0.22$ \\
             all\_$\rho$\_mid &    $-105\pm 89$ &    $   9\pm 25$ &    $ 163\pm 73$ &    $  55\pm 22$ &   $0.048\pm0.020$ &   $0.007\pm0.005$ &   $0.044\pm0.023$ &   $0.008\pm0.005$ &   $13.08\pm 0.18$ &  $12.32\pm 0.26$ \\
             all\_$\rho$\_high$^{d}$ &    $ -29\pm 31$ &    $ -43\pm 37$ &    $ 123\pm 29$ &    $  86\pm 41$ &   $0.089\pm0.042$ &   $0.008\pm0.006$ &   $0.094\pm0.045$ &   $0.010\pm0.006$ &   $13.34\pm 0.20$ &  $12.39\pm 0.27$ \\
             iso\_$\rho$\_low &    $ -66\pm 22$ &    $  43\pm 62$ &    $  77\pm 22$ &    $ 112\pm 62$ &   $0.065\pm0.028$ &   $0.010\pm0.008$ &   $0.065\pm0.030$ &   $0.011\pm0.008$ &   $13.21\pm 0.19$ &  $12.45\pm 0.32$ \\
             iso\_$\rho$\_mid &    $  89\pm 23$ &    $  23\pm 36$ &    $  47\pm 19$ &    $  38\pm 33$ &   $0.022\pm0.022$ &   $0.003\pm0.004$ &   $0.040\pm0.027$ &   $0.005\pm0.005$ &   $12.73\pm 0.43$ &  $12.11\pm 0.40$ \\
             iso\_$\rho$\_high &    $ -86\pm 65$ &    $ -10\pm 23$ &    $ 149\pm 69$ &    $  44\pm 19$ &   $0.074\pm0.054$ &   $0.006\pm0.005$ &   $0.087\pm0.058$ &   $0.005\pm0.006$ &   $13.27\pm 0.31$ &  $12.11\pm 0.46$ \\
     iso\_$z_{\rm peak}$\_low$^{a}$ &    $ -22\pm117$ &    $  46\pm 34$ &    $ 188\pm144$ &    $  83\pm 31$ &   $0.067\pm0.033$ &   $0.005\pm0.003$ &   $0.069\pm0.034$ &   $0.006\pm0.003$ &   $13.22\pm 0.22$ &  $12.15\pm 0.22$ \\
     iso\_$z_{\rm peak}$\_high$^{a}$ &    $ -29\pm 87$ &    $  12\pm 51$ &    $ 131\pm 75$ &    $ 105\pm 48$ &   $0.061\pm0.027$ &   $0.007\pm0.005$ &   $0.061\pm0.036$ &   $0.009\pm0.006$ &   $13.18\pm 0.19$ &  $12.33\pm 0.33$ \\
iso\_$L({\rm Ly\alpha})$\_low$^{e}$ &    $-131\pm207$ &    $   7\pm 36$ &    $ 127\pm142$ &    $  70\pm 33$ &   $0.041\pm0.027$ &   $0.004\pm0.004$ &   $0.037\pm0.032$ &   $0.005\pm0.005$ &   $13.01\pm 0.28$ &  $12.07\pm 0.44$ \\
iso\_$L({\rm Ly\alpha})$\_high$^{f}$ &    $ -11\pm 55$ &    $  23\pm 57$ &    $ 183\pm 73$ &    $  94\pm 70$ &   $0.082\pm0.035$ &   $0.007\pm0.004$ &   $0.091\pm0.036$ &   $0.008\pm0.004$ &   $13.31\pm 0.18$ &  $12.28\pm 0.23$ \\
               iso\_FWHM\_low$^{a}$ &    $  25\pm 48$ &    $  25\pm 32$ &    $  84\pm 45$ &    $  60\pm 29$ &   $0.049\pm0.024$ &   $0.006\pm0.005$ &   $0.064\pm0.032$ &   $0.008\pm0.006$ &   $13.08\pm 0.22$ &  $12.28\pm 0.31$ \\
               iso\_FWHM\_high &    $  21\pm 61$ &    $  20\pm 42$ &    $ 174\pm 71$ &    $ 108\pm 42$ &   $0.082\pm0.038$ &   $0.006\pm0.004$ &   $0.075\pm0.038$ &   $0.006\pm0.004$ &   $13.31\pm 0.20$ &  $12.16\pm 0.29$ \\
                all\_SFR\_non-det &    $ -10\pm 29$ &    $  13\pm 17$ &    $  87\pm 25$ &    $  56\pm 15$ &   $0.036\pm0.015$ &   $0.007\pm0.003$ &   $0.035\pm0.017$ &   $0.008\pm0.003$ &   $12.96\pm 0.18$ &  $12.29\pm 0.17$ \\
                all\_SFR\_low &    $ -12\pm 24$ &    $ -39\pm 53$ &    $  58\pm 19$ &    $  85\pm 37$ &   $0.049\pm0.030$ &   $0.007\pm0.006$ &   $0.047\pm0.033$ &   $0.007\pm0.007$ &   $13.08\pm 0.27$ &  $12.22\pm 0.46$ \\
                all\_SFR\_high &    $  85\pm 63$ &    $ -20\pm 45$ &    $  82\pm 57$ &    $  56\pm 42$ &   $0.034\pm0.017$ &   $0.005\pm0.007$ &   $0.043\pm0.029$ &   $0.009\pm0.008$ &   $12.92\pm 0.22$ &  $12.34\pm 0.39$ \\
             all\_$EW_0$\_low &    $ -28\pm 22$ &    $ -12\pm 24$ &    $  58\pm 21$ &    $  45\pm 22$ &   $0.043\pm0.028$ &   $0.009\pm0.009$ &   $0.056\pm0.038$ &   $0.012\pm0.010$ &   $13.03\pm 0.28$ &  $12.48\pm 0.37$ \\
             all\_$EW_0$\_high &    $  82\pm 22$ &    $   5\pm 60$ &    $  53\pm 17$ &    $ 101\pm 45$ &   $0.030\pm0.019$ &   $0.004\pm0.003$ &   $0.033\pm0.021$ &   $0.005\pm0.004$ &   $12.87\pm 0.27$ &  $12.06\pm 0.41$ \\
\hline
\end{tabular}
\label{tab:CIVsummary}
Notes--  The same as Table~\ref{tab:HIsummary} but for \CIV. A velocity range $\rm [-250, +250]$~\kms\ is used for single--component Gaussian fits, unless specified otherwise. \\ 
$^{a}$ The velocity range used for single--component Gaussian fitting to the mean absorption profile is: $[-150.0, +250.0]$~\kms. \\  
$^{b}$ The velocity range used for single--component Gaussian fitting to the median absorption profile is: $[-150.0, +250.0]$~\kms.  \\   
$^{c}$ The velocity range used for single--component Gaussian fitting to the median absorption profile is: $[-150.0, +200.0]$~\kms.  \\  
$^{d}$ The velocity range used for single--component Gaussian fitting to the median absorption profile is: $[-200.0, +200.0]$~\kms.  \\ 
$^{e}$ The velocity range used for single--component Gaussian fitting to the mean absorption profile is: $[-200.0, +250.0]$~\kms.  \\   
$^{f}$ The velocity range used for single--component Gaussian fitting to the median absorption profile is: $[-150.0, +150.0]$~\kms.  \\ 
\end{threeparttable}
\end{center}
\end{landscape}


\section{Additional figures (online only)}

\begin{figure*}
\centerline{\vbox{
\centerline{\hbox{ 
\includegraphics[trim=0.0cm 0.0cm 0.0cm 2.5cm,width=0.44\textwidth,angle=00]{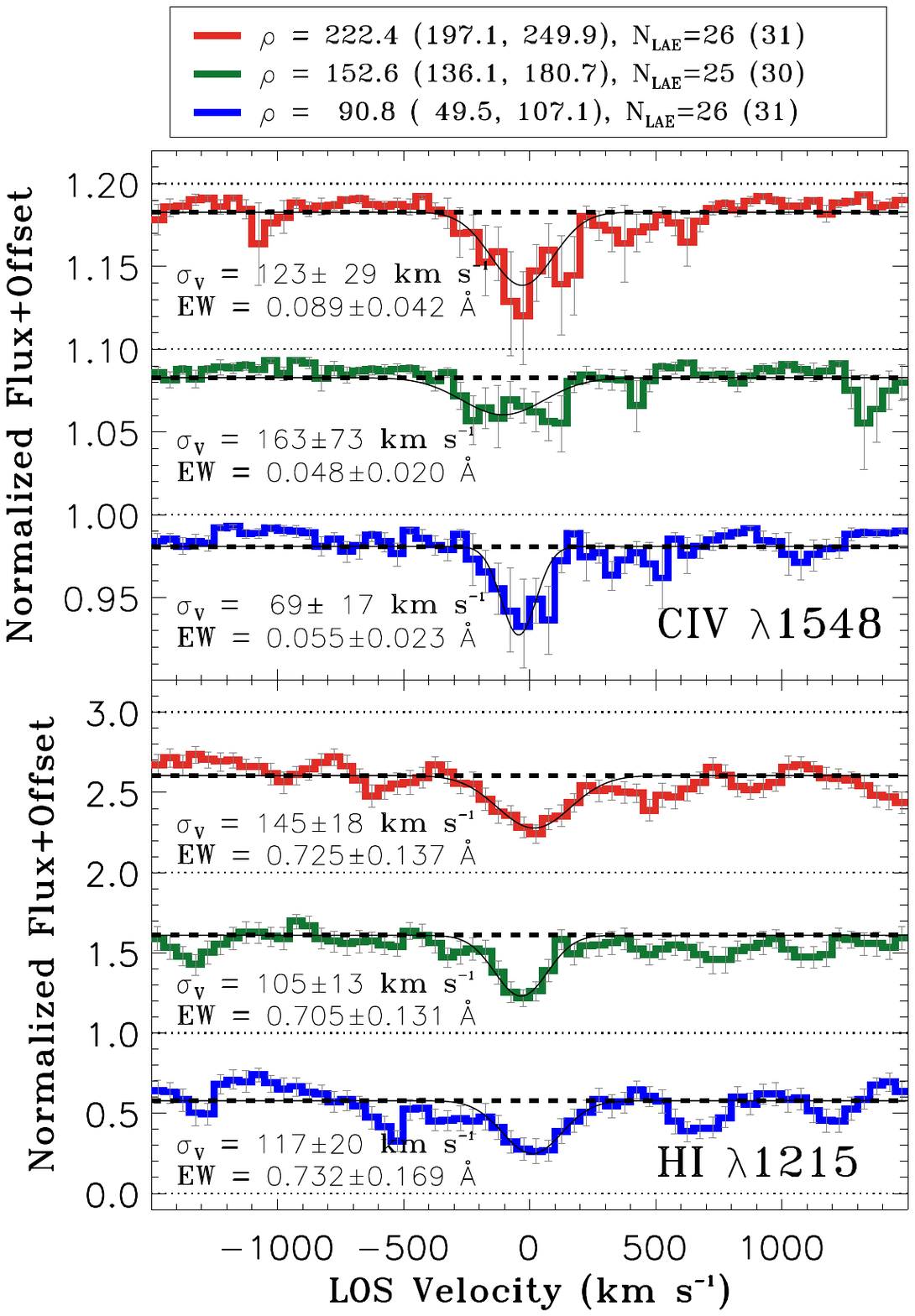} 
\includegraphics[trim=0.0cm 0.0cm 0.0cm 2.5cm,width=0.44\textwidth,angle=00]{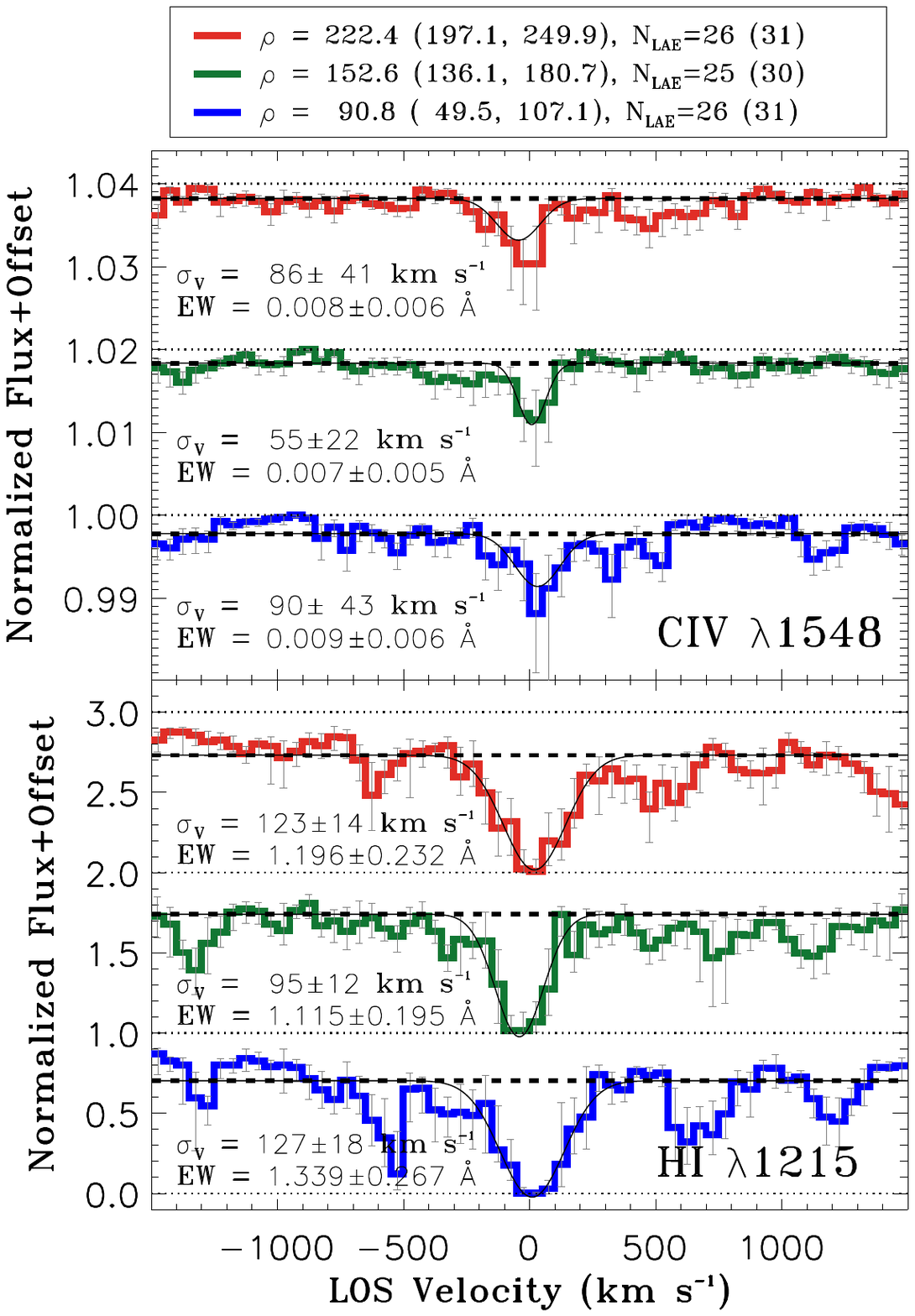} 
}}
}}   
\caption{{\tt Left:} Similar to the left panel of Fig.~\ref{fig:flux_environment} but for the three impact parameter bins: all\_$\rho$\_low, all\_$\rho$\_mid, and all\_$\rho$\_high as defined in Table~\ref{tab:sample_summary}. The median impact parameters and 68 percentile ranges in the parentheses in the legends are for the LAEs contributing to the \HI\ stack. The corresponding values for \CIV\ stack can be found in Table~\ref{tab:sample_summary}. {\tt Right:} Similar to the right panel of Fig.~\ref{fig:flux_environment} but for the same three impact parameter bins as in the left panel. No significant dependence on impact parameter is seen in either \HI\ or \CIV\ absorption.}                 
\label{fig:rhodep_full}     
\end{figure*}

\begin{figure*}
\centerline{\vbox{
\centerline{\hbox{ 
\includegraphics[trim=0.0cm 0.0cm 0.0cm 2.5cm,width=0.44\textwidth,angle=00]{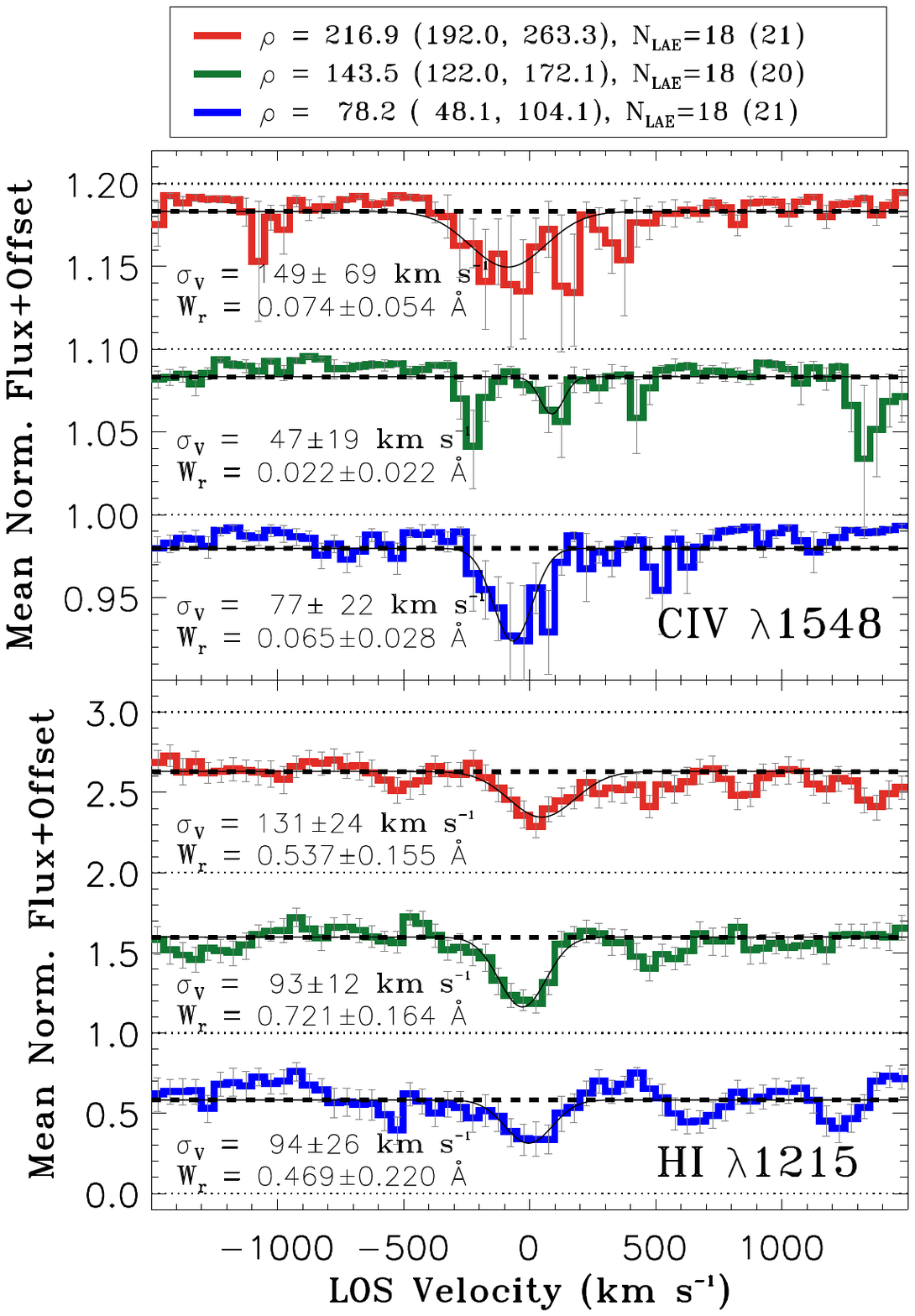} 
\includegraphics[trim=0.0cm 0.0cm 0.0cm 2.5cm,width=0.44\textwidth,angle=00]{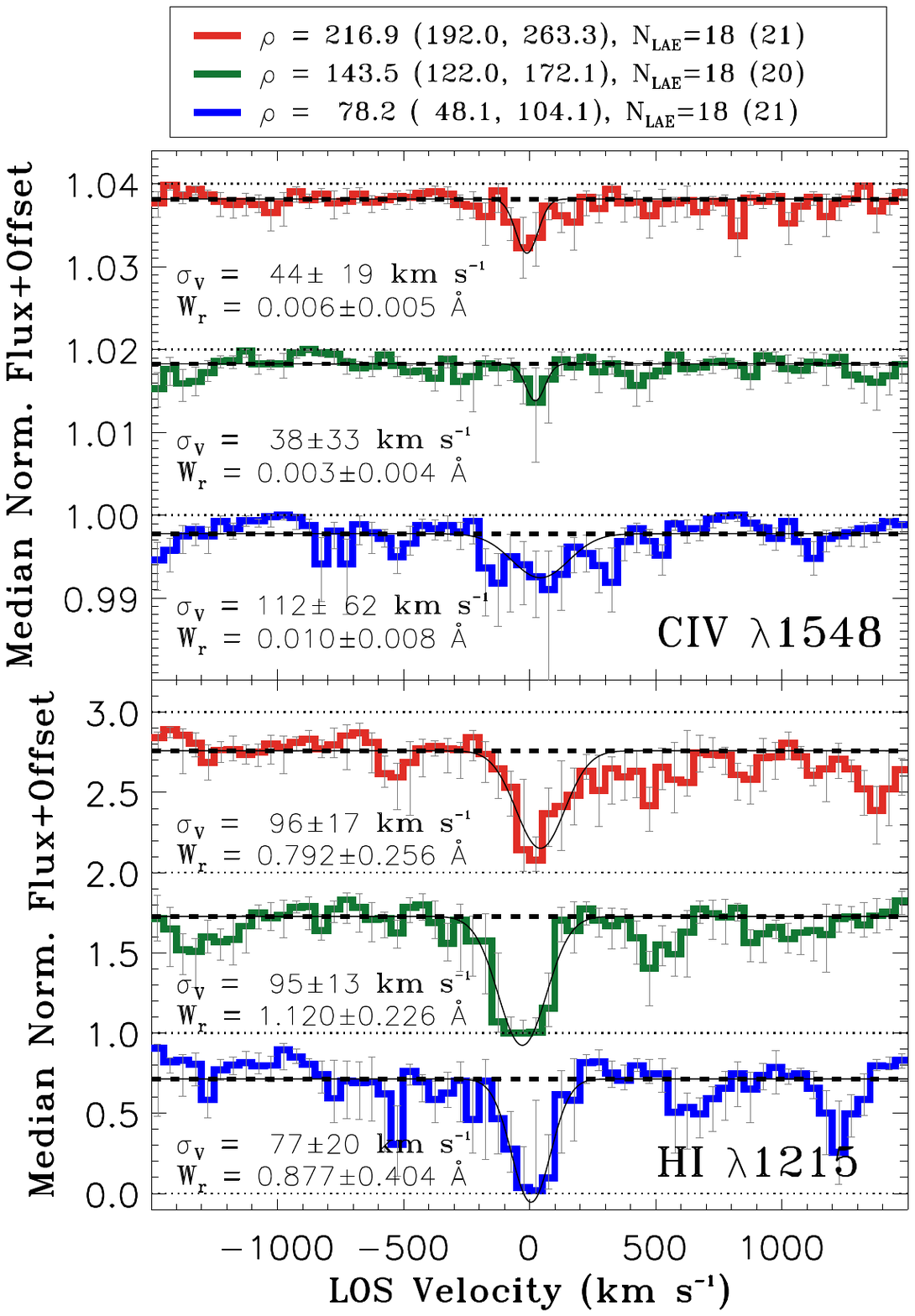} 
}}
}}   
\caption{Same as Fig.~\ref{fig:rhodep_full} but for the iso\_$\rho$\_low, iso\_$\rho$\_mid, and iso\_$\rho$\_high subsamples (see Table~\ref{tab:sample_summary}). No significant dependence on impact parameter is seen in either \HI\ or \CIV\ absorption.}     
\label{fig:rhodep_isolated}     
\end{figure*}

\begin{figure*}
\centerline{\vbox{
\centerline{\hbox{ 
\includegraphics[trim=0.5cm 1.0cm 0.5cm 3.0cm,width=0.44\textwidth,angle=00]{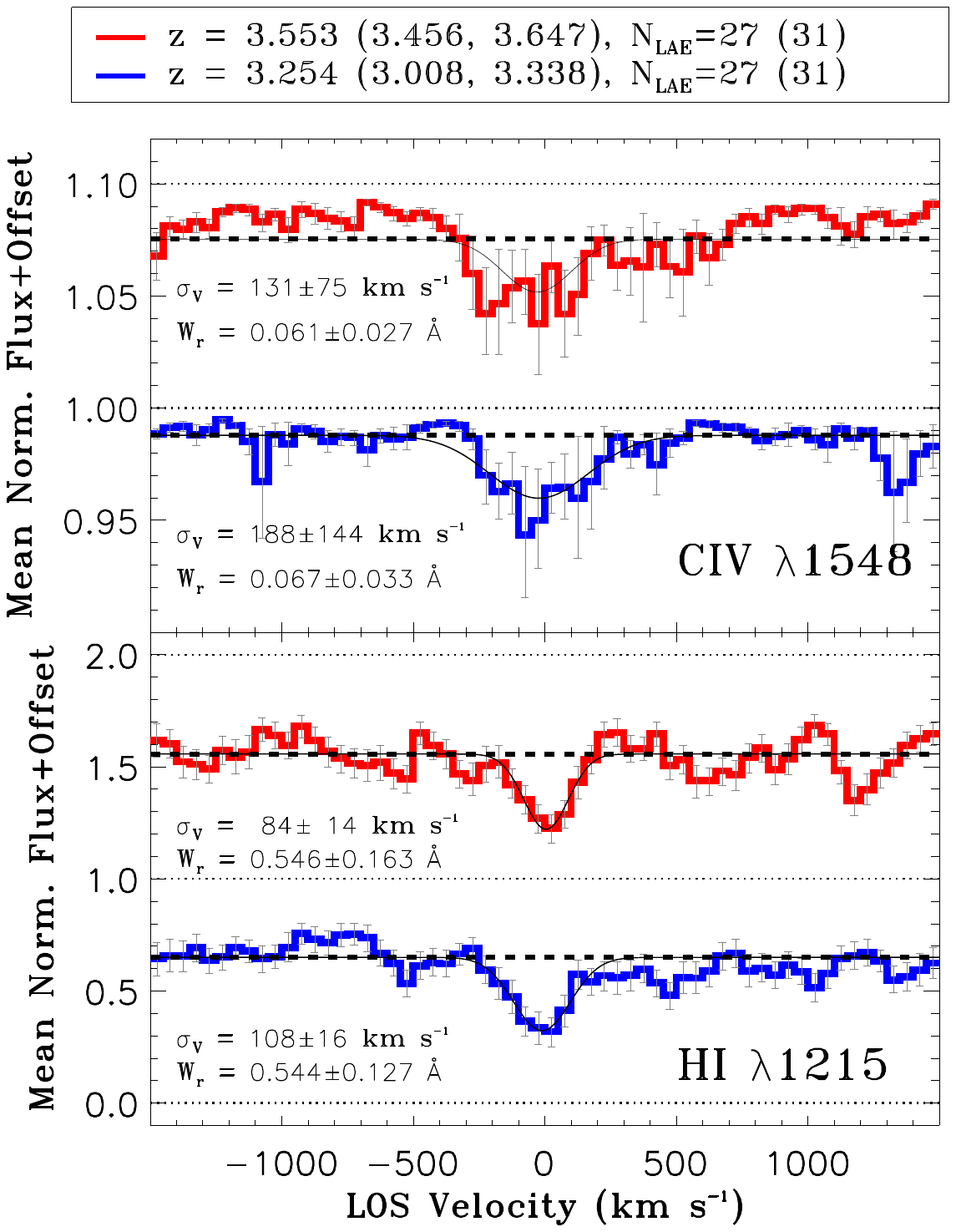} 
\includegraphics[trim=0.5cm 1.0cm 0.5cm 3.0cm,width=0.44\textwidth,angle=00]{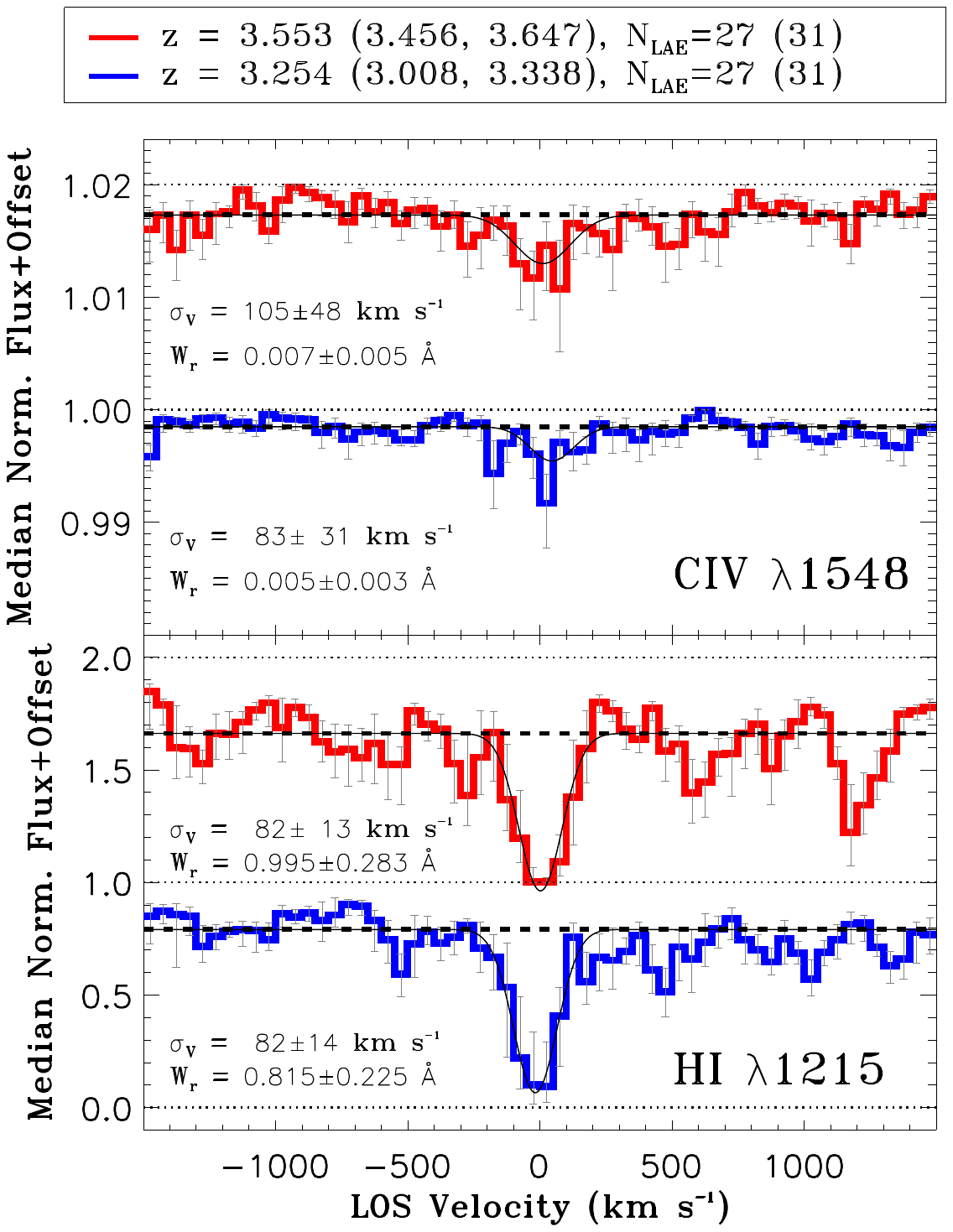} 
}}
}}   
\caption{Similar to Fig.~\ref{fig:flux_environment} but for two different redshift bins (i.e., iso\_\zlae\_low and iso\_\zlae\_high). The median redshifts and 68 percentile ranges in the parentheses in the legends are for the LAEs contributing to the \HI\ stack. The corresponding values for \CIV\ stack can be found in Table~\ref{tab:sample_summary}. No significant dependence on redshift is seen in either \HI\ or \CIV\ absorption. 
}  
\label{fig:redshiftevolution_appendix}     
\end{figure*}

\begin{figure*}
\centerline{\vbox{
\centerline{\hbox{ 
\includegraphics[trim=0.5cm 1.0cm 0.5cm 3.0cm,width=0.44\textwidth,angle=00]{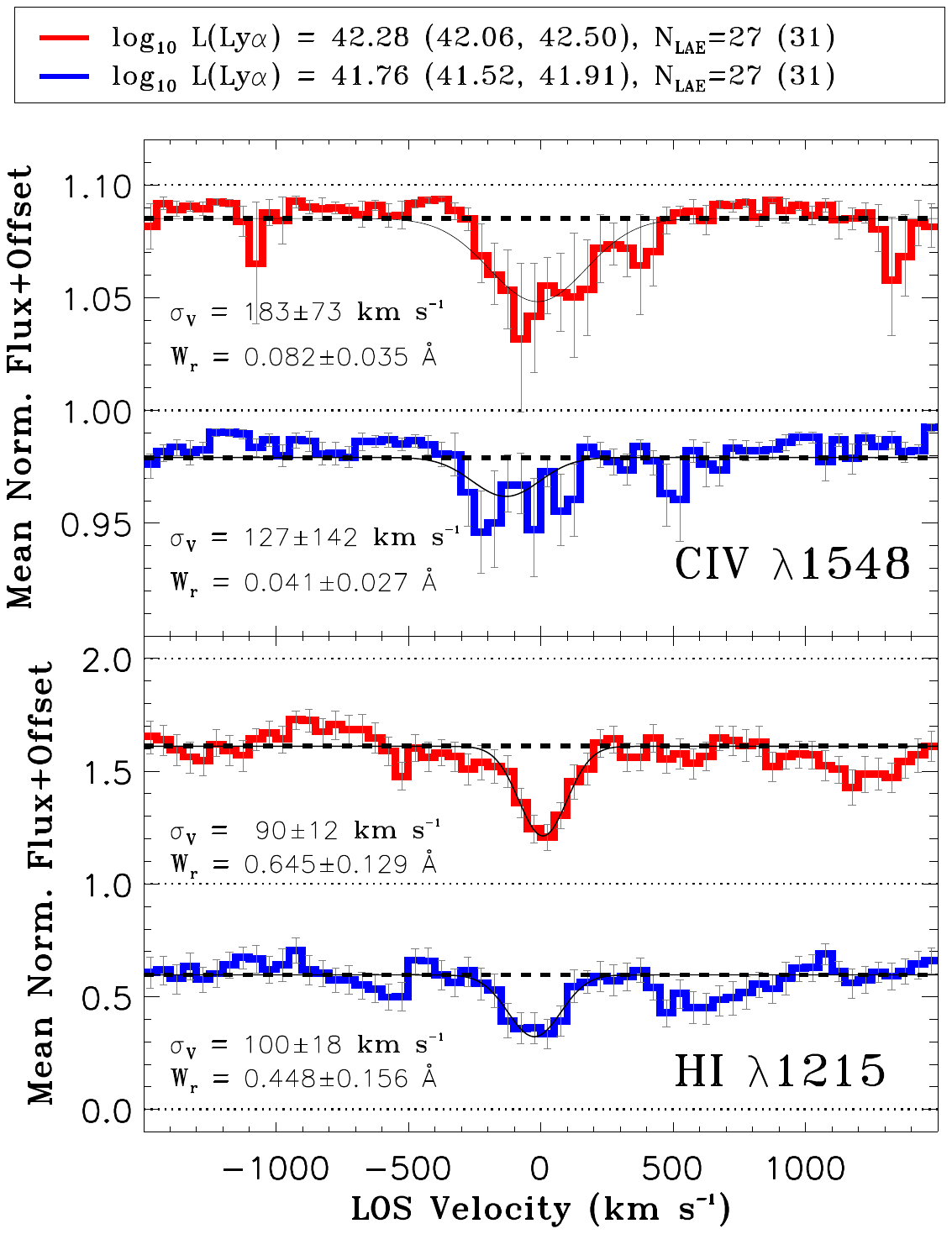} 
\includegraphics[trim=0.5cm 1.0cm 0.5cm 3.0cm,width=0.44\textwidth,angle=00]{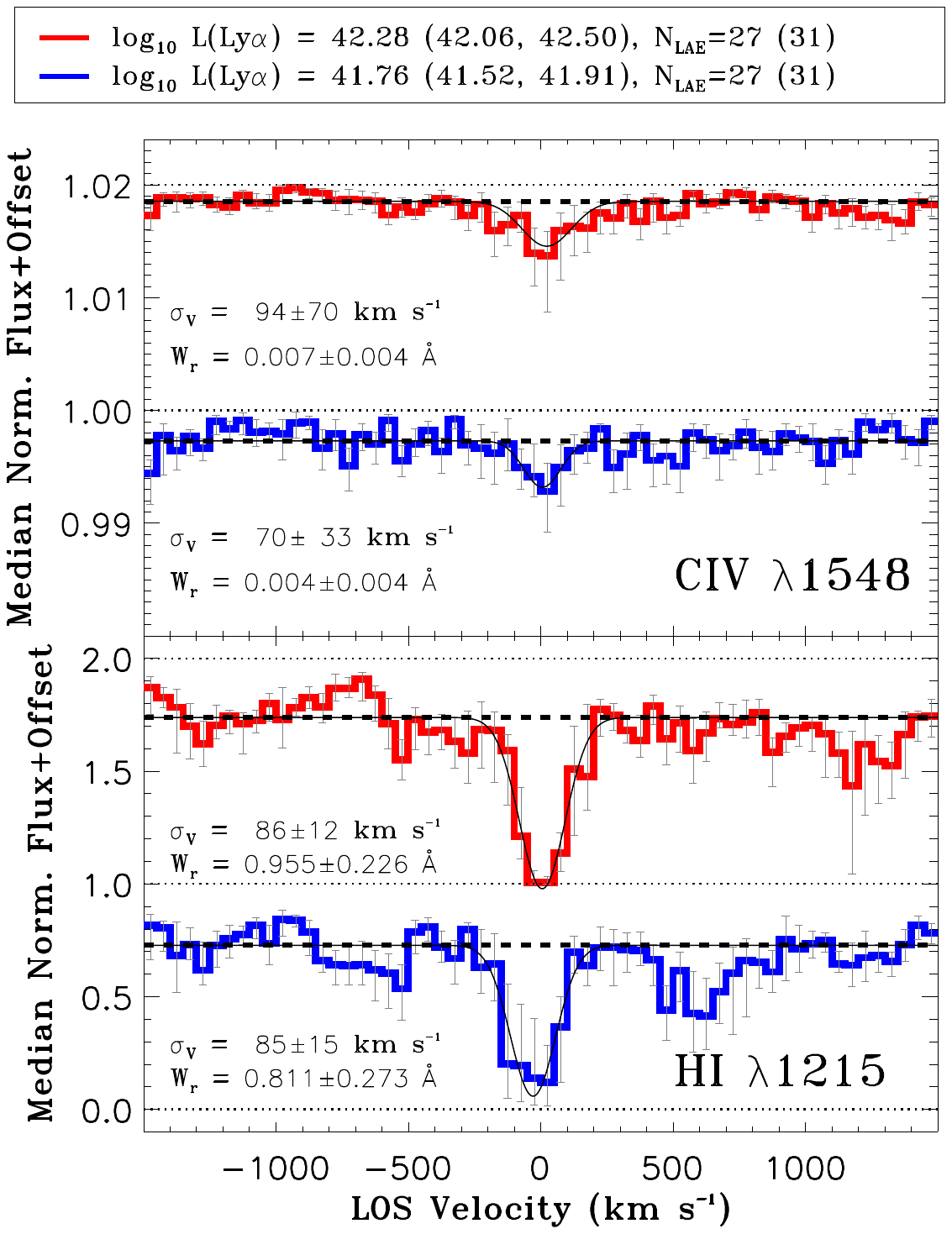} 
}}
}}   
\caption{Similar to Fig.~\ref{fig:flux_environment} but for two different luminosity bins (iso\_$L\rm(Ly\alpha)$\_low and iso\_$L\rm(Ly\alpha)$\_high). The median luminosities and 68 percentile ranges in the parentheses in the legends are for the LAEs contributing to the \HI\ stack. The corresponding values for \CIV\ stack can be found in Table~\ref{tab:sample_summary}. No significant dependence on \lya\ line luminosity is seen in either \HI\ or \CIV\ absorption.}  
\label{fig:lumdependence_appendix}    
\end{figure*}

\begin{figure*}
\centerline{\vbox{
\centerline{\hbox{ 
\includegraphics[trim=0.5cm 1.0cm 0.5cm 3.2cm,width=0.44\textwidth,angle=00]{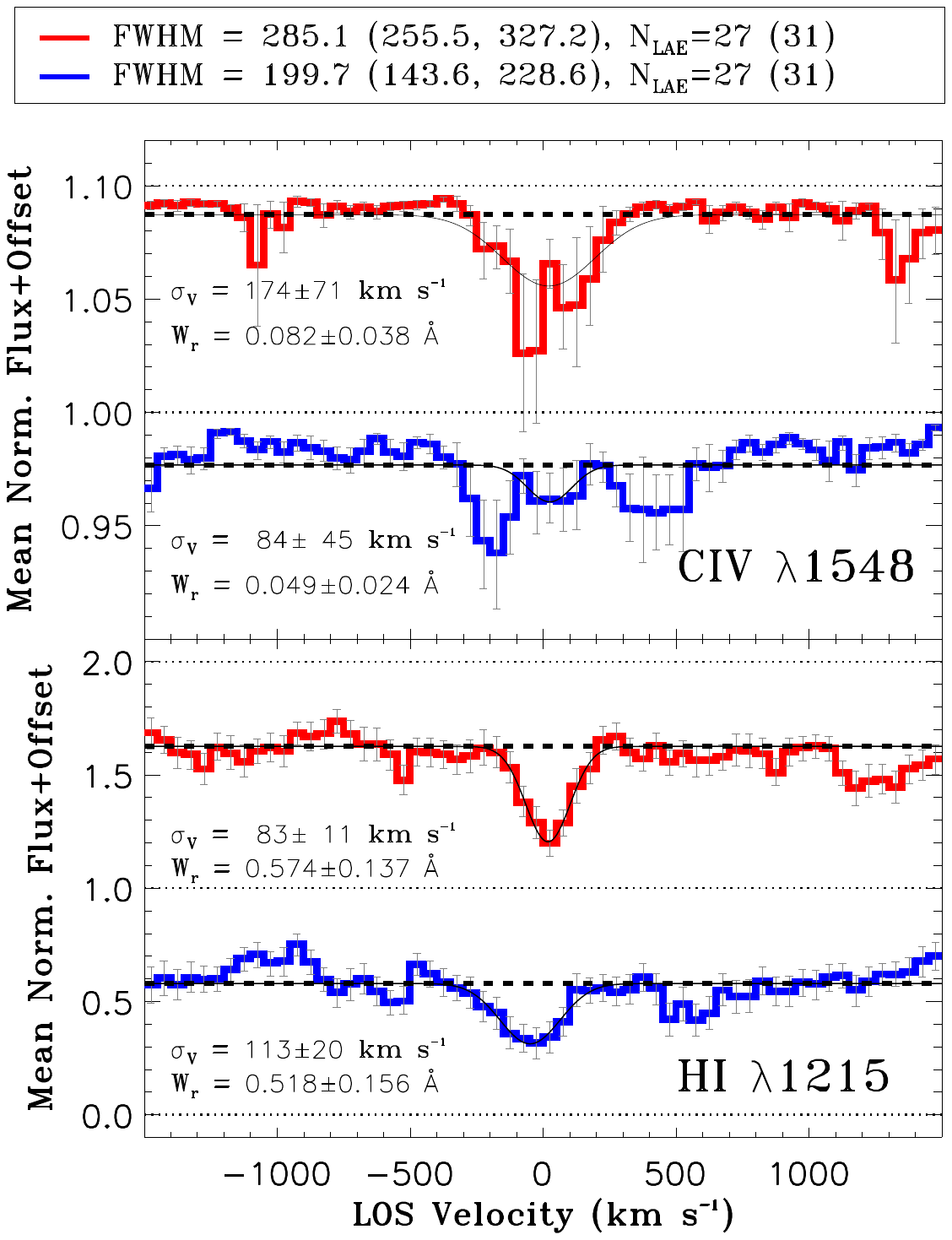} 
\includegraphics[trim=0.5cm 1.0cm 0.5cm 3.2cm,width=0.44\textwidth,angle=00]{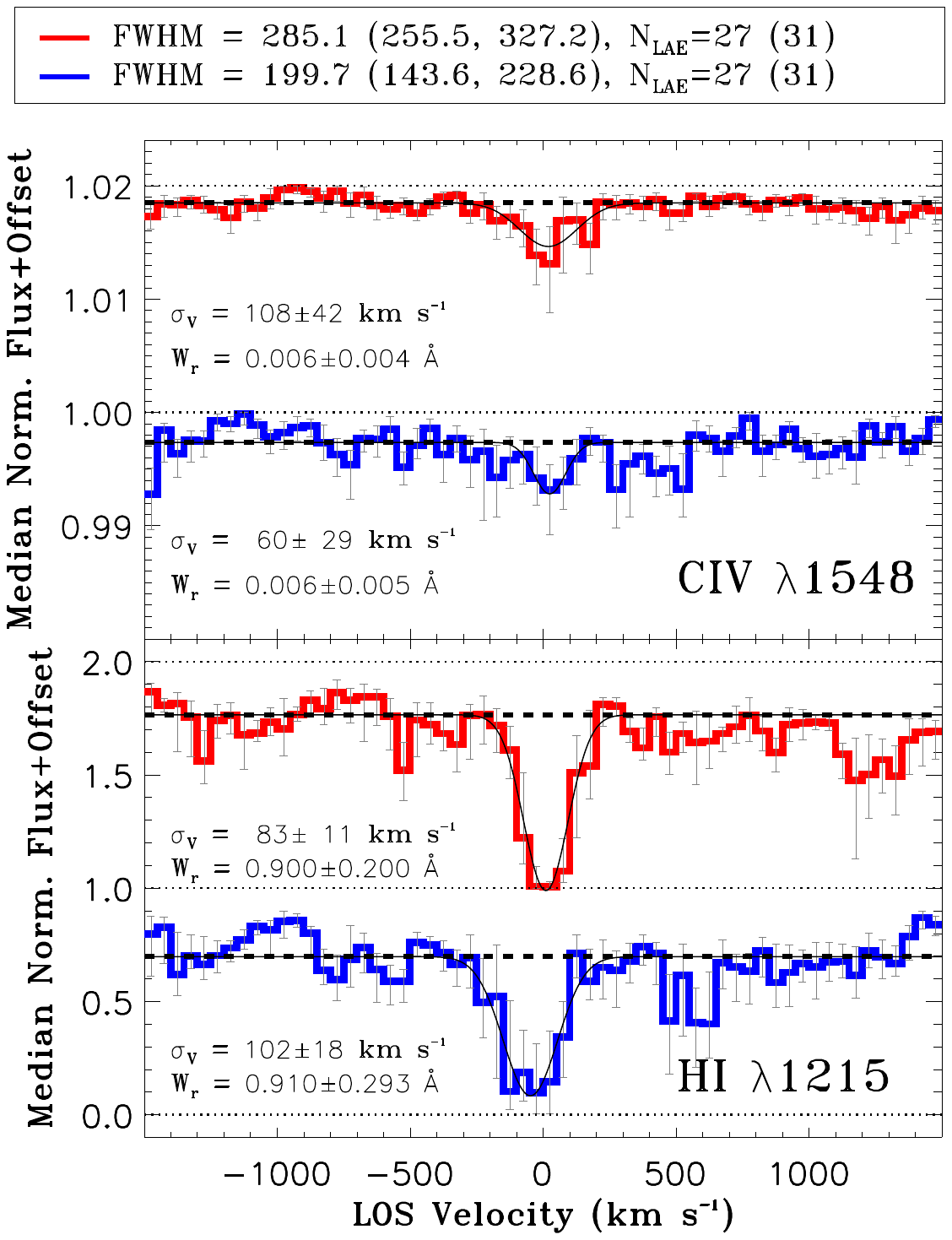} 
}}
}}   
\caption{Similar to Fig.~\ref{fig:flux_environment} but for the two different FWHM bins: iso\_FWHM\_low and iso\_FWHM\_high. The median FWHM values and 68 percentile ranges in the parentheses in the legends are for the LAEs contributing to the \HI\ stack. The corresponding values for \CIV\ stack can be found in Table~\ref{tab:sample_summary}. No significant dependence on FWHM of \lya\ emission line is seen in either \HI\ or \CIV\ absorption. However, both the mean and median stacked \HI\ profiles show somewhat higher widths for the iso\_FWHM\_low sample.}  
\label{fig:fwhmdependence_appendix}     
\end{figure*}

\begin{figure*}
\centerline{\vbox{
\centerline{\hbox{ 
\includegraphics[trim=0.5cm 0.7cm 0.5cm 3.0cm,width=0.44\textwidth,angle=00]{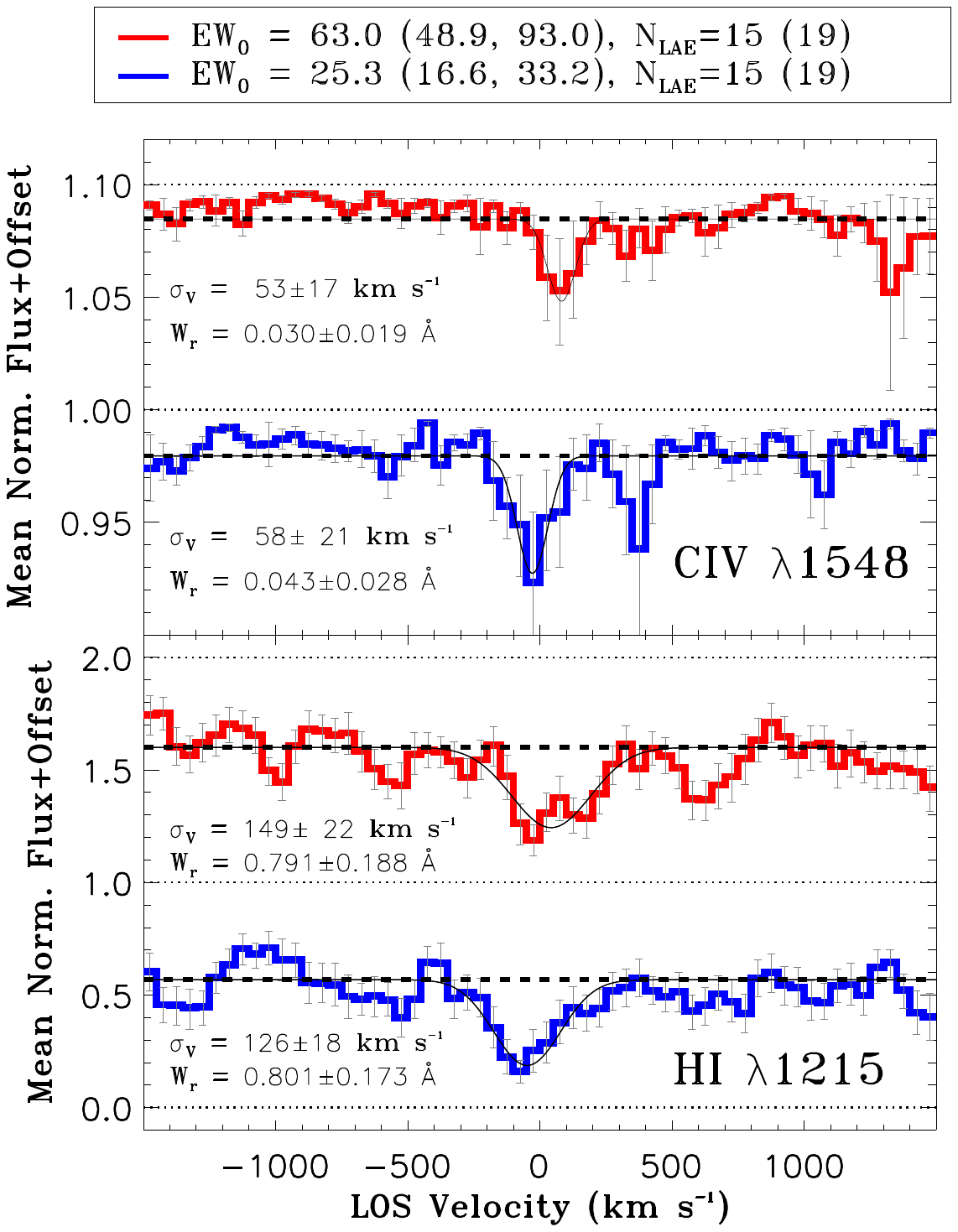} 
\includegraphics[trim=0.5cm 0.7cm 0.5cm 3.0cm,width=0.44\textwidth,angle=00]{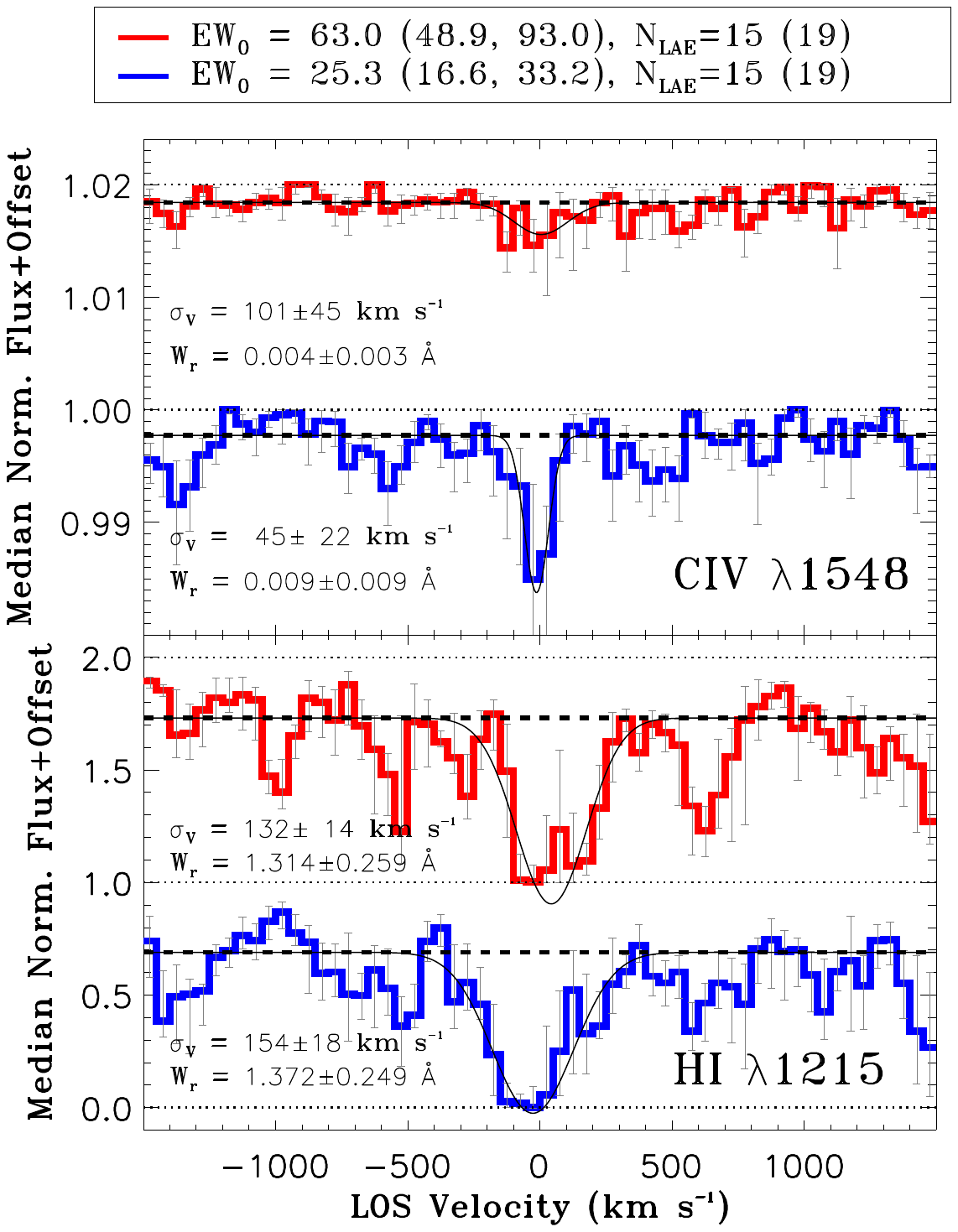} 
}}
}}   
\caption{Similar to Fig.~\ref{fig:flux_environment} but for two different $EW_0$ bins (all\_$EW_0$\_low and all\_$EW_0$\_high). The median  $EW_0$ values and 68 percentile ranges in the parentheses in the legends are for the LAEs contributing to the \HI\ stack. The corresponding values for \CIV\ stack can be found in Table~\ref{tab:sample_summary}. No significant dependence on $EW_0$ of \lya\ emission is seen in either \HI\ or \CIV\ absorption.}  
\label{fig:ewdependence_appendix}    
\end{figure*}

Fig.~\ref{fig:rhodep_full}: shows $\rho$-dependence for the full sample.  \\   
Fig.~\ref{fig:rhodep_isolated}: shows $\rho$-dependence for the `isolated only' sample.  \\    
Fig.~\ref{fig:redshiftevolution_appendix}: shows redshift-dependence for the `isolated only' sample. \\  
Fig.~\ref{fig:lumdependence_appendix}: shows $L$(\lya)-dependence for the `isolated only' sample.  \\ 
Fig.~\ref{fig:fwhmdependence_appendix}: shows FWHM-dependence for the `isolated only' sample.    \\ 
Fig.~\ref{fig:ewdependence_appendix}: shows $EW_0$-dependence for the full sample.  \\

\label{lastpage} 
\end{document}